\documentclass[twocolumn, times]{aastex631} 
\usepackage{graphicx}
\usepackage{booktabs}
\usepackage{hyperref}
\usepackage{amsmath}

\newcommand{\oiii}{[O\,\textsc{iii}]~}
\newcommand{\feiii}{[Fe\,\textsc{iii}]~}
\newcommand{\feii}{[Fe\,\textsc{ii}]~}
\newcommand{\nii}{[N\,\textsc{ii}]~}
\newcommand{\sii}{[S\,\textsc{ii}]~}
\newcommand{\oii}{[O\,\textsc{ii}]~}
\newcommand{\oi}{[O\,\textsc{i}]~}
\newcommand{\orl}{O\,\textsc{ii}~}
\newcommand{\nrl}{N\,\textsc{ii}~}
\newcommand{\hii}{H\,\textsc{ii}~}
\newcommand{\ha}{H$\alpha$~}
\newcommand{\hb}{H$\beta$~}
\newcommand{\hg}{H$\gamma$~}
\newcommand{\hd}{H$\delta$~}
\newcommand{\he}{H$\eta$~}
\newcommand{\hi}{H$\iota$~}
\newcommand{\hk}{H$\kappa$~}
\newcommand{\wave}{$\lambda$}

\begin{document}

\title {SDSS-V LVM: A spatially resolved study of the physical conditions and the chemical abundance discrepancy in the Lagoon Nebula (M\,8)}

\author[0009-0000-3962-103X]{Amrita Singh} 
\newcommand{\UChile}{\affiliation{Departamento de Astronom\'{i}a, Universidad de Chile, Camino del Observatorio 1515, Las Condes, Santiago, Chile}} 
\UChile

\author[0000-0003-4218-3944]{Guillermo A. Blanc}
\affiliation{Observatories of the Carnegie Institution for Science, 813 Santa Barbara Street, Pasadena, CA 91101, USA}
\UChile

\author[0000-0002-5320-2568]{Nimisha Kumari}
\newcommand{\stsci}{\affiliation{Space Telescope Science Institute, 3700 San Martin Drive, Baltimore, MD 21218, USA}}
\stsci

\author[0000-0002-6972-6411]{J. E. M\'endez-Delgado}
\affiliation{Instituto de Astronom\'ia, Universidad Nacional Aut\'onoma de M\'exico, Ap. 70-264, 04510 CDMX, M\'exico}

\author[0000-0001-6444-9307]{Sebastián F.\ Sánchez}
\affiliation{Instituto de Astronom\'ia, Universidad Nacional Auton\'oma de M\'exico, A.P. 106, Ensenada 22800, BC, Mexico}
\affiliation{Instituto de Astrof\'\i sica de Canarias, La Laguna, Tenerife, E-38200, Spain}

\author[0000-0001-5801-6724]{Christophe Morisset}
\affiliation{Instituto de Astronomía (IA), Universidad Nacional Autónoma de México, Apdo. Postal 106, C.P., 22800 Ensenada, Baja California, México}
\affiliation{Instituto de Ciencias Físicas, Universidad Nacional Autónoma de México, Av. Universidad s/n, 62210 Cuernavaca, Morelos, México}

\author[0000-0002-8549-4083]{Enrico Congiu}
\affiliation{European Southern Observatory, Alonso de C\'ordova 3107, Casilla 19, Santiago 19001, Chile}

\author[0000-0001-6551-3091]{Kathryn Kreckel}
\newcommand{\heidelberg}{\affiliation{Astronomisches Rechen-Institut, Zentrum f\"{u}r Astronomie der Universit\"{a}t Heidelberg, M\"{o}nchhofstra\ss e 12-14, D-69120 Heidelberg, Germany}}
\heidelberg

\author[0000-0002-1379-4204]{Alexandre Roman-Lopes}
\affiliation{Department of Astronomy, Universidad de La Serena, Av. Raul Bitran 1302, La Serena, Chile}

\author[0000-0002-4755-118X]{Oleg Egorov}
\heidelberg

\author[0000-0002-7339-3170]{Niv Drory}
\affiliation{McDonald Observatory, The University of Texas at Austin, 1 University Station, Austin, TX 78712-0259, USA}

\author[0000-0001-8858-1943]{Ravi Sankrit}
\affiliation{Space Telescope Science Institute, 3700 San Martin Drive, Baltimore, MD 21218, USA}

\author[0000-0002-8931-2398]{Alfredo Mej\'ia-Narv\'aez}
\UChile

\author[0000-0003-2717-8784]{Evgeniya Egorova}
\heidelberg

\author[0000-0002-2262-8240]{Amy M.\ Jones}
\stsci

\author[0000-0002-3601-133X]{Dmitry Bizyaev}
\affiliation{Apache Point Observatory and New Mexico State
University, P.O. Box 59, Sunspot, NM, 88349-0059, USA}

\author[0000-0002-8883-6018]{Natascha Sattler}
\heidelberg 

\author[0000-0002-2368-6469]{Evelyn J.\ Johnston}
\affiliation{Department of Astronomy / Departamento de Astronomía, Universidad de La Serena, La Serena, Chile}

\author[0000-0002-7064-099X]{Dante Minniti}
\affiliation{ Instituto de Astrofísica, Depto. de Ciencias Físicas, Facultad de Ciencias Exactas, Universidad Andres Bello, Av. Fernandez Concha 700, Las Condes, Santiago, Chile}
\affiliation{Vatican Observatory, V-00120 Vatican City State, Italy}

\author[0009-0009-0081-4323]{Rodolfo de J. Zermeño}
\affiliation{Instituto de Astronom\'ia, Universidad Nacional Aut\'onoma de M\'exico, Ap. 70-264, 04510 CDMX, M\'exico}

\author[0000-0003-3526-5052]{Jos\'e G. Fern\'andez-Trincado}
\affiliation{Universidad Cat\'olica del Norte, N\'ucleo UCN en Arqueolog\'ia Gal\'actica, Av. Angamos 0610, Antofagasta, Chile}
\affiliation{ Universidad Cat\'olica del Norte, Departamento de Ingenier\'ia de Sistemas y Computaci\'on, Av. Angamos 0610, Antofagasta, Chile}

\author[0000-0001-9852-1610]{Juna A.\  Kollmeier}
\affiliation{Canadian Institute for Theoretical Astrophysics, University of Toronto, Toronto, ON M5S-98H, Canada}
\affiliation{Observatories of the Carnegie Institution for Science, 813 Santa Barbara Street, Pasadena, CA 91101, USA}

\correspondingauthor{Amrita Singh}
\email{amrita@das.uchile.cl}

\begin{abstract}

The abundance discrepancy problem refers to the systematic differences observed between chemical abundances derived from collisionally excited lines (CELs) and recombination lines (RLs) of heavy-ions. It remains a major unsolved problem in the study of ionized nebulae and is quantified by the abundance discrepancy factor (ADF). In this work, we present a deep integral field spectroscopic dataset of the entire Lagoon nebula (M\,8), obtained by the SDSS-V Local Volume Mapper project, at a spatial resolution of 0.21 pc/spaxel. This unique dataset allows us, for the first time, to investigate spatially resolved maps of oxygen RL intensities (\orl V1), together with maps of HI RLs, heavy-ion CELs, and dust attenuation across a whole \hii region. We map the electron temperature using CELs and RLs of $O^{2+}$, CELs of $N^{+}$, and the electron density using CELs of $S^{+}$. We derive CELs-based ionic and elemental oxygen abundances and, for the first time, a spatially resolved map of the RL-based $O^{2+}$ abundance in an \hii region. These measurements enable construction of the first spatially resolved ADF($O^{2+}$) map of an \hii region and yield a global mean ADF of $\sim$0.47 $\pm$ 0.02 dex. Focusing on the central region of M\,8, where ionization is dominated by the O-type star, Her\,36, we find radial variations in the ADF, ranging between $\sim$0.35-0.50~dex. Our findings provide novel constraints on the spatial behavior and origin of the abundance discrepancy in the \hii regions.

\end{abstract}


\section{Introduction} 
\label{sec:intro}

The chemical composition of gas in local star-forming regions provides a direct probe of the present-day conditions in the interstellar medium (ISM), and offers valuable insight into the processes governing galactic chemical evolution across both spatial and temporal scales \citep{osterbrock06, kewley19, maiolino19}. \hii regions, in particular, are volumes of gas surrounding newly formed massive stars, whose intense ultraviolet (UV) radiation maintains the gas in an ionized phase, in which ions emit a plethora of collisionally excited lines (CELs) and recombination lines (RLs). These regions serve as cosmic laboratories to study the interplay between stellar evolution, feedback, and chemical enrichment in galaxies. As stars produce elements heavier than He (hereafter metals) via nucleosynthesis, and subsequently disperse them into the ISM \citep{kobayashi20}, analyzing these regions provides insight into the chemical enrichment of galaxies over time. 

\hii regions are often modeled as homogeneous, spherically symmetric, and ionization-bounded systems \citep[e.g.,][]{sutherland13, ferland17}, providing useful first-order frameworks. However, observations increasingly highlight their complex and dynamic structures, including clumps, filaments, density condensations \citep{dyson68}, photoevaporating dust pillars \citep{hester96}, and low density channels that permit the escape of ionizing photons \citep{oey97, castellanos02, ramambason22}. 
Recent photoionization modeling efforts have moved beyond simple geometries to incorporate more realistic and complex internal structures and geometries \citep{Jin22, Xing26}. These studies highlight the need for increasingly sophisticated models and physically realistic frameworks.
Such structural complexity leads to pronounced spatial variations in physical conditions, making their detailed spatially-resolved characterization crucial for deriving reliable chemical abundances.

Chemical abundances in ionized nebulae are commonly estimated using the direct method (DM), which relies on the use of CELs to ''directly'' derive the electron temperature ($T_e$) and electron density ($n_e$) \citep{montero17}. Since CEL emissivities depend strongly on temperature, accurate knowledge of $T_e$ is essential to constrain ionic abundances from the ratios of forbidden lines to hydrogen RLs \citep{osterbrock06}.  To account for the range of ionization states present in the \hii regions, the DM often adopts a three-zone model, in which $T_e$ is estimated separately for low, intermediate, and high ionization species \citep{garnett92}. The derived $T_e$ and $n_e$ are then used to derive the gas-phase chemical abundances across these different ionization zones. The DM is widely regarded as one of the most robust techniques for measuring gas-phase metallicities \citep{berg15, peimbert17}.

However, the strong $T_e$-dependence of CEL emissivities makes them sensitive to internal fluctuations in the gas conditions, introducing systematic biases in luminosity-weighted measurements \citep[e.g.,][]{peimbert67}. Consequently, in the presence of temperature inhomogeneities, $T_e$ values are overestimated, and abundances underestimated \citep{esteban04, osterbrock06}. These biases can be especially pronounced in extragalactic studies with limited spatial resolution \citep{berg20}, although even in spatially resolved studies, the integration of the gas emissivity along the line of sight is also expected to introduce these effects. A promising approach to constraining this bias is to compare the CEL-based abundances with those derived from metal RLs.

The recombination line method relies on the detection of extremely faint metal RLs, whose ratios to hydrogen RLs directly trace ionic abundances, as the dependence on $T_e$ and density is almost effectively canceled in these ratios \citep{osterbrock06}. Due to the low abundance of metals ($10^3–10^4$ times lower than H), metal RLs are intrinsically faint, and the application of this method has been typically limited to bright \hii regions in the Milky Way (MW) and the Local Group \citep[e.g.,][]{rojas07, toribio16, skillman20, esteban20, eduardo23Natur, yuguang23}.

Since the pioneering work of \citet{wyse42}, it has been well established that RL-based ionic abundances are systematically higher than those derived from CELs. In \hii regions, these differences are typically a factor of 2-4 \citep[e.g.,][]{peimbert93, esteban99, rojas07, esteban09, blanc15, eduardo23Natur}, while in planetary nebulae (PNe), they can reach factors of 5-700 \citep[e.g.,][]{ wesson03, tsamis04, corradi15, ali&dopita19, rojas22}, giving rise to the long-standing abundance discrepancy (AD) problem. This discrepancy, which is quantified by the abundance discrepancy factor (ADF), defined as the logarithmic difference between RL- and CEL-based abundances, remains a major challenge in nebular astrophysics. Understanding the physical reason of this discrepancy is essential for improving the accuracy of chemical abundance determinations across diverse cosmic environments in the Universe. 

Multiple hypotheses have been proposed to explain the AD problem over the years. These include $T_e$ fluctuations \citep[e.g.,][]{peimbert67} and $n_e$ inhomogeneities \citep{viegas94, bergerud19} in a chemically homogeneous medium, the presence of multiple chemically and thermally inhomogeneous components \citep[primarily in PNe; e.g.,][]{torres-peimbert80, liu00, liu06, ali&dopita19, rojas22, gomez+llanos24}, shock waves and turbulence \citep{peimbert91} and non-Maxwellian electron energy distributions and fluorescent excitation \citep{nicholls12}.

Spatially resolved ADF studies are essential to directly link the observed discrepancies with changes in local physical conditions. To date, most of such studies have focused on PNe, which have higher surface brightness and compact angular sizes \citep{tsamis08, ali&dopita19, rojas22, llanos24}. The limited field-of-view (FOV) of existing integral field spectrographs, such as Multi Unit Spectroscopic Explorer (MUSE) \citep[$\sim1$~arcmin$^2$;][]{bacon10}, has precluded similar studies in more extended nearby Galactic \hii regions. Nearly all RL-based abundance studies of Galactic and Local Group \hii regions have primarily utilized slit or echelle spectroscopy, which samples only small selected areas within nebulae \citep[e.g.,][and references therein]{peimbert_storey93, esteban95, tsamis03, delgado10, eduardo23Natur, yuguang23}. This has constrained our ability to assess spatial variations in the ADF within \hii regions, and to investigate how these correlate with potential physical drivers.

The advent of ultra-wide-field integral field spectroscopy (IFS), which is being pioneered by the Local Volume Mapper (LVM) project \citep{drory24}, as part of the Sloan Digital Sky Survey-V (SDSS-V) \citep{kollmeier19, kollmeier26}, overcomes these limitations. The LVM project is using the LVM-Instrument \citep[LVM-I][]{konidaris24}, a dedicated facility at Las Campanas Observatory in Chile, to obtain deep spatially resolved spectra across thousands of square degrees on the MW plane, the Magellanic Clouds, and other Local Volume galaxies \citep{drory24}, with an unprecedented spatial resolution of $<1$~pc per spaxel in the MW. The LVM sensitivity enables the detection of both faint CELs and extremely faint metal RLs across the extended structures of nearby bright Galactic \hii regions. For the first time, we can systematically map spatial ADF variations across these nebulae and explore their underlying causes.

The LVM project recently published similar datasets to the one presented here, for the Orion star-forming region \cite{kreckel24}, the Rosette nebula \citep{villa25}, the Trifid nebula \citep{sattler25} and the supernova remnant RCW86 \citep{sumit25}.

This paper is the first in a series addressing the AD problem in \hii regions using LVM data. Here, we investigate ADF spatial variations by mapping both CEL and RL-based $O^{2+}$ ionic abundances across the Lagoon Nebula (M\,8). 

This paper is organized as follows. Section~\ref{sec:lagoon} introduces the M\,8 nebula. The LVM observations are described in Section~\ref{sec:obs}, and the data reduction process in Section~\ref{sec:dr}. Section~\ref{sec:methods} outlines the methodology used to fit and measure emission-line fluxes, and to derive maps of line intensities and physical conditions across the nebula. We present results in Section~\ref{sec:results}, followed by a discussion in Section~\ref{sec:discussion}. We summarize the main conclusions in Section~\ref{sec:conclusion}. Supplementary plots are provided as an online-only figure set in the journal.

\section{The Lagoon Nebula (M\,8)}
\label{sec:lagoon}

\begin{figure*}[t]
\centering
\gridline{
    \parbox[b]{0.50\textwidth}{
        \centering
        \includegraphics[height=0.40\textheight]{\detokenize{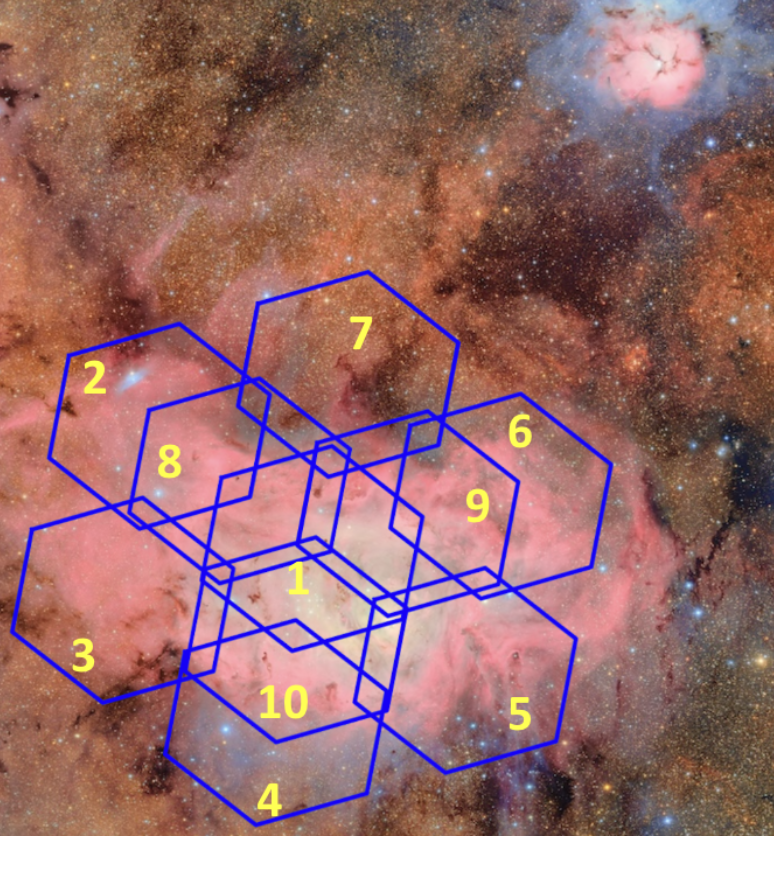}}\\
    }
    \hfill
    \parbox[b]{0.50\textwidth}{
        \centering
        \includegraphics[height=0.40\textheight]{\detokenize{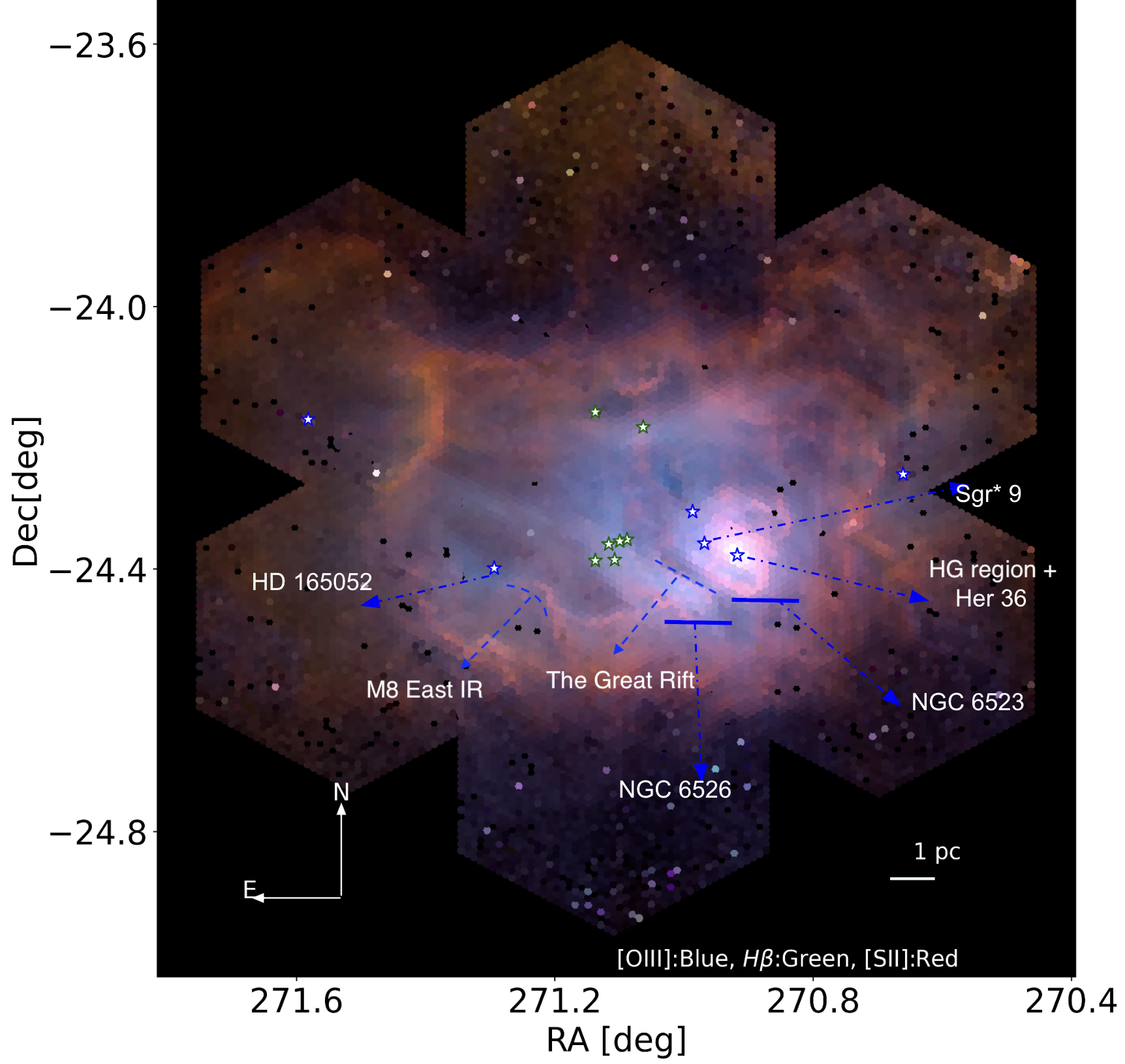}}\\
    }
}
\caption{Left panel: The LVM footprint for M\,8 overlaid on the recently released \href{https://rubinobservatory.org/news/rubin-first-look/trifid-lagoon}{first images of Vera Rubin Observatory}. Tile numbers are annotated in yellow, on the upper right of the panel is the Trifid nebula (M\,20, see one tile coverage of LVM in \citet{sattler25}). Right panel: Full LVM mosaic of the M\,8 nebula, consisting of 10 tiles and 17,492 fibers. The RGB composite maps emission from \oiii$\lambda\lambda$4959,5007 (blue), \hb (green), and \sii$\lambda\lambda$6717,6731 (red), highlighting the filamentary structures. Black dots denote the dead fibers and the masked spaxels excluded due to contamination from bright stars (see Section~\ref{subsec:stellar_cont_removal}). The blue horizontal solid and dashed lines indicate the approximate projected width of the annotated \hii regions (NGC~6526 and NGC~6523), the Great Rift and M\,8 East IR; these widths are schematics and not drawn to scale relative to the physical dimensions of the nebula.} Blue and green stars represent O- and B-type stars in the FOV, respectively, which are the primary ionizing sources of the nebula (see Table 1 of \citealt{wright19}). The Hourglass region and Herschel 36 are in the center, shown with a star here.
\label{fig:figure1}
\end{figure*}

\begin{table}
    \caption{M\,8 parameters}
    \begin{tabular}{cc} 
    \hline
    Parameters & M\,8 \\
    \hline
    Other designations & M\,8, Sh~25, RCW\,146, Gum\,72\\
    Type & Emission nebula\\
    Distance (kpc) & $\sim 1.25$  \tablenotemark{a} \\
    Apparent magnitude  & $6.0$ \tablenotemark{b} \\
    RA (J2000) & 271.1038 \\
    Dec (J2000) & -24.2767 \\
    Galactic Arm & Sagittarius-Carina arm \\
    \hline
    \end{tabular}
    \label{tab:table1}
    
\tablenotemark{a}{\citet{eduardo22}}
\tablenotemark{b}{\citet{tothill08}}

\end{table}

M\,8 is one of the largest and brightest nebulae in the MW. It is located in front of the Sagittarius-Carina arm \citep{tothill08}, at a Galactocentric radius of $\approx7$ kpc \citep{eduardo22}, and a distance of $\sim1.25$ kpc from the Sun \citep{prisinzano05, eduardo22}. Table~\ref{tab:table1} summarizes the key parameters of M\,8. It comprises multiple bright star-forming regions, \citep[most prominently M\,8-Main and M\,8-East, ][]{kahle24}, leading to its general classification as an \hii region in nebular diagnostics studies. The ionizing sources of M\,8 are associated with a young open star cluster, NGC~6530 \citep[$\sim$2-4 Myr old, ][]{chen07}, which contains approximately half a dozen O-type and several dozen B-type stars \citep{tothill08}. 

Panels~(a) and (b) of Figure~\ref{fig:figure1} present the nebula and its surroundings as seen in the Vera Rubin Observatory first release image and the LVM RGB composite showing internal details, respectively. These massive stars ionize three separate \hii regions in M\,8: NGC~6523 (the core of the nebula, $\approx 4\arcmin$ in radius, \citealt{tothill08}), NGC~6526, and the region around the binary star HD 165052 (O6.5+ and O7.5+) \citep{arias02}. 

NGC~6523 is a face-on ionized cavity formed by stellar winds, and it includes the famous compact Hourglass region (hereafter, HG). The region is powered by the brightest component of the trinary star Herschel 36 \citep{arias10}, (hereafter Her\,36), an O7.5 star that photoionizes the edges of a background molecular cloud, also referred to as M\,8-Main in \citet{kahle24}. NGC~6526, in contrast, is ionized by the spectroscopic binary 9 Sagittarii (hereafter Sag\,9), located $\approx 1.5-1.8$ pc along the line of sight toward us \citep{damiani17}, and the NGC~6530 cluster. Finally, the last \hii region is the largest and faintest one, which is ionized by HD 165052. It is located to the east of a bright, highly illumination young stellar object, also known as M\,8 East-IR in \citet{kahle24}, which is embedded in a star-forming molecular cloud complex \citep{tiwari20}.

The NGC~6523 and NGC~6526 regions are separated by a molecular cloud complex, known as the Great Rift \citep{tothill08}. Studies such as \citet{kumar10} have confirmed the detection of polycyclic aromatic hydrocarbons and other molecules in the rift. Additionally, a tenuous trench-like structure located between the Rift and the HG region could be seen; this may have been sculpted by stellar winds from Sgr\,9.

The LVM RGB composite, in Panel~(b) of Figure~\ref{fig:figure1} is created by fitting Gaussian profiles to the emission lines, see Section~\ref{subsec:measure_fluxes} for details. The primary ionizing sources and the main \hii regions in M\,8 are also annotated in the image.

Due to its extreme brightness, M\,8 has long been a benchmark in studies of metal RLs and the AD problem in photoionized nebulae \citep[e.g.,][] {peimbert93, esteban99, rojas07}. However, all previous work in this region has been conducted using slit and echelle spectroscopy, and has been focused mainly on the bright HG region (which is contained approximately within a single LVM-I Integral Field Unit (IFU) spaxel; see Figure~\ref{fig:figure1}). This has limited the spatial scope of abundance determinations from metal RLs and ADFs \citep[e.g.,][] {peimbert93, esteban99, rojas07, rodriguez10, delgado10}. Our LVM-I observations overcome this limitation by providing, for the first time, coverage of the entire nebula with a spatial resolution of $ 0.21$ pc/spaxel, and a spectral depth that allows a comprehensive analysis of the ADF over extended areas of the nebula.

\section{Observations}
\label{sec:obs}

The M\,8 IFS observations are taken with the LVM-I. The core of the LVM-I is an ultra-wide-field IFU with a 0.165 deg$^2$ hexagonal FOV that is 30.2$\arcmin$across. The IFU is composed of 1801 circular 35.3$\arcsec$ diameter spaxels arranged in a hexagonal pattern with a 37$\arcsec$ pitch, \citep[see][for more details]{herbst24}. The light collected by the spaxels is transmitted through optical fibers to three dedicated spectrographs (Spec1, Spec2, and Spec3). Each spectrograph splits light into three spectral channels covering the blue ($b$; 3600-5800~\AA), red ($r$; 5750-7570~\AA), and near infrared ($z$; 7520-9800~\AA) parts of the optical wavelength range, with an average spectral resolution R=2700, 4000, and 4600 in the $b$, $r$, and $z$ channels, respectively, \cite[see][]{drory24}. 

The science IFU is fed by a 16.1~cm aperture telescope. Two identical telescopes feed two small IFUs with 60 and 59 fibers each, which are used to obtain simultaneous spectra of background sky region away from the source for sky subtraction. During the science observations, a fourth telescope of the same aperture size is used to feed a sparse array of optical fibers with light from a set of spectrophotometric calibration stars near the science field, which are used for flux calibration. For a detailed description of the instrument and the observing strategy, we refer the reader to \cite{konidaris24, drory24}, and \cite{blanc25}.

Observations of M\,8 were obtained over several nights between August 22nd and October 19th, 2023, as part of the commissioning and early science campaign of the SDSS-V LVM project. During that period, the $z$ channel of the third LVM-I spectrograph (Spec3) warmed up, rendering data in that channel unavailable for some of the frames in the dataset. To compensate for this, re-observations of some of these early-science frames were carried out on July 2nd, 2024. 

The entire nebula can be covered by tiling 7 pointings of the LVM-I science IFU. However, the use of an incorrect position angle (PA) during these commissioning observations resulted in gaps in the tiling pattern. To address this, we observed 3 additional pointings, ultimately covering the nebula with a total of 10 tiles. Panel~(a) of Figure \ref{fig:figure1} presents a Vera Rubin Observatory first release image of M\,8 with the position of the LVM tiles, observed for this work, overlaid. In this work, we number tiles as shown by the yellow numbers in the figure (note that this numbering will be different from the the tile ID numbers used in the main LVM survey).

For each tile, we obtained 900~s exposures, as planned for the LVM survey. All tiles were observed at least twice in these two modes, to achieve reliable $S/N$ to map faint lines, with a few of them receiving more (up to 5) exposures. Summing the two types of exposures, the total exposure times on each tile range from 1800~s to 4500~s. These are longer than the standard 900~s LVM survey exposures \citep{drory24, blanc25}. Table ~\ref{tab:table2} summarizes the observations. We list the coordinates and the parameters of the exposures obtained for each tile. Calibration frames required for the reduction and calibration of the data (including bias, flat-fields, arc-lamp frames, and standard stars) were also obtained.
 
The total spatial coverage of the dataset is $\sim$ 1.1 sq. degrees centered on the M\,8 nebula.  This ultra-wide-field IFU dataset, providing full optical spectral coverage of the entire M\,8 region at sub-pc resolution, is among the first of its kind, allowing a detailed study of the resolved structure of the ionized ISM in this object. 

\begin{table*}[t]
\centering
\caption{Observational Log. Column 1 lists the tile ID of the M\,8 mosaic, columns 2 and 3 show the central RA and Dec of these tiles, column 4 contains the observation date (MJD), columns 5, 6, and 7 list the exposure ID, exposure time for each exposure, and the total exposure time for each tile.}
\label{tab:table2}

\begin{tabular}{c c c c c c c}
\hline
Tile & RA (HH:MM:SS) & Dec (DD:MM:SS) & MJD (J2000) & Exposure ID & Total exposure time (s) \\ 
\hline
\hline
1  & $18:04:24.91$ & $-24:16:36.12$ & {60178, 60186} & {3501, 3503, 3505, 3989, 3991} & 4500 \\ 
2  & $18:06:01.90$ & $-24:03:27.36$ & {60179, 60493} & {3595, 2504, 2505} & 2700 \\ 
3  & $18:06:03.48$ & $-24:29:11.04$ & {60188, 60191} & {4160, 4285} & 1800 \\ 
4  & $18:04:26.18$ & $-24:42:19.80$ & {60191} & {4289, 4291} & 1800 \\ 
5  & $18:02:47.59$ & $-24:29:40.92$ & {60203, 60493} & {5020, 2508, 2509} & 2700 \\ 
6  & $18:02:46.68$ & $-24:03:57.24$ & {60228} & {6425, 6426} & 1800 \\ 
7  & $18:04:23.66$ & $-23:50:52.44$ & {60228} & {6421, 6422} & 1800 \\ 
8  & $18:05:12.00$ & $-24:09:00.00$ & {60231} & {6587, 6588} & 1800 \\ 
9  & $18:03:36.00$ & $-24:09:00.00$ & {60232} & {6659, 6660} & 1800 \\ 
10 & $18:04:24.00$ & $-24:30:00.00$ & {60493} & {20500, 20501} & 1800 \\ 
\hline
\end{tabular}

\end{table*}

\section{Data Reduction} 
\label{sec:dr}

We reduced the data using the  LVM Data Reduction Pipeline \citep [{\tt lvmdrp} version 1.1.2dev,][] {mejia25}, hereafter the DRP. This pipeline produces flat-fielded, wavelength-corrected, flux-calibrated, and sky-subtracted row-stacked spectra (RSS) for each exposure. A high-level overview of the LVM data processing will be available in \cite{blanc25}.

The main outputs of the DRP are the {\tt{lvmSFrame}} multi-extension {\tt{FITS}} files, which are produced for each individual exposure. These files contain the spectra of all IFU spaxels in RSS format in the {\tt{FLUX}} extension, the error spectra (in inverse variance units) in the {\tt{IVAR}} extension, pixel masks that include detector artifacts and cosmic rays in the {\tt{MASK}} extension, and the wavelength array in the {\tt{WAVE}} extension. Additional information for each fiber, including its right ascension (RA) and declination (Dec), is provided in binary table format in the {\tt{SLITMAP}} extension. A quality flag for each fiber ({\tt{fibstatus}}) is also provided here. 
In these files, the spectra and their errors are stored in flux density units of erg~s$^{-1}$~cm$^{-2}$~\AA$^{-1}$ per spaxel, with the wavelengths in units of \AA. Of the 1801 science spaxels in each tile, only those with {\tt{fibstatus = "good"}} are selected, thereby excluding dead and low throughput fibers. This selection reduced the fiber count to 1754; a number consistent across all LVM exposures.

Given that data are acquired at different epochs, we apply a barycentric correction to the flux and IVAR of each {\tt{lvmSFrame}} to account for the Earth's motion. The correction uses the barycentric velocity provided in the header keyword ({\tt{WAVE $HELIORV\_SCI$}}) of each frame. 

To mitigate low-level systematic errors in flux calibration across the multiple exposures of the different tiles (see Table~\ref{tab:table2}), we implement a two-step normalization procedure. First, we normalize all exposures of the same tile relative to a reference exposure. This reference exposure is chosen as the one whose average spectrum is closest to the median of all exposures for that tile. 

For each exposure, we then fit the average spectrum and measure the line fluxes of \hb, \nii$\lambda$6584, and P\,9$\lambda$9229 in the $b$, $r$, and $z$ channels, respectively, following the line fitting procedure outlined in Section~\ref{subsec:measure_fluxes}. Scaling factors are derived by taking the ratio of the line fluxes in the average exposure of each exposure to those of the reference exposure, and these factors are used to scale the flux and error spectra of each spectral channel independently. The normalized exposures of each tile are then combined using an inverse-variance weighting scheme, to produce a single frame per tile.

 
In the second normalization step, we correct for tile-to-tile variations. We compare the average spectra of overlapping regions between adjacent tiles, again fitting the same key emission lines, and deriving per channel scaling factors to match the absolute flux calibration level of all tiles with that of the central tile (i.e. tile 1).

After normalizing, all tiles are merged into a single RSS file comprising 17,540 spectra from 10 mosaic tiles (see panel~(a) of Fig.~\ref{fig:figure1}). We convert the flux units into surface brightness units of erg~s$^{-1}$~cm$^{-2}$~\AA$^{-1}$~arcsec$^{-2}$ by dividing the flux and error spectra by the spaxel area (978.18 arcsec$^{2}$). 
All further analysis in this paper is performed using this merged RSS file.

\section{Methods}
\label{sec:methods}

The primary objective of this paper is to perform a spatially resolved analysis of ADF(O$^{2+}$). In this section, we outline the methodologies adopted to estimate all relevant physical quantities from the LVM spectra to achieve our primary goal.

\begin{figure*} [t]
    \centering
    \includegraphics[width=0.96\textwidth]{\detokenize{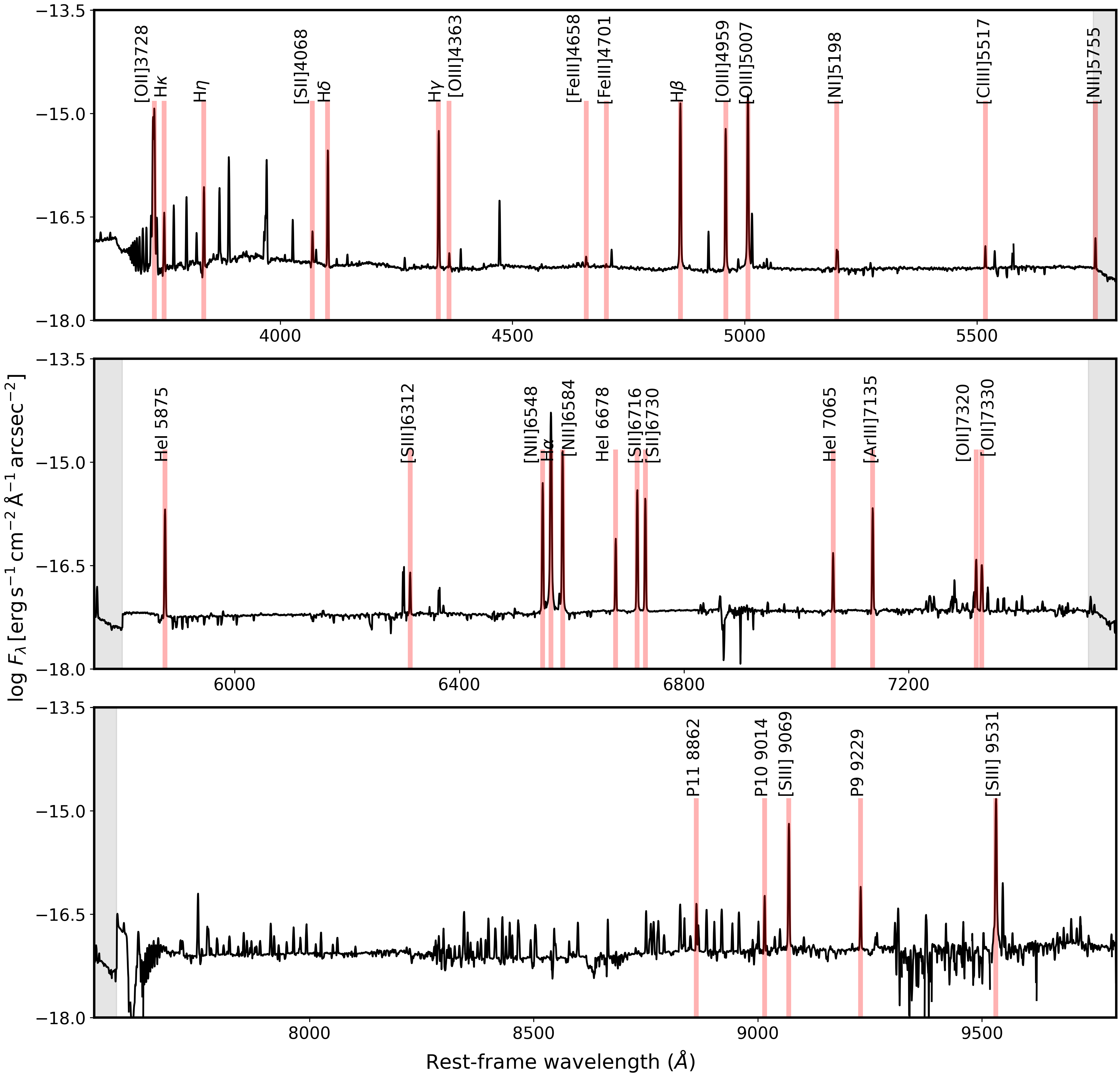}}
\caption{Average spectrum of M\,8 in blue (b, upper panel), red (r, middle panel) and near-infrared (z, bottom panel) channels. Grey vertical bands highlight regions where channels overlap. Prominent emission lines are marked with vertical red lines. In the z channel, only a few lines are labeled, since most features correspond to sky emission lines that are not used in this work.}
\label{fig:figure2}
\end{figure*}

\begin{figure}[t]
    \centering
    \includegraphics[width=0.99\columnwidth]{\detokenize{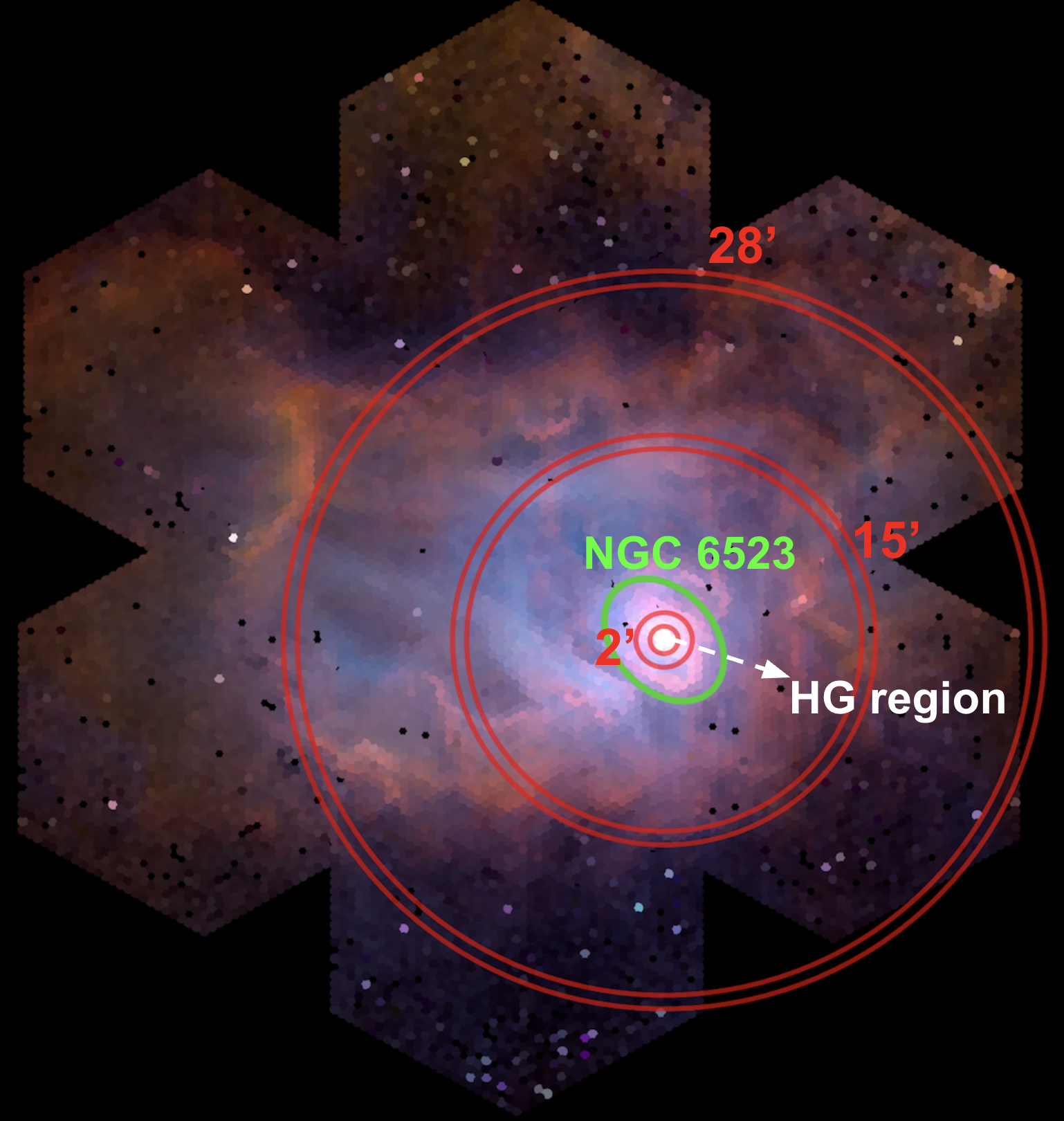}}
    \caption{Color composite map of M\,8 (extended from Figure\ref{fig:figure1}), highlighting the central HG region powered by Her\,36 (both in the central bin). The red circles mark three 1 arcmin wide annular bins at a distance of $2\arcmin$, $15\arcmin$, and $28\arcmin$ from Her\,36. The green ellipse marks the NGC\,6523 \hii region ionized by Her\,36, while the area within the $15\arcmin$ annulus encompasses the NGC\,6533 \hii region.}
    \label{fig:figure3a}
\end{figure}

This analysis is carried out on different spatial scales: (1) individual spaxels, providing fully resolved maps across the FOV; (2) spectra binned in 1 $\arcmin$ wide annuli, centered at Her\,36, across the FOV; (3) the average spectrum of the entire nebula (see Figure \ref{fig:figure2}); and (4) both annular binned (also 1 $\arcmin$ wide) and the average spectrum of the elliptical region, NGC~6523 (see Figure \ref{fig:figure3a}), whose ionization is dominated by Her\,36.

Since M\,8 hosts multiple hard ionizing sources, the observed spectrum in any given spaxel may arise from a mix of gas ionized by different sources, each with distinct spectral characteristics and located at varying distances from the emitting gas. This complexity can obscure physical interpretations of the spatial trends in emission and gas properties. To mitigate this, we identify and isolate spaxels whose emission is dominated by gas ionized by Her\,36 (the main ionization source of the NGC~6523 region on the west side of M\,8) by defining an elliptical aperture based on \citet{tothill08}. 
For these spaxels, we repeat both the annuli binning and the average spectrum analysis, enabling a more focused interpretation of the nebular conditions under a single dominant ionizing source, Her~36. At each level of analysis, we distinguish between the entire region and the Her\,36 subset, which corresponds to NGC~6523 \citep{tothill08}.

The presence of these stars across the FOV \citep[see][]{wright19} produces detectable stellar continua in the observed LVM spectra, which need to be separated from the nebular emissions. 

In this section, we first describe how we subtract the stellar spectra from each spaxel in the FOV, and how we build the binned and average spectra mentioned above. We then detail the procedures to fit and measure the emission line fluxes, followed by the methods for dust attenuation correction, and finally the determination of physical and chemical conditions using \texttt{PyNeb} (version 1.1.18, \citealt{luridiana15}). The data sources adopted are summarized in Table~\ref{tab:table3}. To estimate the associated uncertainties of the physical and chemical quantities, such as $n_e$, $T_e$ and ionic abundance, we perform 50 Monte Carlo (MC) iterations, perturbing the emission line fluxes to their errors. Each perturbed realization is used to recompute the diagnostic, and the final measurements are stored as the mean and standard deviation of the resulting distribution.


\begin{table*}[t]
\centering
\caption{Atomic data sources used in this study for CEL and RL diagnostics}

\begin{tabular}{cccc}
\hline
\textbf{Ion} & \textbf{Transition Type} & \textbf{Data Type} & \textbf{Reference} \\
\hline
\hline
S$^{+}$  & CEL (\sii 6716, 6731) & Transition Probabilities & \citet{podobedova09} \\
         &                                              & Collision Strengths      & \citet{tayal10} \\
O$^{+}$  & CEL (\oii 3726, 3729) & Transition Probabilities & \citet{zeippen82} \\
         &                                              & Collision Strengths      & \citet{kisielius09} \\
N$^{+}$  & CEL (\oii 5755; 6548, 6584) & Transition Probabilities & \citet{fischer04} \\
         &                                              & Collision Strengths      & \citet{tayal11} \\
O$^{2+}$ & CEL (\oiii 4959, 5007; 4363)      & Transition Probabilities & \citet{storey00} \\
         &                                              & Collision Strengths      & \citet{aggarwal99} \\
H$^{0}$  & RL (\ha, \hb, \hg, \hd, etc.)     & Recombination Coefficients & \citet{storey95} \\
O$^{2+}$  & RL (\orl Multiplet 1)               & Recombination Coefficients & \citet{storey17} \\
N$^{+}$  & (\nrl Multiplet 3)               & Recombination Coefficients & \citet{fang11} \\
\hline
\end{tabular}
\label{tab:table3}
\end{table*}

\subsection{Stellar Continuum Subtraction and Masking}
\label{subsec:stellar_cont_removal}

All the LVM exposures are automatically processed by the LVM Data Analysis Pipeline \citep[hereafter DAP,][]{sanchez25}, which performs a full spectral modeling, including stellar continuum and emission line fitting. The DAP models the stellar continuum using a linear combination of templates from its resolved stellar population (RSP) library, which currently includes 108 models for stars with $T_{eff}< 15,000$ K. An extension to incorporate hotter stellar templates, required for accurate modeling of \hii regions, is currently underway. These models account for redshift, velocity dispersion, and dust attenuation, and are optimized to match the observed LVM spectra. 

We adopt the best-fit continuum model of DAP to subtract the underlying stellar contribution from each LVM spectrum prior to the emission line fitting. This subtraction yields a residual spectrum which retains the nebular emissions with significantly reduced contamination from stellar continua. 

However, since the default DAP RSP template library does not currently include spectra for massive O- and B-type stars, some spaxels exhibit significant stellar continuum residuals. These residuals can bias the determination of the local continuum level during the line fitting. To avoid such systematics, we exclude from the analysis any spaxel whose continuum brightness at \ha$\lambda$6562.8 exceeds $6.6\times10^{-16}$ erg s$^{-1}$ cm$^{-2}$ $\AA^{-1}$ arcsec$^{-2}$. Visual inspection of the \ha$\lambda$6562.8 continua map confirms that these bright-continuum spaxels are primarily associated with O-and B-type stars. This threshold effectively removes only the most strongly stellar contaminated spaxels, without significantly reducing the usable dataset. After masking these spaxels, which constitute less than 0.3\% of the full FOV, the final combined RSS is restored. This RSS, which comprises 17,492 spaxels, serves as the basis of all subsequent analyses, including the spatially resolved measurements and the construction of average and annular binned spectra.


\subsection{Spatially resolved, average, and binned spectra}

To study the resolved structure of the ionized ISM on sub-parsec scales, we use the spatially resolved spectra of individual spaxels stored in the final combined RSS dataset (see Section \ref{subsec:stellar_cont_removal}). To emulate spatially unresolved observations, like those typically obtained in extragalactic contexts, we construct an average spectrum of M\,8 by computing the mean flux and error across the FOV (see Figure~\ref{fig:figure2}). An integrated spectrum for the full FOV can be obtained by multiplying the average surface brightness spectrum by the total area covered by the mosaic.

\begin{figure}[t]
    \centering
    \includegraphics[width=0.999\columnwidth]{\detokenize{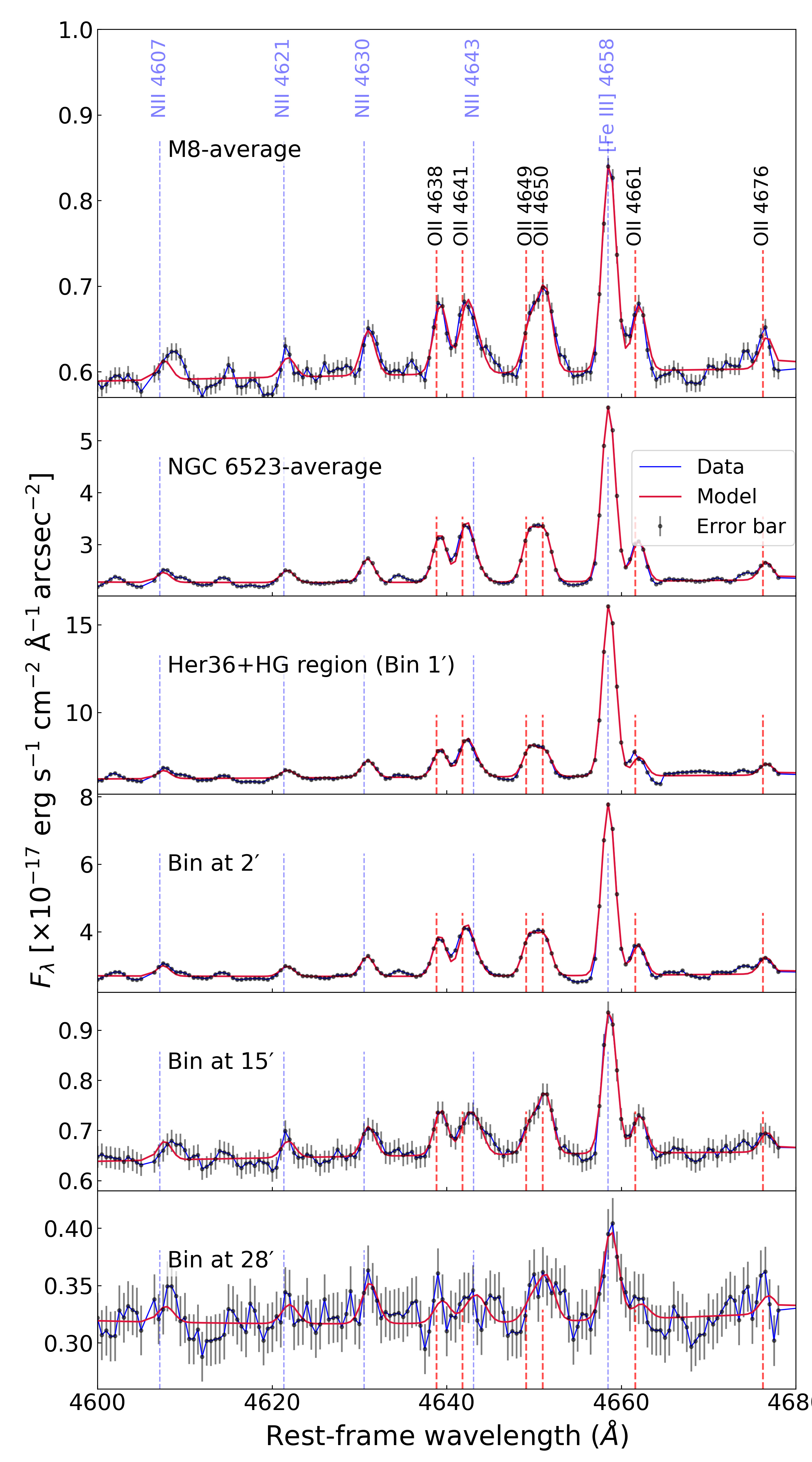}}
    \caption{Spectral fits to the \orl $\lambda\lambda$4600–4700 region are presented in six panels: (i) first -- for the M\,8 average spectrum, second -- the average spectrum of NGC~6523, third -- for the central bin covering the HG region, last three are -- representative spectra at $2\arcmin$ (top), $15\arcmin$ (middle), and $28\arcmin$ bins, respectively. In the spectra, observed data are in blue, and the best-fit model is in red. Black lines show error bars associated with the flux. Vertical red dashed lines indicate key \orl transitions of the V1 multiplet; blue dashed lines mark multiplet 4 \nrl permitted lines and the \feiii$\lambda$4658 CEL. 
    }
    \label{fig:figure3b}
\end{figure}

To enhance the $S/N$ and improve the detectability of faint lines in the spectra, especially towards the outer low surface brightness parts of the nebula, we bin spaxels in concentric annuli centered on the Her\,36 star. We do so by computing the mean flux and error spectra of the contributing spaxels in annular 1$\arcmin$ wide bins, propagating the flux uncertainties using Gaussian error propagation. This results in 53 radially binned spectra, enabling the detection and measurement of faint lines across the FOV (see Figure \ref{fig:figure3b}).

These concentric annuli are centered on the main ionizing source of the NGC~6523 \hii region, so at radii $>2$~pc they start encompassing flux from the neighboring NGC 6526 region, and at radii $\sim$8~pc they start sampling flux from the \hii region associated with HD 165052 (see Figure \ref{fig:figure1}), complicating the physical interpretation of the observed radial trends. To overcome these complications associated with the complex structure of the M\,8 nebula, we define an elliptical aperture centered on Her\,36. It  has semi-major and semi-minor axes of 1.82~pc and 1.31~pc respectively, and a position angle of 50$^{\circ}$. By inspecting different line tracers (mainly the high ionization \oiii lines), we expect all spaxels within this aperture to be largely dominated by emission associated with direct ionization from Her\,36. We refer to this as the NGC\,6523 aperture, although we caution the reader that this aperture mainly covers the higher ionization central parts of the region, and it does not extend out of NGC\,6523's boundary, which is poorly defined anyway.

The NGC~6523 aperture is used for two purposes. The first is for building an average spectrum by computing the mean of the flux and error spectra of all spaxels within the aperture. The second is to repeat the binning over 1$\arcmin$ wide concentric annuli described above, but masking all spaxels that fall outside this aperture. The latter allows us to study radial trends that are directly associated with the radiative transfer of the Her\,36 ionizing radiation as it propagates through the gas.



\begin{table}
    \caption{Nebular emission lines used in this work and associated physical parameters derived from them.}
    \resizebox{0.99\columnwidth}{!}
{%
    \begin{tabular}{ccc}
    \hline
    Line & $\lambda_0 \tablenotemark{c} (\AA)$ & Parameter \\ 
    \hline
    \hline
    \oii & 3726.03 & $n_{e} (O^{+})$, $(O^{+}/H^{+})_{CEL}$ \\ 
    \oii & 3728.81 & $n_{e} (O^{+})$, $(O^{+}/H^{+})_{CEL}$\\ 
    \hk & 3751.5 & $R_V$, E$(B-V)$\tablenotemark{d}\\ 
    \hi & 3770.63 & $R_V$, E$(B-V)$\tablenotemark{d}\\ 
    \he & 3835.38 & $R_V$, E$(B-V)$ \tablenotemark{d}\\ 
    \hd & 4101.74 & $R_V$, E$(B-V)$\\ 
    \hg & 4340.46 & $R_V$, E$(B-V)$\\     
    \oiii & 4363.2 & $T_e^{CEL}(O^{2+})$\\ 
    \nrl& 4607.16 &  To remove contamination\\
    \orl & 4638.86 & $T_e^{RLs/CEL}(O^{2+})$, $(O^{2+}/H^{+})_{RL}$\\ 
    \orl & 4641.81 & $T_e^{RLs/CEL}(O^{2+})$, $(O^{2+}/H^{+})_{RL}$\\ 
    \nrl& 4643.16 &  To remove contamination\\
    \orl & 4649.13 & $T_e^{RLs/CEL}(O^{2+})$, $(O^{2+}/H^{+})_{RL}$\\ 
    \orl & 4650.84 & $T_e^{RLs/CEL}(O^{2+})$, $(O^{2+}/H^{+})_{RL}$\\ 
    \orl & 4661.63 & $T_e^{RLs/CEL}(O^{2+})$, $(O^{2+}/H^{+})_{RL}$\\ 
    \orl & 4676.23 & $T_e^{RLs/CEL}(O^{2+})$, $(O^{2+}/H^{+})_{RL}$\\ 
    \hb & 4861.32 & $R_V$, E$(B-V)$\\ 
    \oiii & 4958.91 & $T_e^{CEL}(O^{2+})$, $(O^{2+}/H^{+})_{CEL}$\\ 
    \oiii & 5006.84 & $T_e^{CEL}(O^{2+})$, $T_e^{RLs/CEL}(O^{2+})$, $(O^{2+}/H^{+})_{CEL}$\\ 
    \nii & 5754.64 & $T_e^{CEL}(N^{+})$\\
    \nii & 6548.04 & $T_e^{CEL} (N^{+})$\\ 
    \ha & 6562.80 & $R_V$, E$(B-V)$ \tablenotemark{d}\\ 
    \nii & 6583.46 & $T_e^{CEL}(N^{+})$\\ 
    \sii & 6716.44 &  $n_{e} (S^{+})$\\ 
    \sii & 6730.82 &  $n_{e} (S^{+})$\\ 
    P\,11 & 8862.78 & $R_V$, E$(B-V)$\\
    P\,10 & 9014.91 & $R_V$, E$(B-V)$ \tablenotemark{d}\\
    P\,9  & 9229.01 & $R_V$, E$(B-V)$\\
    
    \hline
    \label{tab:table4}
    \end{tabular}%
}
\tablenotetext{c}{Rest-frame wavelength in air.}
\tablenotetext{d}{These lines are not used in calculation, see Section~\ref{subsec:dust}.}
\end{table}


\subsection{Measurement of emission line fluxes}
\label{subsec:measure_fluxes}

For all spectra (individual spaxels, radially binned, or average), we measure emission line fluxes listed in Table~\ref{tab:table4}, by fitting Gaussian profiles to these lines using a custom script. 

We do not use the fluxes measured by the DAP, but instead use a custom fitting routine to process the metal RLs, this gives us more precise control over the fitting of blended lines, such as the neighboring faint \nrl RLs and \feiii CEL, which require a custom fitting approach. For the sake of uniformity, we also fit all other emission lines, including bright HI lines, auroral, and strong metal CELs, using the same custom algorithms. We have validated the consistency with the DAP results by comparing the fluxes of bright emission lines measured with both approaches, finding an excellent agreement.


Gaussian profiles are fitted to emission lines of interest using their rest-frame air wavelengths, adopted from the National Institute of Standards and Technology \citep[see][for more details]{nist} as initial centroid estimates. The fitting is performed over spectral windows that include both the emission lines and adjacent continuum regions, using a model composed of a constant continuum for all strong lines such as Balmer lines, Paschen lines, and nebular CELs. For auroral and metal RLs, we fit a quadratic function to the continua to avoid erroneous measurements in the outer regions with low $S/N$. We adopt a fitting window of $\pm3~\AA$ around each line centroid, along with the continuum regions identified via visual inspection. 

The final model consists of a continuum plus one or more Gaussian components, each Gaussian characterized by three free parameters: central wavelength, velocity width ($\sigma$), and amplitude. These parameters are optimized using the \texttt{curve\_fit} routine, which minimizes the residuals between the observed and model spectra. Uncertainties in the fitted parameters are derived from the diagonal elements of the covariance matrix and are propagated via 50 MC realizations in subsequent analyses of nebular diagnostics.

Most emission lines listed in Table \ref{tab:table4} are fit independently with a single Gaussian. However, for blended or closely spaced lines, we adopt multiple Gaussian models; for instance, a double Gaussian model (with a shared yet free centroid and velocity width, but independent amplitudes) is used for blended doublets of \oii $\lambda$$\lambda$3726, 3729. A similar approach is used for closely spaced lines of \feii $\lambda$4360 and \oiii $\lambda$4363, to avoid possible contamination of the former line in the high metallicity regime \citep[see][for more details]{curti17}. The spectral region around \oiii $\lambda$4363 suffers from DRP issues; we tie the velocity shift of this line to \oiii $\lambda$5007, to avoid any erroneous flux measurements of the line in outer regions, where $S/N$ declines.

The V1 multiplet is the strongest of the \orl RL multiplets in the optical regime and is predominantly emitted via recombination, with a negligible contribution from fluorescence processes \citep{peimbert93}. In close proximity, the \nrl multiplet 5 lines are likely affected by resonance fluorescence \citep{esteban99}, potentially enhancing their observed intensities. 
The \orl V1 multiplet consists of 8 lines: 4638.86, 4641.81, 4649.13, 4650.84, 4661.63, 4673.73, 4676.23, and 4696.36 $\AA$; we detect six of these eight, see Table \ref{tab:table4}, along with the forbidden \feiii$\lambda$4658 line, which is blended with \orl$\lambda$4661.63. We also detect four of the six permitted lines of the \nrl multiplet 5 (\nrl 4643.16 is blended with \orl$\lambda$4641.81). We fit a multi-Gaussian model with 11 components over the 4600\AA-4690\AA\ region to measure line fluxes of all the aforementioned lines together, to account for blending.

As shown in Figure ~\ref{fig:figure3b}, the \nrl$\lambda$4643.09 is significantly blended with \orl$\lambda$4641.81, and must be accurately accounted for to measure the precise flux of the latter. The \nrl$\lambda$4643.09 maintains a constant intensity ratio of 1.37 relative to \nrl$\lambda$4607.16 across a wide range of $T_e$ and $n_e$ \citep{fang11}. We therefore fix the relative amplitude of the two lines using this ratio while fitting the lines. Furthermore, we assume a common velocity shift and velocity width for all 11 Gaussian components in this spectral region. Figure~\ref{fig:figure3b} shows representative examples of this spectral region and the corresponding best-fit spectral models for different spectra across M\,8.  


\subsection{Dust Attenuation Correction}
\label{subsec:dust}

Interstellar dust along the line of sight introduces attenuation and reddening that can significantly affect the observed emission line fluxes. This causes significant biases in line ratios based on transitions that are widely separated in wavelength. To estimate the amount of dust attenuation and reddening, a standard approach is to compare the observed flux ratio of two HI RLs (typically \ha and \hb) to their theoretical values under Case B recombination. By assuming a particular attenuation law, one can quantify the amount of reddening needed to explain the observed difference. This is known as the Balmer decrement technique and is widely discussed in the literature \citep[e.g.,][]{groves12, xiao12, maheson24}. 

An attenuation law implicitly assumes a specific spatial distribution of dust relative to the stellar populations, as well as a fixed distribution of dust grain properties such as size, composition, and optical characteristics \citep{draine11}. However, due to the structural complexity of \hii regions, such as the presence of strong radiation fields from massive stars, which can alter the dust properties through photodissociation, the geometry between the dust and gas can vary significantly across the nebula. 

Over the wavelength coverage of the LVM observations, the extinction among commonly used Galactic extinction laws is insignificant. We therefore adopt the parameterized extinction law of \citet{fitzpatrick99} (hereafter, F99), which provides a widely used and physically motivated framework, flexible in both the shape and normalization of the extinction curve. This model is defined by two parameters: (a) the total-to-selective extinction ratio, $R_V$, and (b) the color-excess, $E(B-V)$. By varying these parameters, we can constrain the spatial variations in the dust properties across the nebula.

Although the Balmer decrement \ha/\hb is typically used for extinction correction, the \ha line is saturated in the brightest region of the nebula in our data, precluding its use in those areas. Thus, we avoid using the H$\alpha$ emission line in further analysis. Also, using a single line ratio, it is not possible to constrain two parameters at the same time ($R_V$ and $E(B-V)$). Instead, we use a multi-line approach based on the \hd, \hg, \hb, Paschen 11 $\lambda$8863 (P\,11) and Paschen 9 $\lambda$9229 (P\,9) lines, spanning a wide wavelength range. These lines are selected on the basis of their high $S/N$, absence of saturation, blending, telluric absorption, and contamination via the neighboring sky lines. Higher-order Balmer lines are excluded because their fluxes are overestimated due to their high-sensitivity to deviations from the ``Case B'' recombination conditions, an issue already reported in \citet{luridiana09, delgado09, guzman22}, while additional Paschen lines, such as P\,10 $\lambda$9015, are avoided to reduce potential skyline residual contamination. The broad wavelength coverage increases the sensitivity to the wavelength dependence of extinction, which allows us to jointly constrain both $R_V$ and $E(B-V)$.

Generally, a canonical value of $R_V = 3.1$ has been measured for the diffuse ISM in the Milky Way \citep[see][]{cardelli89, draine03, schlafly11, pottasch13, siebenmorgen23}. By default, many studies of \hii regions and the ionized ISM in galaxies use this same value \citep[for e.g., ][]{rojas05, rojas06, kreckel16, griffith19, andrade25, pathak25}. However, $R_V$ can deviate significantly from the standard value in \hii regions such as M\,42 and M\,8 \citep[see][]{mathis90}. In particular, several works focused on the HG region of M\,8, have shown that values as high as $R_V \sim 5.0$ better reproduce the observed line ratios \citep{hecht82, sanchez91}, and therefore such values have been adopted in subsequent analyses by \citet{peimbert93, esteban99, rojas07}.

To test this in our data, we compute the theoretical intrinsic ratios for the selected HI lines using \texttt{PyNeb}. These ratios depend mildly on the $T_e$ and $n_e$, and are evaluated under Case B recombination conditions. We then determine the extinction curve that best reproduces the observed HI ratios in the LVM spectra. 

We model the observed ratios by attenuating the intrinsic ratios with the F99 law in the {\tt dust extinction} package. Both $R_V$ and $E(B-V)$ are treated as free parameters, but are constrained in the ranges $R_V \in [2, 6]$, as per the limits of the model, and $E(B-V) \in [0, 5]$. The best-fit values are obtained by minimizing the $\chi^2$ between the observed ratios and the intrinsic ratios reddened according to the F99 model. In this formulation, we first determine  the values of $R_V$ and $E(B-V)$ that best reproduce the observed ratios, and only after the fit do we apply F99 extinction law to compute the modeled line ratios.

Because the theoretical HI ratios depend on $T_e$ and $n_e$, and the physical conditions in turn depend on dust corrected line intensities, we adopt an iterative procedure. We first compute the intrinsic ratio assuming $T_e=10^4$K and $n_e=10^2$ cm$^{-3}$ and fit for $R_V$ and $E(B-V)$ using the average spectrum of M\,8. Once the best-fit values of $R_V$ and $E(B-V)$ are obtained, they are used to calculate the extinction at each wavelength $A_{\lambda}$ using:
\begin{equation}
    A_\lambda = k(\lambda, R_V) \times E(B-V)
\end{equation} 
where $k(\lambda, R_V)$ is the extinction curve defined by the F99 law for a given $R_V$. The intrinsic line intensities are then calculated from the observed values using the following relation.

\begin{equation}
    F_{\text{int}}(\lambda) = F_{\text{obs}}(\lambda) \times 10^{0.4 A_\lambda}
\end{equation}

where $F_{\text{obs}}(\lambda)$ is the observed line intensity. The uncertainties in $F_{\text{obs}}(\lambda)$ and $E(B-V)$ are propagated through this relation to estimate errors on $F_{\text{int}}(\lambda)$. The resulting extinction parameters are then used to correct the observed line fluxes, from which we derive updated values of $n_e$ and $T_e$ (see Sections~\ref{subsec:ne} and \ref{subsec:te}). With these revised physical conditions we recompute the intrinsic HI ratios with \texttt{PyNeb}, refit the extinction curve, and iterate this cycle until convergence is achieved in $n_e$, $T_e$, $R_V$ and $E(B-V)$, in this case three times.

\begin{figure}[t]
    \centering
    \includegraphics[width=0.99\columnwidth]{\detokenize{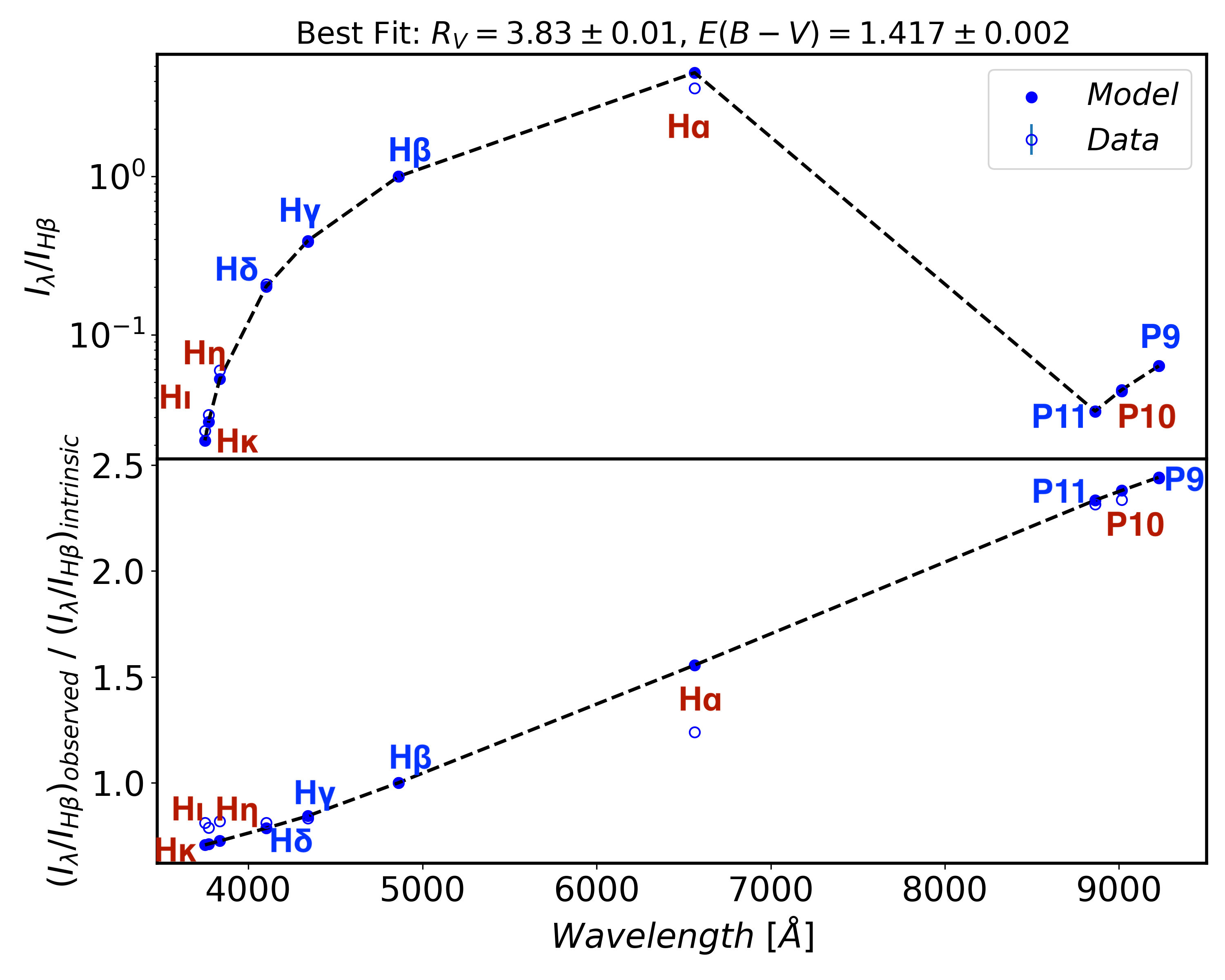}}
    \caption{Upper panel: Best-fit $R_V$ and $E(B-V)$ obtained using Balmer and Paschen lines to \hb ratios for the average spectrum. Lower panel: Observed and modeled line ratios of HI lines w.r.t. the \hb relative to theoretical line ratios without dust. Solid circles represent the modeled ratios, and hollow circles represent the observed ratios. The best fit is calculated using the \hg, \hd, \hb, P\,11 and P\,9; these line labels are annotated in blue, while the rest are in red.}
    \label{fig:figure4}
\end{figure}

The resulting extinction parameters, along with the observed and dust-corrected line intensities for the average spectrum, are presented in Table~\ref{tab:table5}. The best-fit extinction model is shown in Figure \ref{fig:figure4} and discussed in Section~\ref{sec:results}.

\begin{table*}[t]
\centering
\caption{Observed and dust corrected intrinsic intensities with uncertainties (in units of percentage of \hb), for the average spectrum of the entire M\,8 region, average spectrum of NGC~6523 and the central bin containing HG region and the Her~36 star. Observed and intrinsic \hb intensities are in $erg\,s^{-1} cm^{-2} arcsec^{-2}$.}
\label{tab:table5}
\resizebox{0.99\textwidth}{!}
{
\begin{tabular}{ccccccc}
\hline
Line & M\,8-average & & NGC~6523-average & & HG region & \\
\hline
\hline
 & $F_{\mathrm{obs}} \pm \sigma_{\mathrm{obs}}$ & $F_{\mathrm{int}} \pm \sigma_{\mathrm{int}}$& $F_{\mathrm{obs}} \pm \sigma_{\mathrm{obs}}$ & $F_{\mathrm{int}} \pm \sigma_{\mathrm{int}}$& $F_{\mathrm{obs}} \pm \sigma_{\mathrm{obs}}$ & $F_{\mathrm{int}} \pm \sigma_{\mathrm{int}}$ \\
\hline
\oii$\lambda$3726.03      & $66.945   \pm  0.009$ & $95.4 \pm 0.4$ & $57.337 \pm 0.001$ & $79.98 \pm 0.03$ & $74.69 \pm 0.04$ & $102.24\pm0.05$\\
\oii$\lambda$3728.81      & $88.61  \pm 0.01$ & $126.2 \pm 0.5$ & $61.715 \pm 0.001$ & $86.02 \pm 0.03$ & $68.39\pm0.05$ & $93.54\pm0.05$\\
\hk$\lambda$3750.15       & $2.459   \pm 0.007$ & $3.48 \pm 0.02$ & $2.5410 \pm 0.0008$ & $3.521 \pm 0.002$ & $2.574\pm0.008$ & $3.50\pm0.01$\\
\hi$\lambda$3770.63       & $3.109   \pm 0.007$ & $4.37 \pm 0.02$ &  $3.1993 \pm 0.0008$ & $4.408 \pm 0.002$ & $3.197\pm0.009$ & $4.33\pm0.01$ \\
\he$\lambda$3835.38       & $5.928   \pm 0.007$ & $8.17 \pm 0.03$ & $6.0100 \pm 0.0008$ & $8.133 \pm 0.003$ & $6.067\pm0.009$ & $8.08\pm0.01$ \\
\hd$\lambda$4101.74       & $20.823 \pm 0.006$ & $26.50 \pm 0.10$ & $20.9666 \pm 0.0007$ & $26.37 \pm 0.01$ & $21.32\pm0.01$ & $21.33\pm0.01$ \\
\hg$\lambda$4340.46       & $38.707  \pm 0.007$ & $45.89 \pm 0.17$ & $39.0979 \pm 0.0008$ & $46.01 \pm 0.02$ & $39.40\pm0.02$ & $46.06\pm0.02$ \\
\oiii$\lambda$4363.2      & $0.309   \pm 0.009$ & $0.36 \pm 0.01$ & $0.441 \pm 0.001$ & $0.516 \pm 0.001$ & $0.460\pm0.004$ & $0.535\pm0.004$ \\
\orl$\lambda$4638.86      & $0.068  \pm 0.005$ & $0.073  \pm 0.006$ & $0.0877 \pm 0.0006$ & $0.0943 \pm 0.0007$ & $0.078\pm0.002$ & $0.084\pm0.002$ \\
\orl$\lambda$4641.81      & $0.063  \pm 0.006$ & $0.068 \pm 0.006$ & $0.1037 \pm 0.0007$ & $0.1114 \pm 0.0007$ & $0.097\pm0.002$ & $0.104\pm0.002$ \\
\orl$\lambda$4649.13      & $0.049  \pm 0.006$ & $0.052 \pm 0.006$ & $0.0918 \pm 0.0006$ & $0.0984 \pm 0.0007$ & $0.080\pm0.002$ & $0.086\pm0.002$ \\
\orl$\lambda$4650.84      & $0.081  \pm 0.006$ & $0.087 \pm 0.006$ & $0.0909 \pm 0.0006$ & $0.0974 \pm 0.0007$ & $0.069\pm0.002$ & $0.073\pm0.002$ \\
\orl$\lambda$4661.63      & $0.066  \pm 0.005$ & $0.069 \pm 0.005$ & $0.0760 \pm 0.0006$ & $0.0811 \pm 0.0007$ & $0.074\pm0.01$ & $0.079\pm0.002$ \\
\orl$\lambda$4676.23      & $0.034  \pm 0.005$ & $0.036 \pm 0.006$ & $0.0377 \pm 0.0006$ & $0.0400 \pm 0.0006$ & $0.039\pm0.002$ & $0.041\pm0.002$ \\
\hb$\lambda$4861.32       & $100.000 \pm 0.004$ & $100.00  \pm 0.03$ & $100.000 \pm 0.001$ & $100.00 \pm 0.04$ & $100.00\pm0.03$ & $100.00\pm0.03$ \\
\oiii$\lambda$4958.91     & $42.074  \pm 0.007$ & $40.69  \pm 0.14$ & $59.6686 \pm 0.0009$ & $57.78 \pm 0.02$ & $56.09\pm0.02$ & $54.38\pm0.02$ \\
\oiii$\lambda$5006.84     & $134.87 \pm 0.01$ & $128.34 \pm 0.44$  & $187.975 \pm 0.002$ & $179.24 \pm 0.06$ & $173.74\pm0.04$ & $165.95\pm0.04$ \\
\nii$\lambda$5754.64     & $0.802 \pm 0.006$ & $0.619 \pm 0.005$ & $0.7545 \pm 0.0007$ & $0.5911 \pm 0.0006$ & $1.020\pm0.003$ & $0.811\pm0.002$ \\
\nii$\lambda$6548.04     & $32.312 \pm 0.006$ & $20.84 \pm 0.06$ & $27.3147 \pm 0.0007$ & $17.728 \pm 0.006$ & $35.87\pm0.01$ & $23.326\pm0.007$ \\
\ha$\lambda$6562.80 \tablenotemark{e} & $361.08 \pm 0.03$ & $231.18 \pm 0.70$ & $290.717 \pm 0.002$ & $187.98 \pm 0.06$ & $192.66\pm0.04$ & $124.76\pm0.03$ \\
\nii$\lambda$6583.46     & $96.537 \pm 0.009$ & $61.79 \pm 0.19$ & $78.3409 \pm 0.0009$ & $50.39 \pm 0.02$ & $99.23\pm0.02$ & $63.87\pm0.02$ \\
\sii$\lambda$6716.44      & $26.621  \pm 0.006$ & $16.55  \pm 0.05$ & $16.5914 \pm 0.0007$ & $10.316 \pm 0.003$ & $17.578\pm0.007$ & $10.876\pm0.004$ \\
\sii$\lambda$6730.82      & $20.046  \pm 0.006$ & $12.42  \pm 0.04$ & $15.2494 \pm 0.0007$ & $9.447 \pm 0.003$ & $18.667\pm0.007$ & $11.500\pm0.004$ \\
P\,10$\lambda$8862.78     & $3.257  \pm 0.007$ & $1.396   \pm 0.005$ & $3.6581 \pm 0.0008$ & $1.4261 \pm 0.0005$ & $4.147\pm0.003$ & $1.445\pm0.001$ \\
P\,11$\lambda$9014.91     & $4.394  \pm 0.007$ & $1.846   \pm 0.006$ & $4.9853 \pm 0.0008$ & $1.8950 \pm 0.0006$ & $5.628\pm0.003$ & $1.900\pm0.001$ \\
P\,9$\lambda$9229.01      & $6.334  \pm 0.007$ & $2.594   \pm 0.007$ & $7.0318 \pm 0.0008$ & $2.5839 \pm 0.0008$ & $7.935\pm0.004$ & $2.567\pm0.001$ \\
\hline
$R_V$ & $3.83\pm0.01$ & & $4.781\pm0.001$& &  $6.00\pm$0.01\\
$E(B-V)$ & $1.416\pm0.002$ & & $1.6963\pm0.0002$& & $2.013\pm 0.001$\\
$A_V$ & $5.42\pm0.02$& & $8.109\pm0.002$ & & $12.08\pm0.09$ &\\
\hline
\hline
\hb Intensity & $2.3942 \pm 0.0002 \times 10^{-15}$ & $1.052  \pm 0.003 \times 10^{-14}$ & $2.04646 \pm 0.00002 \times 10^{-14}$ & $1.1512 \pm 0.0003 \times 10^{-13}$ &$4.478\pm0.001 \times 10^{-14}$ & $3.350\pm0.001 \times 10^{-13}$\\
\hline
\end{tabular}%
}
\tablenotetext{e}{Saturated in the central region.}
\end{table*}


Using the spatially resolved $n_e$ and $T_e$ across the region produces a nearly identical spatial distribution of $R_V$ and $E(B-V)$ to those obtained using average physical conditions. Because the differences are negligible, we use the global values of $T_e = 8025$ K and $n_e =116$~cm$^{-3}$ for all spatially resolved and annular-binned spectra, ensuring consistency across the FOV and providing greater robustness in the low $S/N$ outskirts of the nebula.

\begin{figure}[t]
    \centering
    \includegraphics[width=0.999\columnwidth]{\detokenize{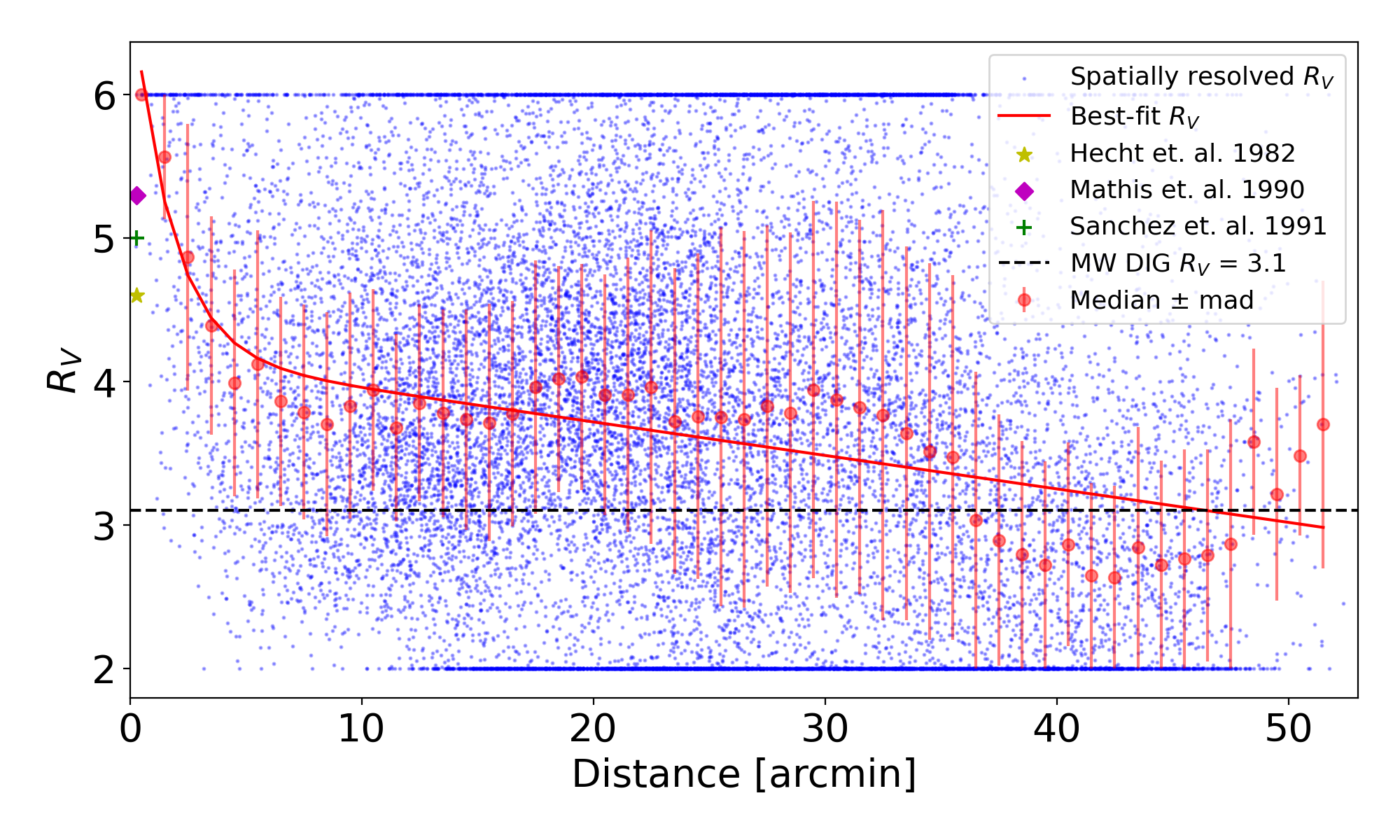}}
    \caption{Radial variation of $R_V$. Spatially resolved data are in blue; the best fit $R_V$(r) is drawn in red follows the median measurements of $R_V$ in 1$\arcmin$ bins. The black dashed lines show the standard $R_V$ of 3.1, used for the Milky Way \hii regions. Previous measurements of $R_V$ in the HG region are also added for comparison. Since the F99 model limits the $R_V$ between 2-6, we see a concentration of data points at the upper and lower limits of the dust extinction model.}
    \label{fig:figure5}
\end{figure}

Following the methodology described above to determine $R_V$, we note that the spaxel-by-spaxel variation of best-fit $R_V$ is noisy. The concentration of data points at the upper and lower bounds of the F99 model reflects the imposed limits of the extinction parameter space rather than the intrinsic physical boundaries. However, a clear radial trend emerges when taking the median in 1$\arcmin$ wide bins, starting at a radius of 0.5$\arcmin$, similar to the size of an LVM spaxel. The values near the central region are in approximate consistency with previously measured values in the HG region \citep[such as][]{hecht82, mathis90, sanchez91}. To capture this behavior, we adopt a fixed radial profile by combining an exponential function with a linear function, for $R_V$ as a function of the projected distance from Her\,36 (distance in units of arcminutes, see Figure~\ref{fig:figure5}). 

The spatially resolved $R_V$ distribution shows a central peak, reaching the limit of $\approx R_V=6$, set by the implementation of F99 extinction model used in this work. In fact, the resulting best-fit function $R_V(r)$, $R_v(r)= 2.663 * exp(-r / 1.704) -0.023 * r + 4.185$ formally extrapolates to values above this limit (i.e., $R_V > 6$) in radii, r$<0.65\arcmin$. We adopt $R_V = 6$ as a representative value for radii$<0.65\arcmin$. 

Beyond $0.65\arcmin$, we adopt the $R_V(r)$ profile derived from the fitting (see the best-fit equation $R_V(r)$ above and in Figure~\ref{fig:figure5}), which yields values significantly higher than the standard Galactic average of 3.1. The resulting $R_V(r)$ profile is then used to assign the $R_V$ values to individual spaxels based on the projected distance from the ionizing star, after which $E(B-V)$ is refitted at the corresponding fixed $R_V(r)$.

The increase of $R_V$ toward the ionizing star in the plane of the sky suggests that the intense radiation field from the central massive star modifies the dust properties along the line of sight. Large $R_V$ values indicate a dust population dominated by large grains, producing a flatter extinction curve and a reddening that is less dependent on wavelength \citep{draine11, stasinska23}. Such conditions are expected in dense, highly irradiated environments, where small grains can be selectively destroyed by hard ionizing photons, leading to an enhancement of the large-grain population \citep[e.g.][]{hecht82}. This interpretation is further supported by independent dust tracers, such as enhanced Fe depletion towards the ionizing source  \citep{mendez-delgado24}.

This hybrid strategy leverages the flexibility of our method while improving robustness in noisy regions. The resulting radially adjusted $R_V$ values and the corresponding spaxel-wise $E(B-V)$ distribution are used to determine $A_{\lambda}$ for each spaxel. We then apply the dust correction procedure described above to all emission lines in the resolved spectra. 

For the binned analyses, we derive $R_V$ and $E(B-V)$ from the binned spectra themselves. In the first case, we construct annular bins centered on Her\,36, each 1$\arcmin$ wide, to trace the radial variation of dust properties across the FOV. In the second case, we apply the same procedure to spectra extracted within the elliptical aperture encompassing the bright \hii region, NGC~6523. For the first case, as previously mentioned, we adopt the global values of $T_e$ and $n_e$ to compute the $R_V$ and $E(B-V)$. In contrast, for the NGC~6523 aperture, which is concentrated towards the nebular center, we use a representative temperature derived from its average spectrum, as it provides a more appropriate characterization of the aperture. In both cases, the resulting best-fit radial profile of $E(B-V)$ is consistent with the spatially resolved trend. The $R_V$ and $E(B-V)$ profiles are then used to derive the $A_{\lambda}$ and obtain dust-corrected line intensities for the binned spectra.

The final spatial distributions of the $E(B-V)$ and dust-corrected emission line maps are shown in Sections~\ref{sec:results} and in the online-only figure set, respectively. 


\subsection{Electron Density ($n_e$) from CELs}
\label{subsec:ne}

Electron density CEL diagnostics are based on the measurement of line ratios involving transitions of the same ion with similar upper energy levels but differing in collisional excitation or de-excitation rates. These ratios are sensitive to the electron density because the ratio of collisional de-excitation rates from the two upper levels varies with $n_e$, while being independent of the ionic abundance and largely independent of the electron temperature \citep{osterbrock06}.

In this work, we study two well-established density-sensitive diagnostics based on the \sii$\lambda$6717/\sii$\lambda$6731 and \oii$\lambda$3726/\oii$\lambda$3729 ratios. Both of these ions are sensitive to a similar range of $10^{2} < n_{e} < 10^{4}$ cm$^{-3}$ \citep[see][]{eduardo23}, as they emit from the same volume of gas and they also have similar transitions, metastable rates and critical densities (of order $10^3$ cm$^{-3}$). 

Although these ratios are mostly sensitive to $n_e$, the collisional excitation rates have a mild dependence on $T_e$, especially at low densities. Therefore, a better approach is to use a true representative $T_e$ of the gas. To achieve this, we follow a similar iterative procedure outlined in Section \ref{subsec:dust}. We first estimate $n_e$ from the dereddened \sii $\lambda$6731/\sii $\lambda$6717 and \oii$\lambda$3726/\oii$\lambda$3729 line ratios using the \texttt{getTemDen} routine in \texttt{PyNeb}, applied to the M\,8 average spectrum, assuming an initial $T_e=10^4$K. This initial $n_e$ is then used to determine the $T_e$ following the methodology described in Section~\ref{subsec:te}. The updated $T_e$ is subsequently used to recompute $n_e$, and we achieve a convergence toward an accurate $n_e$ value after three iterations.

The final global $T_e$ value derived from the average spectrum is adopted for computing the electron densities in the individual spaxels and in the annular binned spectra across the FOV. For binned spectra within the elliptical aperture encompassing NGC~6523, we instead use the $T_e$ derived from its average spectrum and perform three iterations to verify convergence in the resulting $n_e$ values.


However, given spatial systematics affecting the \oii$\lambda\lambda$3726, 3729 ratio map in the current data reduction, we adopt $n_e$ derived from the \sii$\lambda\lambda$6717, 6731 diagnostic as input for the determination of other nebular properties, such as $T_e$, ionic and elemental abundances in this work.

The spatial variation of $n_e(S^{+})$ ion across the FOV is presented and discussed in Section \ref{sec:results}.

\subsection{Electron Temperature ($T_e$) from CELs}
\label{subsec:te}
CELs have emissivities that depend exponentially on the $T_e$, and linearly on the ion and electron densities \citep[see][]{osterbrock06}. 

Electron temperature is determined from the intensity ratios of CELs of the same ion but involving transitions from energy levels with significantly different excitation energies \citep[see][]{osterbrock06}. These ratios are highly sensitive to $T_e$ but only weakly dependent on density in the low-density limit, where collisional de-excitations can be neglected. Accurate determination of $T_e$ is critical to derive gas-phase abundances via the DM, which relies on CEL emissivities. 

In our analysis, we adopt a two-zone ionization scheme, in which the gas contributing to each spectrum is assumed to be in two separate low-ionization and high-ionization regimes. Each region is characterized by a representative $T_e$, determined from CELs emitted by dominant ionic species, based on their ionization potentials. In this work, we focus on determining $T_e$ in the low-ionization zones using singly ionized species ($O^{+}$ and $N^{+}$), and for the high-ionization zone using doubly ionized oxygen ($O^{2+}$). Neutral oxygen ($O^{0}$) lines, such as \oi$\lambda$6300 and $\lambda$6363, are not detected in our data because at the spectral resolution of LVM, they blend with bright \oi sky lines. 
 
We compute $T_e$ for both ionization regimes using the \texttt{getTemDen} routine from \texttt{PyNeb}. 
For the calculations, we use $n_e(S^{+})$ and the auroral-to-nebular CELs ratios: \oiii$\lambda$4363/$\lambda$5007 ($T_e^{CEL}(O^{2+})$) for the high ionization zone , and \oii$\lambda\lambda$7320,7330/$\lambda\lambda$3726,3729 ($T_e^{CEL}(O^{+})$) and \nii$\lambda$5755/$\lambda\lambda$6548,6584 ($T_e^{CEL}(N^{+})$) for the low ionization zone. Uncertainties in the dereddened line intensities and input $n_e$ are propagated through, and the resulting distributions are used to estimate the uncertainties in $T_e$. 

The $T_e$ measurement procedure is applied independently to the spatially resolved, annular binned, and average spectra, each using its corresponding $n_e$ estimate. Since the \oii$\lambda\lambda$7320,7330/$\lambda\lambda$3726,3729 ratio is more affected by dust extinction and sky line contamination, due to its wide baseline and proximity to strong sky lines in the red channel, as well as its sensitivity to $n_e$ \citep{eduardo23}, in the following analysis we adopt $T_e^{CEL}(N^{+})$ as the representative temperature for the low-ionization zone.

\subsection{Electron Temperature ($T_e$) from RLs/CEL}
\label{subsec:terl}

\begin{figure}[t]
    \centering
    \includegraphics[width=0.99\columnwidth]{\detokenize{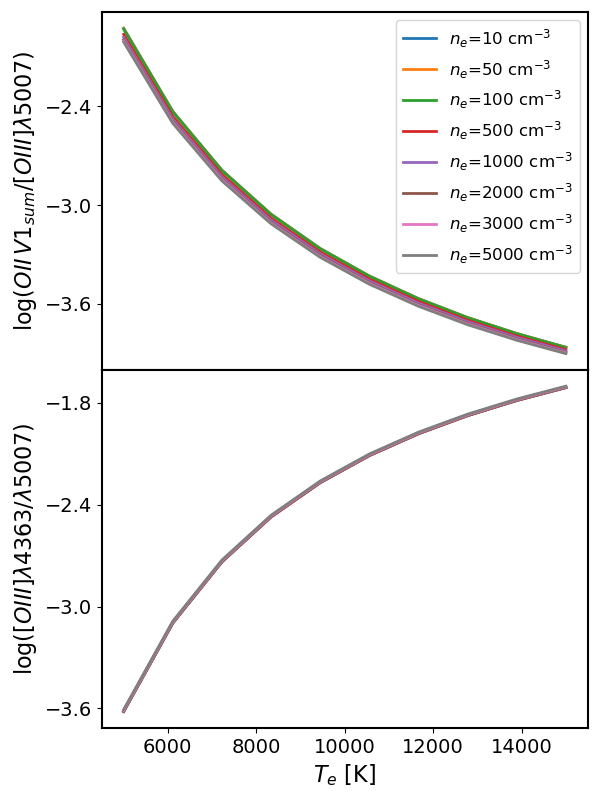}}
    \caption{Upper panel: Variation of the ratio of \orl V1 sum to the \oiii$\lambda$5007 with respect to $T_e$ for $n_e$ values ranging from 10-5000 cm$^{-3}$; at lower $T_e$ the \orl intensities dominate, as the $T_e$ increases, the CEL intensity dominates, therefore the ratio decreases. Lower panel: \oiii$\lambda$4363/\oiii$\lambda$5007 ratio shows a negligible dependence on the $n_e$ in comparison to the former ratio, as both are CELs, leading to a monotonic increase of the ratio with $T_e$. It is evident that both these ratios are extremely sensitive to the $T_e$. The y-axes are on logarithm scale.}
    \label{fig:figure6}
\end{figure}

For the high ionization zone, we also compute the electron temperature using a diagnostic that combines \orl RLs and \oiii CEL. We refer to this quantity as $T_e^{RLs/CEL}(O^{2+})$. In particular, we use the ratio of the intensity of \orl V1 multiplet recombination emissions, spanning 4635-4680$\AA$, relative to \oiii $\lambda$5007. The total \orl intensity is obtained as the sum of six resolved components of \orl at 4638.86, 4641.81, 4649.13, 4650.84, 4661.63 and 4676.23 $\AA$.  This ratio is sensitive to $T_e$ due to the very different dependence that RLs and CELs have on temperature, while the emissivity of RLs decreases roughly linearly with temperature \citep{osterbrock06}. 

\begin{equation}
    \epsilon_{RL} \propto T_e^{-m}
    \label{eq:rl_emissivity_te}
\end{equation}

where $\epsilon$ is the line emissivity, and $m$ is a constant. This relationship holds approximately over a limited $T_e$ range (5000-20000K), with $m \simeq 1$ for many RLs; however, the precise value of $m$ varies depending on the specific RLs transitions \citep{storey17}.

Thus, at fixed density, the above ratio changes strongly as a function of $T_e$. To quantify this dependence and explore the effects of $n_e$ variations on this method, we compute the theoretical emissivity ratios of the summed \orl RL multiplet to \oiii $\lambda$5007 over a grid of $T_e$ and $n_e$ values using the \texttt{PyNeb} \texttt{getEmissivity} routine. 

The upper panel of Figure ~\ref{fig:figure6} presents the dependence of the \orl V1 sum to \oiii$\lambda$5007 ratio with electron temperature, for several assumed values of $n_e$. The equivalent plot for the auroral to forbidden CEL ratio \oiii$\lambda$4363/$\lambda$5007, used in the classic direct method, is shown in the lower panel of the same figure. The RLs/CEL temperature diagnostic is as sensitive to $T_e$ as the direct method, although in the opposite direction (i.e., the line ratio decreases instead of increasing as the temperature grows). The density dependence of the RLs/CEL diagnostic is larger than that of the direct method, and it therefore requires an independent estimate of the $n_e$ in order to provide accurate estimations of $T_e$.

We then use a two-dimensional interpolation function to infer $T_e$ from this grid, for any combination of $n_e$ and observed \orl V1/\oiii$\lambda$5007 ratio. We estimate $T_e^{RLs/CEL}(O^{2+})$ for all types of spectra (individual spaxels, radially binned, and average), using their corresponding $n_e(S^{+})$, as measured in Section~\ref{subsec:ne}. The resulting map and radially varying physical conditions are presented and analyzed in Section \ref{sec:results}. 

\subsection{Ionic (O$^+$/H$^{+}$, O$^{2+}$/H$^{+}$) and elemental (O/H) abundances from CELs}
\label{subsec:abund_cel}
We derive the ionic abundance of $O^+$ and $O^{2+}$ from CELs using \texttt{PyNeb}. The O$^+$/H$^+$ abundance is derived from the relative intensity of the \oii$\lambda\lambda$3726,3729 doublet to that of \hb, using the $T_e^{CEL}(N^{+})$ and $n_e(S^{+})$. Similarly, the O$^{2+}$/H$^+$ abundance is determined using the relative intensities of \oiii$\lambda\lambda$4959, 5007 to that of \hb, in combination with $T_e^{CEL}(O^{2+})$ derived from \oiii$\lambda$4363/$\lambda$5007, and $n_e(S^{+})$.

The methodology is independently applied to all spectra (individual spaxels, radially binned, and average). As mentioned previously, neutral oxygen \oi is not detected in our data, because its emission lines overlaps with strong sky lines such as \wave5577, \wave 6300. It is therefore not included in the  abundance calculation. Moreover, since the  ionization potential of \oi (12.1~eV) is very close to that of HI (13.6~eV), the large majority of oxygen inside an \hii region will be in the ionized form. As a result, \oi emissions originate primarily in the surrounding photo-dissociation region (PDR). Additionally, $O^{3+}$ emission is absent in our M\,8 data, as its production requires extremely hot ionizing sources ($T_{eff}>100,000$K), typically found only in PNe with very hot central stars. Therefore, the elemental oxygen abundance in an \hii region can be directly calculated by combining O$^+$/H$^{+}$ and O$^{2+}$/H$^{+}$  as \[
\frac{N(O)}{N(H)} = \frac{N(\text{O}^+)}{N(\text{H}^+)}+ \frac{N(\text{O}^{2+})}{N(\text{H}^+)}
\]
The resulting ionic and total oxygen abundance measurements are presented and discussed in Section \ref{sec:results}.

\subsection{Ionic (O$^{2+}/H^{+}$) from RLs and elemental abundances from RLs+CELs}
\label{subsec:abund_rl}
We also derive the O$^{2+}$ ionic abundance from RLs, making use of the six resolved components of Multiplet 1 of \orl at 4638.86, 4641.81, 4649.13, 4650.84, 4661.63 and 4676.23 $\AA$. 

\begin{equation} 
\label{eq:rl_abund}
\frac{N(\mathrm{X}^{i+1})}{N(\mathrm{H}^+)} \;=\; \frac{\lambda}{4861\AA} \, \frac{\gamma^{\text{eff}}(\text{H}\beta)}{\gamma^{\text{eff}}(\lambda)} \, \frac{I(\lambda)}{I(\text{H}\beta)} 
\end{equation}

where $I(\lambda)$ are the dereddened intensities of the RLs at wavelengths $\lambda$ relative to \hb, and $\gamma^{eff}$ denotes the effective recombination coefficient for the given line. In practice, this expression applies to each RL individually. However, the relative intensities of the \orl lines within the multiplet~1 are affected by the departure from the local thermodynamic equilibrium (LTE) in the population of the upper levels. This leads to a density-dependent redistribution of intensity among individual components \citep[e.g., see][]{ruiz03}. For this reason, the abundance derived from a single \orl line can be strongly biased. 

A robust way to mitigate this issue is to use the sum of all observed components of the multiplet~1 to estimate the abundance. While the relative strengths of individual lines within the multiplet are highly sensitive to $n_e$, their combined intensity is comparatively weakly dependent on density, making it a more robust abundance diagnostic \citep[see][]{ruiz03, peimbert13}. We also note that all individual lines of multiplet~1 are independent of $n_e$ in the range of $n_e<100 \,cm^{-3}$; consequently, their sum is also independent (Singh et. al. in prep.). 

We therefore adopt the total measured intensity of the six lines. 

\begin{equation}
    I(\text{V1}) \, = \, \sum_{i=1}^6 I(\lambda_i),
\end{equation}
and use this in the abundance calculation, replacing $I(\lambda)$ in equation~\ref{eq:rl_abund} with $I(\text{V1})$ and the corresponding multiplet recombination coefficient.

Although the absolute emissivities of the RLs retain a strong (roughly linear) dependence on $T_e$, the \orl$/$HI line ratios effectively cancel out this dependence \citep{osterbrock06}. To ensure consistency with our CEL analysis, we adopt $T_e$(\oiii) and $n_e(S^{+})$ to determine O$^{2+}$ abundance with \texttt{PyNeb} using summed \orl V1 intensity relative to \hb as input. 

In addition, we estimate the total oxygen abundance using the O$^{2+}$ RLs. Ideally, oxygen elemental abundance based on RLs is obtained by adding $O^{+}$ and $O^{2+}$ abundance derived from RLs. Since $O^{+}$ lines such as OI $\lambda$6046, $\lambda$7002 are not detected in our dataset, we adopt a hybrid approach: we combine the $O^{+}$ abundance derived from the CELs with the $O^{2+}$ abundance derived from the RLs to compute the total oxygen abundance.

\subsection{Abundance Discrepancy Factor (ADF)}

After calculating both the CELs and RLs-based O$^{2+}/H^{+}$ abundance for the spatially resolved, annular binned, and average spectra, we compute the ADF following the standard definition:

\begin{equation} \text{ADF} = \log\left(\frac{N(\text{X}^{i+1})}{N(\text{H}^+)}\right)_{\text{RL}} - \log\left(\frac{N(\text{X}^{i+1})}{N(\text{H}^+)}\right)_{\text{CEL}} 
\end{equation}

The uncertainties in the ADF are estimated by propagating the uncertainties in both CELs and RLs ionic abundances.

\section{Results}
\label{sec:results}

\begin{figure*}[t]
    \centering
    \includegraphics[width=0.75\textwidth]{\detokenize{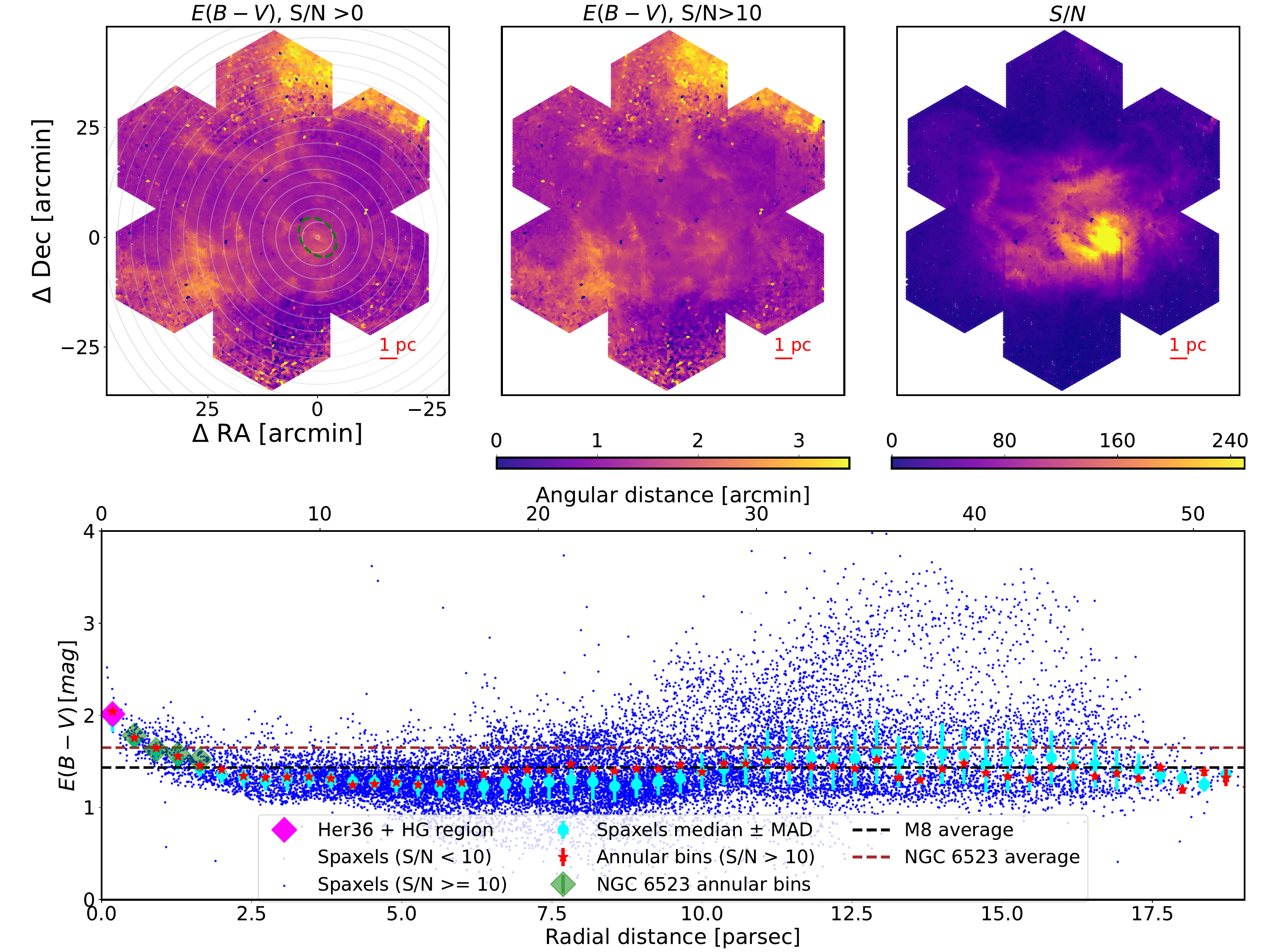}} 
    \caption{Spatially resolved maps and radial profile of $E(B-V)$. Upper panels: (a) $E(B-V)$ map with overlaid annular bins, each $3\arcmin$ wide for clear visualization of structures at relative distance; the dashed green elliptical aperture indicates the NGC~6523 region. (b) $E(B-V)$ map showing only spaxels with $H\beta$ detected above a $10\sigma$ threshold (others masked in white), and (c) corresponding $S/N$ map of $E(B-V)$. Lower panel: Radial profile of $E(B-V)$: blue points show all spaxels with $10\sigma$ $H\beta$ detections (same as in panel b) whereas the grey points show measurements with $S/N$ less than the threshold, cyan points represent the median $\pm$ median absolute deviation (MAD) in radial bins of $1\arcmin$ in which at least $50\%$ of spaxels have $S/N$ higher than the threshold, red star markers correspond to annular binned measurements, and the green rhombus markers show annular binned measurements in the NGC~6523 elliptical region. The black dashed line shows the average $E(B-V)$ across the FOV, and the brown dashed line indicates the average measurement of the elliptical region. The black dashed line denotes the average $E(B-V)$ across the field. The magenta marker shows the central bin containing the Hourglass region and Her\,36.}
    \label{fig:figure7}
\end{figure*}

\begin{figure*}[h]
    \centering
    \includegraphics[width=0.75\textwidth, keepaspectratio]{\detokenize{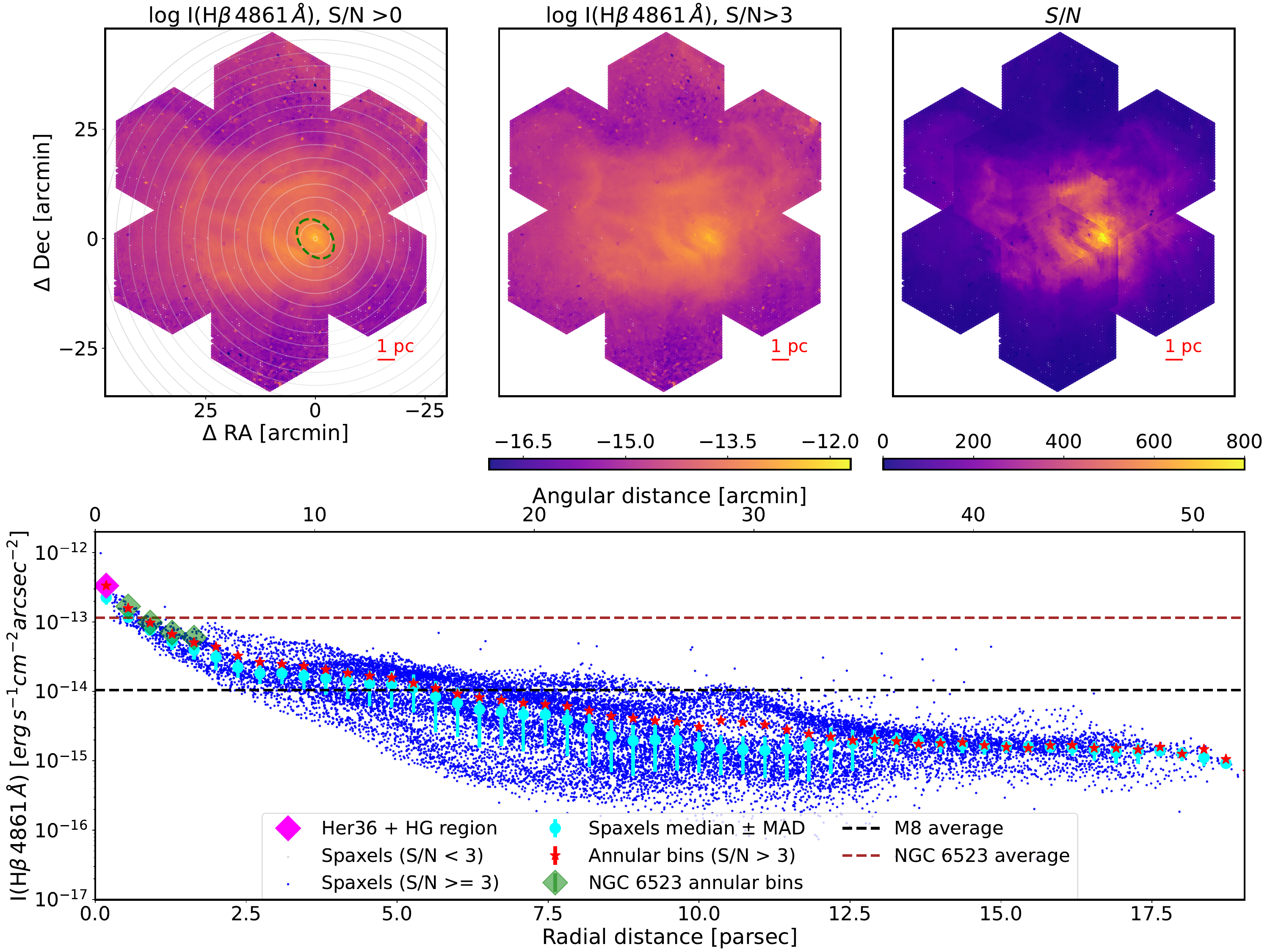}} 
    \caption{Spatially resolved maps and radial profile of \hb intensity (in units of erg s$^{-1}$ cm$^{-2}$ arcsec$^{-2}$). The panels and labels follow the same convention as in Figure~\ref{fig:figure7}, but for \hb intensity after dust correction, with a $S/N$ cut of 3 on the line.}
    \label{fig:figure8}
\end{figure*} 

\begin{figure*}
    \centering
    \includegraphics[width=0.75\textwidth, keepaspectratio]{\detokenize{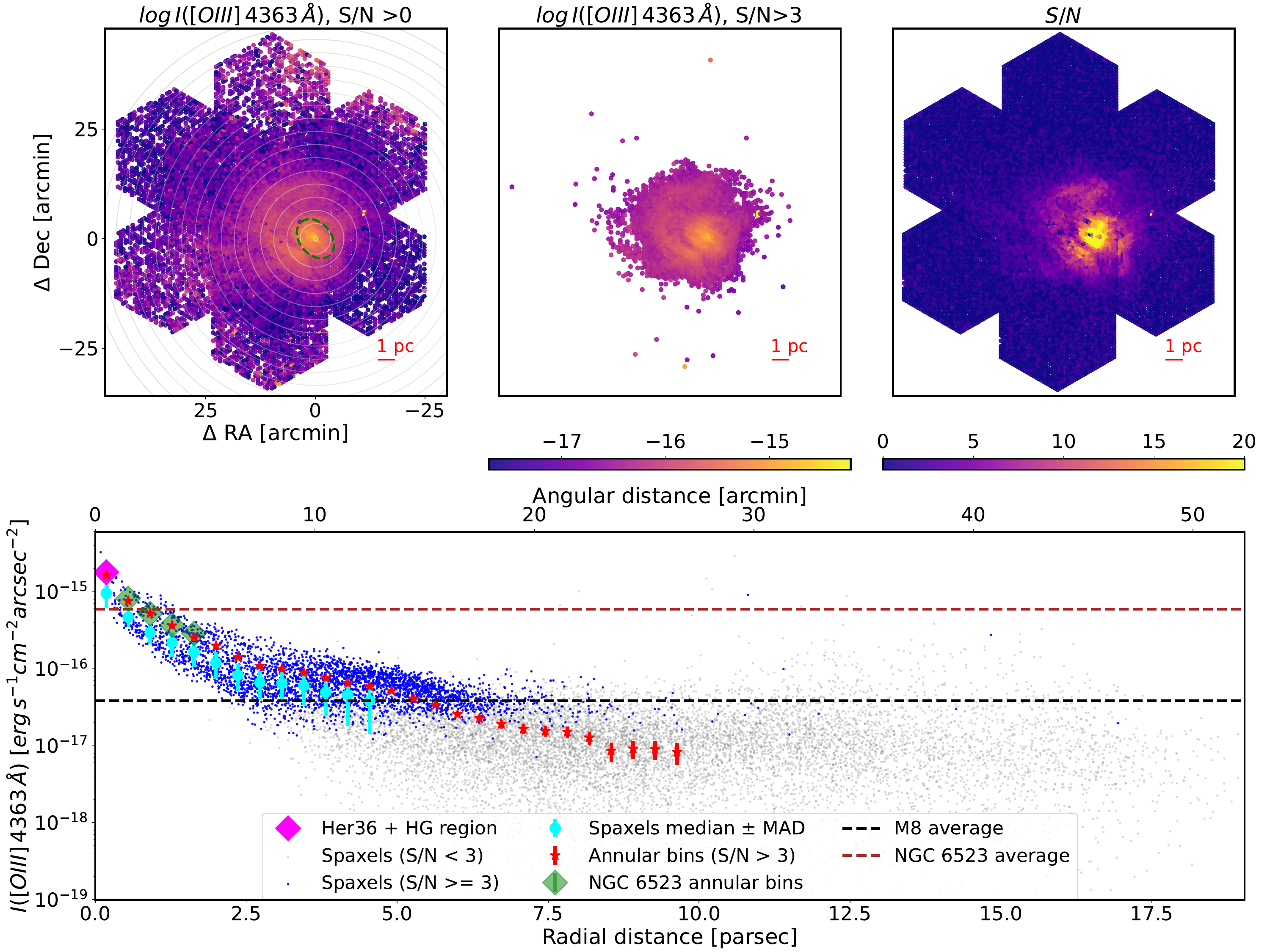}}
    \caption{Spatially resolved maps and radial profile of \oiii$\lambda$4363 intensity (in units of erg s$^{-1}$ cm$^{-2}$ arcsec$^{-2}$). The panels and labels follow the same convention as in Figure~\ref{fig:figure7}, but here the map shows \oiii$\lambda$4363 dust-corrected intensity, with a $S/N$ cut of 3 on the line.}
    \label{fig:figure9}
\end{figure*}

\begin{figure*}
    \centering
    \includegraphics[width= 0.75\textwidth, keepaspectratio]{\detokenize{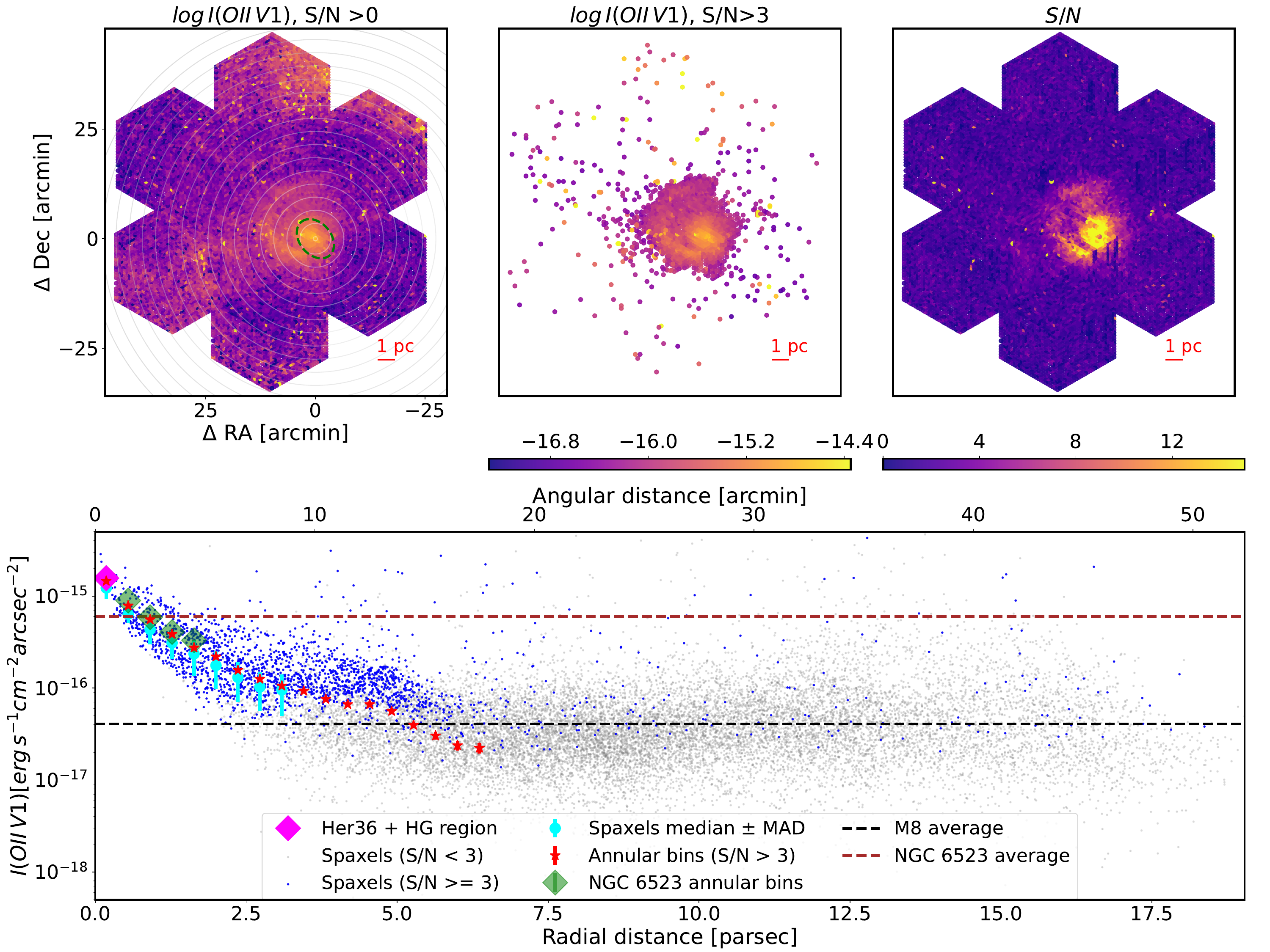}}
    \caption{Spatially resolved map and radial profile of the composite \orl RL intensity across the full field of view, in units of erg s$^{-1}$ cm$^{-2}$ arcsec$^{-2}$. The panels and labels follow the same convention as in Figure~\ref{fig:figure7}, but here the map shows the sum of dust-corrected intensities of the six detected lines of \orl RL V1, see Table~\ref{tab:table4}, in this dataset, with a $S/N$ cut of 3 applied to the brightest pair of lines of the multiplet (\orl$\lambda\lambda$4649+4651).}
    \label{fig:figure10}
\end{figure*}

In this section, we present and analyze the physical quantities derived from the LVM data across the M\,8 nebula. We explore the spatial distribution of the $E(B-V)$, $T_e$, $n_e$, the oxygen ionic and elemental abundances derived from CELs and RLs, and the ADF. 

We examine these quantities in spatially-resolved maps and radial profiles. The radial profiles display  information derived from seven distinct spatial aggregations: (1) individual spaxels, (2) the median $\pm$ MAD of individual spaxels in radial bins of $1\arcmin$ in which at least $50\%$ of spaxels have $S/N$ higher than the threshold, (3) measurements performed on the annular binned spectra in the FOV, and on (4) annular binned spectra within the NGC~6523, elliptical region (i.e., including only gas ionized primarily by Her\,36), (5) measurements conducted on the average spectrum of NGC~6523, (6) the average spectrum of the whole M\,8 region, and (7) the first, central bin containing both the HG region and the Her\,36 system (i.e., the most studied region of M\,8 in terms of nebular diagnostics in the literature).

\subsection{Dust Extinction}
\label{subsec:results_dust}

Figure ~\ref{fig:figure7} presents the spatial distribution of the color excess, $E(B-V)$ across M\,8, ranging from approximately 0~mag to 4.0~mag, with a mean and standard deviation of 1.43 and 0.45~mag, respectively. The annular binned spectra show a narrower range of $E(B-V)$, from 1.19~mag to 2.04~mag, with a mean and standard deviation of 1.40 and 0.14~mag, respectively. The annular bins in the elliptical aperture enclosing NGC~6523 yield a mean and standard deviation of 1.70 and 0.18~mag, respectively, indicating a higher dust content toward the center. This value is consistent with the reddening derived from the average spectrum of this region, $E(B-V)=1.6963 \pm 0.0002$~mag. The average spectrum for the entire FOV gives $E(B-V) = 1.417\pm0.002$~mag. The reddening caused by dust peaks sharply in the dense region around Her\,36, decreases steadily across the inner 2~pc of the region, then flattens, and rises again towards the edge of the nebula, at the transition into the molecular cloud.

In terms of extinction (adopting the best-fit $R_V$ profile presented in Figure \ref{fig:figure5}), we find a mean and standard deviation of $A_V=$ 5.42 $\pm$ 0.02~mag from the average spectrum of the entire FOV. Within the NGC~6523 region, the annular bins yield $A_V =8.109 \pm 0.001$~mag. 

Although the fitted $R_V$ varies substantially with radius, the resulting $A_V$ values remain nearly unchanged because different $R_V$ values lead to correspondingly different best-fit color-excesses; in this case, a mean $\sim$0.3 mag higher than the values obtained when adopting the canonical $R_V=3.1$, preserving the relation $A_V = R_V \times E(B-V)$.

The derived extinction values are used to correct the observed emission line intensities, using the F99 \citep{fitzpatrick99} extinction law, to recover the intrinsic line intensities (see Section~\ref{subsec:dust}), which form the basis of all the following nebular diagnostics. Table~\ref{tab:table5} lists the observed and dust extinction corrected line intensities for all emission lines used in this work, as measured in the average spectrum of the FOV, the average spectrum of NGC~6523, and the central bin spectrum encompassing the HG region and the Her\,36 star.

\subsection{Emission Line Maps and Radial Profiles}
\label{subsec:result_line_maps}

A comprehensive atlas of emission-line maps and radial profiles used in this study is available as an online-only figure set. This section presents only a subset of emission-line maps relevant to the analysis. For all the lines listed in Table \ref{tab:table5}, we provide spatially resolved, extinction-corrected intensity maps and radial profiles. This includes HI RLs such as \hk, \hi, \he, \hd, \hg, \hb (see Figure \ref{fig:figure8}), \ha, P\,11, P\,10, P\,9; nebular CELs such as \oii $\lambda\lambda$3726, 3729, \oiii $\lambda$5007, \nii $\lambda\lambda$6548, 6584 and \sii $\lambda\lambda$6717, 6731, auroral CELs, specifically \oiii $\lambda$4363 (see Figure~\ref{fig:figure9}) and \nii $\lambda$5755; and both individual and summed \orl RLs (see Figure \ref{fig:figure10}).


In these figures, we also present S/N maps, and line intensity maps after applying a $S/N>3$ detection threshold, effectively removing regions with poorly measured intensities. All flux maps are displayed on a logarithmic scale to enhance the visibility of faint structures. Radial line-intensity profiles are presented as a function of projected distance from Her 36, in both arcminutes and parsecs. 

The HI RL intensity maps exhibit smooth variations. The Balmer and Paschen line intensities peak in the NGC~6523 region, reaching a maximum within the HG region, which hosts the primary ionizing source Her\,36, identified as the ionizing center of the nebula. Similarly, the highly-ionized nebular CEL \oiii$\lambda$5007 and the auroral CEL \oiii$\lambda$4363 also show pronounced gradients, with intensities peaking near the ionizing center and decreasing outwards. While \oiii$\lambda$5007 emission is detected at high $S/N$ across the whole nebula, \oiii$\lambda$4363 emission is only detected at $S/N>3$ out to $\approx$ 10 pc. The \oiii$\lambda$5007 line is detected at a significance level greater than $3\sigma$ in 99.96\% of the spaxels, whereas the auroral \oiii$\lambda$4363 line exceeds this threshold in only 22\% of the field, consistent with the significantly lower collisional excitation potential for \oiii$\lambda$5007 compared to \oiii$\lambda$4363.

Low-ionization species such as \nii $\lambda\lambda$6548, 6584, \sii$\lambda\lambda$6717, 6731, and \oii$\lambda\lambda$3726, 3729, along with the auroral line \nii$\lambda$5755, exhibit more extended and filamentary morphologies across the FOV than the high-ionization species (see the online figure set). 
Notably, their radial profiles differ substantially: low-ionization lines trace structures that are not apparent in the high-ionization lines. Such a separation between low- and high-ionization radial behaviors is generally not apparent in IFU studies of nearby galaxies \citep[see][]{kehrig12}; this highlights the importance of high-resolution IFU observations of Galactic \hii regions, such as those provided by LVM. Their enhanced brightness in the HG region further reflects the features associated with the blister around Her\,36 and the adjacent PDRs \citep{kahle24}. Subsequently, they decrease steeply across the inner 2~pc, and show a flatter radial profile across the bulk of the nebula, consistent with them tracing the outer lower ionization shells of the cavities associated with this complex region.

Spatially resolved maps of individual \orl RL from the V1 multiplet are provided in an online figure set. Figure~\ref{fig:figure10} shows the composite map obtained by summing all six detected \orl RLs, tracing the spatial distribution of the V1 multiplet emission. To the best of our knowledge, this is the first study to present spatially resolved maps for individual \orl RLs from multiplet 1, each typically $10^3- 10^4$ dex fainter than \ha, and their combined emission across an entire Galactic \hii region. Together with the work presented in \cite{hilder25} based on LVM observations of the Rosette nebula, these maps reveal the power of ultra-wide-field IFU spectrographs like the LVM-I, to spectrally characterize low surface brightness nebular structures over large areas in the sky, in ways that are inaccessible to previously existing facilities.

The spatially resolved composite map (Figure~\ref{fig:figure10}), along with its corresponding radial profile, reveals detectable ($S/N>3$) \orl emission out to $\sim$ 7~pc from the ionizing center. Individual \orl RL maps show similar spatial extents, except for the faintest of the six detected \orl RLs: $\lambda$4676 line, which is only detected out to $\sim$5~pc. If the chemical composition of the gas is well mixed, we can expect both \orl RLs and the \oiii $\lambda$4363 CELs to originate from the same spatial regions. The 2-D spatial distributions and radial profiles of \oiii $\lambda$4363, $\lambda$5007 and \orl RLs are indeed very similar. The average emission line intensities exhibit exceptionally high $S/N$ (see Figure \ref{fig:figure3b}), with the weakest line, \orl$\lambda$4676, which is $\sim$7000 times fainter than \ha in the average spectrum (Table \ref{tab:table5}), having $S/N \approx 6$.

In general, most of the emission line intensity maps reveal three prominent \hii regions within the FOV: NGC~6523 (extending up to $\sim$2~pc), which also includes the HG region and Her\,36 in its first bin; NGC 6526, ionized primarily by Sgr 9 and the NGC 6530 cluster; and the third region associated with the HD 165052 binary (see Section~\ref{sec:lagoon} for detailed information). The radial profiles of all emission lines exhibit a significantly higher intensity in the HG region, supporting the hypothesis that Her 36 is embedded within the dense, dusty and molecular environment. The observed emission from low-ionization species is primarily associated with the ionization front advancing into this dense gas \citep{tothill08}. 



\begin{table*}[t]
\centering
\caption{Physical conditions and chemical abundances from the average spectra of the FOV and NGC~6523, as well as from the central bin containing the HG region. Systematic uncertainties errors arising from relative flux calibration are shown in brackets, together with the formal errors.}
\begin{tabular}{cccc}
\hline
Parameter  & M\,8-average & NGC~6523-average   & Her\,36+HG region\\
\hline
\hline
Electron density, $n_e$ cm$^{-3}$ & & &\\
\hline

$n_e(S^{+})$  & 116 $\pm$ 3 (49) & 470 $\pm$ 5 (70)& 858$\pm$2(33) \\
\hline
Electron temperature, $T_e$ [K] & & &\\
\hline
$T_e^{CEL}(N^{+})$  & 8628 $\pm$ 22 (70) &9076$\pm$ 4 (70) & 9339$\pm$10 (77)\\
$T_e^{CEL}(O^{2+})$ & 8025 $\pm$64 (19)  & 8027$\pm$ 5 (18) & 8239$\pm$12 (38)\\
$T_e^{RLs/CEL}(O^{2+})$ & 6364 $\pm$10 (48)  &6340$\pm$ 3 (39) & 6354$\pm$12 (30)\\
\hline
Ionic abundances [12+$\log_{10}$(X$^{i+}$/H$^+$)] & & &\\
\hline
CEL - X$^{i+}$ = O$^{+}$      & 8.141 $\pm$ 0.002 (0.004) & 7.9855$\pm$0.0001 (0.005)&  8.0454$\pm$0.0002 (0.0004) \\
CEL - X$^{i+}$ = O$^{2+}$     & 8.021 $\pm$ 0.001 (0.005) & 8.1526$\pm$0.0001 (0.005)&  8.0699$\pm$0.0001 (0.0005)  \\
RL - X$^{i+}$ = O$^{2+}$    & 8.49 $\pm$ 0.02 (0.005) & 8.633$\pm$0.001 (0.005) &  8.594$\pm$0.004 (0.008)   \\
\hline
Elemental abundances [12+$\log_{10}$(X/H)] & & &\\
\hline
CEL - X= O    &   8.39 $\pm$ 0.002 (0.007) & 8.3783$\pm$0.0002 (0.007) & 8.3588$\pm$0.0002 (0.0007)\\
RL+CEL - X= O    &   8.65$\pm$ 0.006 (0.007) & 8.721$\pm$0.007 (0.007)& 8.702$\pm$0.004 (0.007)\\
\hline
\hline
ADF ($O^{2+}$)      & 0.47 $\pm$ 0.02 (0.01)  & 0.481$\pm$0.001 (0.007)& 0.524$\pm$0.004 (0.009) \\
\hline
\end{tabular}

\label{tab:table6}
\end{table*}

\subsection{Physical conditions from average spectra}
\label{subsec:results_intg}

Table \ref{tab:table6} summarizes the physical conditions, chemical abundances and the ADF derived from the average spectrum of the entire M\,8 region, the NGC~6523 subregion, and the first bin centered on the HG region and Her\,36. The formal uncertainties in these measurements are extremely small, reflecting the extremely high $S/N$ of the corresponding spectrum and the associated small errors in the measured line fluxes (see Table \ref{tab:table5}). However, these measurements are also subject to systematic uncertainties, including those arising from atomic data and flux calibration uncertainties, which are not included in the formal errors. To estimate the impact of relative flux calibration systematics, we adopt a $\sim5\%$ calibration uncertainty for the LVM data across the spectral range. To account for this effect, we include an additional wavelength-dependent calibration term in the MC error propagation of line fluxes to determine nebular diagnostics (see Sections~\ref{subsec:ne}--\ref{subsec:abund_rl}). We use a conservative flux calibration uncertainty that increases linearly from $0\%$ at 3600 $\AA$ to $5\%$ at 9800 $\AA$. The uncertainties in the atomic data used for the diagnostics calculation contribute an additional systematic component. These are typically smaller than $10\%$ for the relevant collision strengths and transition probabilities \citep[see][]{morisset20}. We report these systematic errors in parentheses in Table \ref{tab:table6} to indicate the plausible range of variation introduced by flux calibration and atomic data systematics.

The derived $n_e(S^{+})$ across these three regions vary appreciably within a range of $30-860\,$ cm$^{-3}$, with the highest measurement in individual spaxels reaching up to $1840\pm33(258) \,$ cm$^{-3}$ in the brightest spaxel in the HG region. For the NGC~6523 region, we see a steep decline with an average measurement of $\sim 470\pm70\,cm^{-3} $, which drops to $116\pm43\,cm^{-3}$ when averaging across the whole M\,8 nebula.

In all three spectra, the CEL diagnostics show the usual stratification $T_e^{CEL}(N^{+})$ $>$ $T_e^{CEL}(O^{2+})$, consistent with expectations for Galactic \hii regions with solar or super-solar metallicities, where $T_e^{CEL}(O^{2+})$ typically exhibits temperatures below 10,000 K \citep[see Table 1 in both][]{rojas07, rodriguez10}. The derived CEL $T_e$, from $N^{+}$ and $O^{2+}$, ranges between $\sim 7000 - 10000$ K, with substantial differences ($\pm 2500$~K) between M\,8, NGC~6523, and the central bin with HG. The HG bin stands out with a higher low-ionization temperature; we attribute this to the local ionization fronts that dominate the emission in the dense, dusty and molecular environment immediately surrounding Her\,36. The HG region is still transitioning from being embedded in a molecular gas to the ionized gas, due to a progressing ionization front; thus, the low-ionization gas lies much closer to the heating source and is consequently hotter than in the larger-scale regions. 

In contrast, the RLs+CEL based hybrid $T_e$ for $O^{2+}$ are remarkably uniform across all three regions. We find a mean and standard deviation of $T_e^{RLs/CEL}$($O^{2+}) = 6352\pm5 (23)$ K across these regions, fully consistent with the individual measurement for these regions within their uncertainties. The stability of the hybrid $T_e$ compared to the CELs-based $T_e$ likely reflects the higher sensitivity of the latter to the presence of temperature fluctuations. The temperature difference between these two diagnostics implies an offset of $\Delta T_e$ of $\sim$1600 K.

In terms of abundance derived from CELs, the average M\,8 spectrum shows $O^{+}/H^+$ exceeding $O^{2+}/H^+$, indicating that the singly ionized state dominates the overall oxygen budget in M\,8. In contrast, both NGC~6523 and the HG region exhibit higher $O^{2+}/H^+$ fractions, consistent with their proximity to the main ionizing sources, where the stronger radiation field favors higher ionization conditions. The RL-based $O^{2+}/H^+$ abundance is systematically higher than its CEL counterpart for the FOV. 

The CEL-based elemental abundance determined by summing $O^{+}/H^+$ and $O^{2+}/H^+$ is $\sim$0.26 times lower than the hybrid elemental abundance calculated by summing CEL-based $O^{+}/H^+$ and RL-based $O^{2+}/H^+$. We caution the reader that this quantity is different from the classical ADF, which is the logarithmic difference between RL and CEL abundances of the $O^{2+}$ ion.


\begin{figure*}[t]
    \centering 
    \includegraphics[width=0.75\textwidth, keepaspectratio]{\detokenize{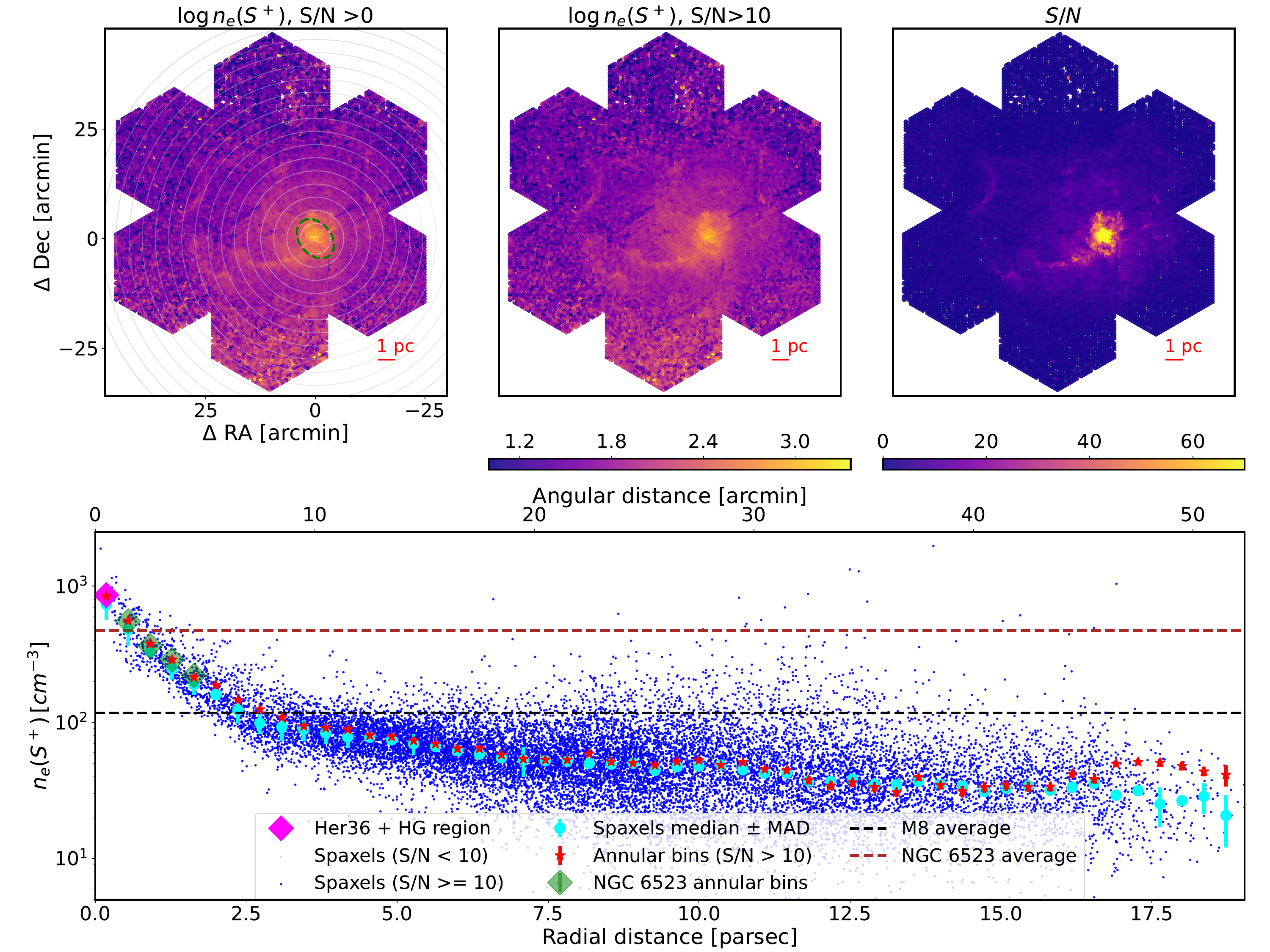}}
    \caption{Spatially resolved map and radial profile of $n_e(S^{+})$. Y-axis and color bars are in logarithmic scale. Labels on the plot are the same as in figure~\ref{fig:figure7} but for $n_e(S^{+})$, with a $S/N > 10$ is applied on \sii$\lambda\lambda$6717,6731 in this plot. 
    }
    \label{fig:figure11}
\end{figure*}

The average ADF is found to be 0.47$\pm$0.02 dex for entire M\,8. Similar high ADFs are measured in the other two sub-regions analyzed separately, with ADFs of 0.481$\pm$0.007 in NGC~6523 and 0.524$\pm$0.004 in the HG region. These values lie at the upper end of the ADFs reported for Galactic \hii regions, \citep[see Table 8 of][]{rojas07}.


\subsection{Spatially resolved physical conditions: maps and radial profiles}
\label{subsec:results_phys_cond}

To investigate the origin of the O abundance discrepancy across M\,8 at sub-pc scales, we analyze the spatially resolved physical and chemical conditions of the gas, such as $T_e$, $n_e$, and the ionic and elemental abundances derived from both CELs and RLs. Here we present spatially resolved maps and radial profiles for these quantities. All maps and profiles shown in this section are subjected to a $S/N$ threshold, applied to: (i) the relevant auroral lines in the case of CEL diagnostics; (ii) the brightest pair of \orl V1 lines, $\lambda\lambda$4649+4651, in the case of RL based diagnostics; (iii) the \sii nebular CEL in case of the density; and (iv) the \oii nebular CEL for the $O^{+}$ abundance. For diagnostics where both CELs and RLs are involved, we apply the $S/N$ threshold to both auroral CEL and \orl V1 $\lambda\lambda$4649+4651 lines.

\subsubsection{Electron Density}

Figure~\ref{fig:figure11} shows the spatially resolved electron density map and its radial profile derived from the \sii line ratio. In the LVM data, the electron density is measured at high significance across the entire field. It exhibits a sharply peaked, $\sim$5~pc wide, central overdense structure at the core of NGC~6523, with $n_e(S^{+})$ reaching its peak in the central bin. The electron density falls by roughly one order of magnitude across this central structure, reaching $\sim100$  cm$^{-3}$ at a radial distance of 2.5~pc. Outside the central structure, the electron density traces a complex superposition of ionization fronts at the edges of different bubbles, which merge and overlap across the larger nebula. In these outer regions, the electron density continues to decrease, although following a shallower radial profile, until leveling at around $\sim$35 cm$^{-3}$ at radii $>12$~pc.

\begin{figure*}
    \centering
    \includegraphics[width=0.75\textwidth]{\detokenize{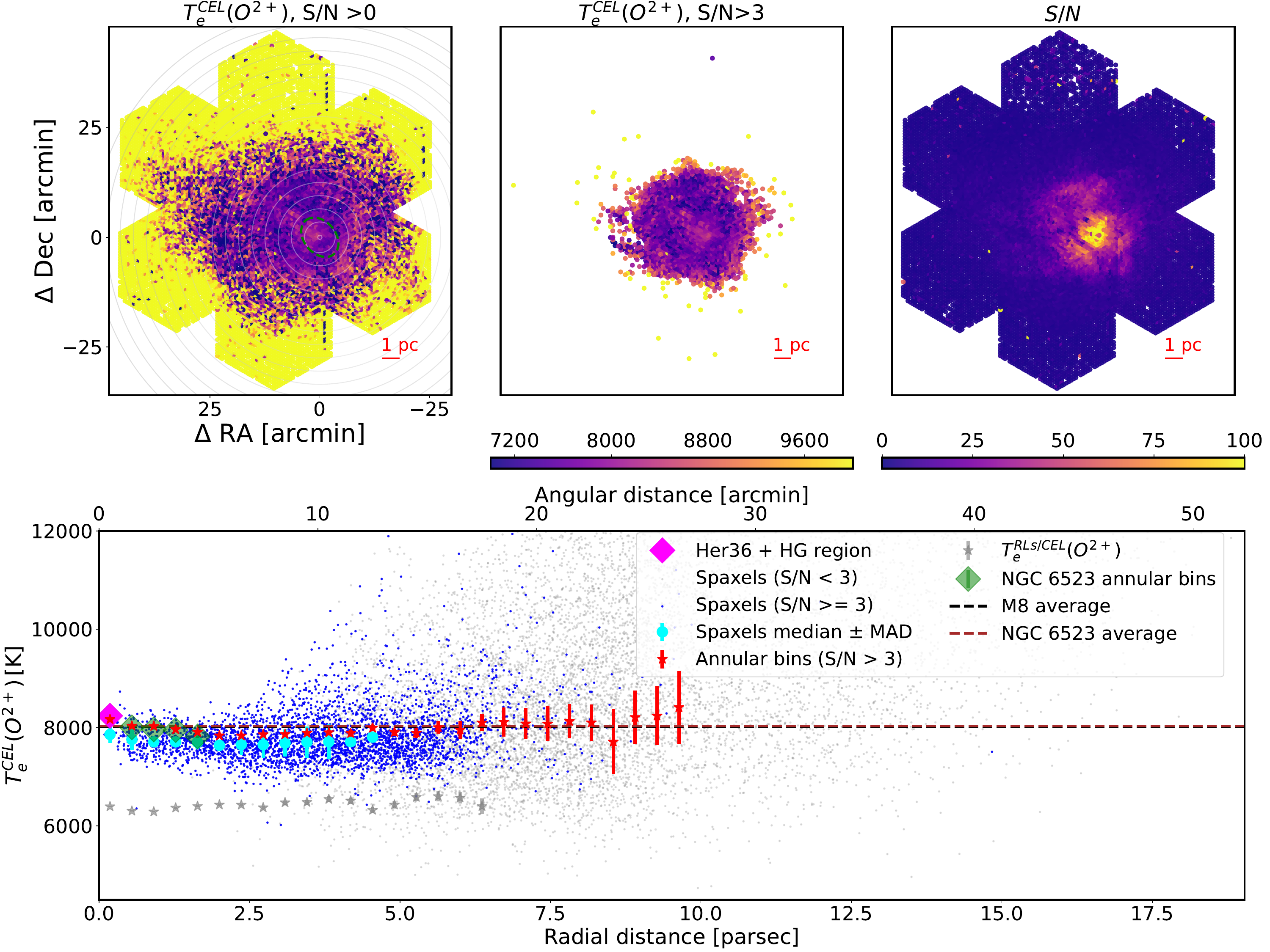}}
    \caption{Spatially resolved map and radial profile of $T_e^{CEL}(O^{2+})$. Labels on the plot are the same as in figure~\ref{fig:figure7} but for $T_e^{CEL}(O^{2+})$, with a $S/N> 3$ cut on \oiii$\lambda$4363 in this plot. Measurements derived from the complimentary diagnostic are shown with grey '*' for comparison.}
    \label{fig:figure12}
\end{figure*}
\begin{figure*}
    \centering
    \includegraphics[width=0.75\textwidth]{\detokenize{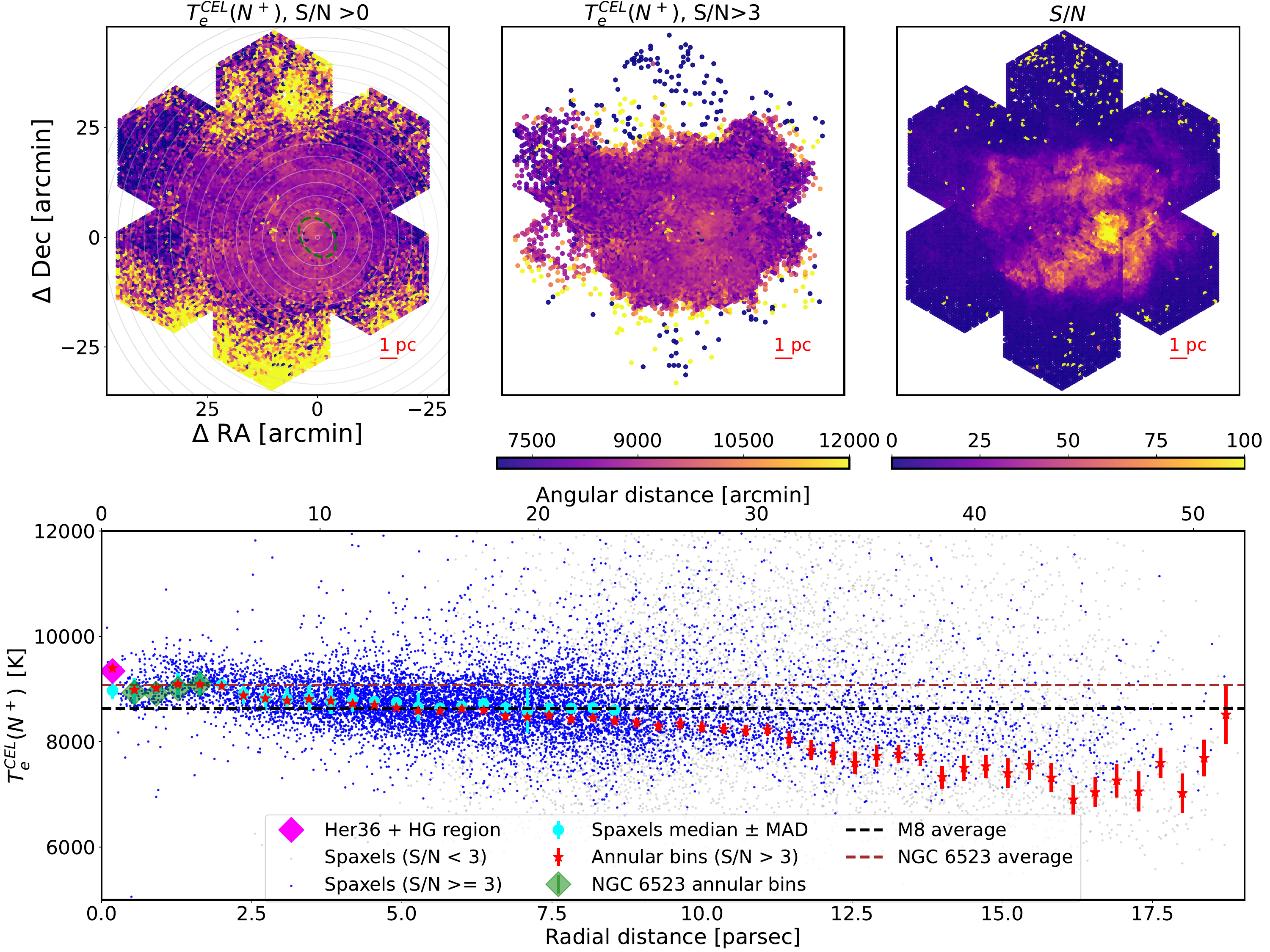}}
    \caption{Spatially resolved map and radial profile of $T_e^{CEL}(N^{+})$. Labels on the plot are the same as in figure~\ref{fig:figure7} but for $T_e^{CEL}(N^{+})$, with a $S/N> 3$ cut on \nii$\lambda$5755 in this plot.
    }
    \label{fig:figure13}
\end{figure*}

\begin{figure*}
    \centering
    \includegraphics[width=0.75\textwidth]{\detokenize{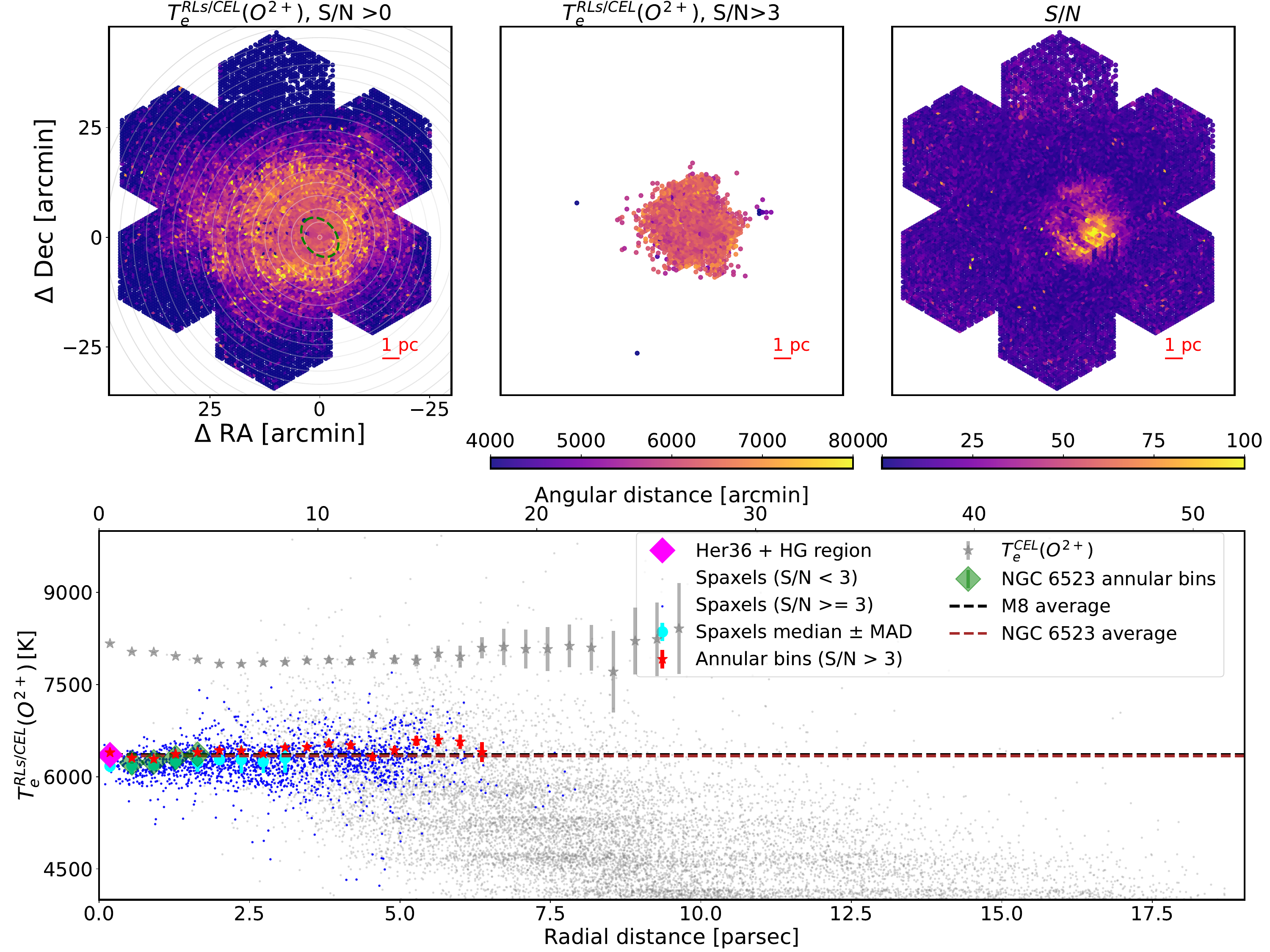}}
    \caption{Spatially resolved map and radial profile of $T_e^{RLs/CEL}(O^{2+})$. Labels on the plot are the same as in figure~\ref{fig:figure12} but for $T_e^{RLs/CEL}(O^{2+})$, with a $S/N> 3$ cut on both \orl $\lambda\lambda$4649+4650 and \oiii $\lambda$4363 in this plot.}
    \label{fig:figure14}
\end{figure*}

\begin{figure*}
    \centering
    \includegraphics[width=0.75\textwidth]{\detokenize{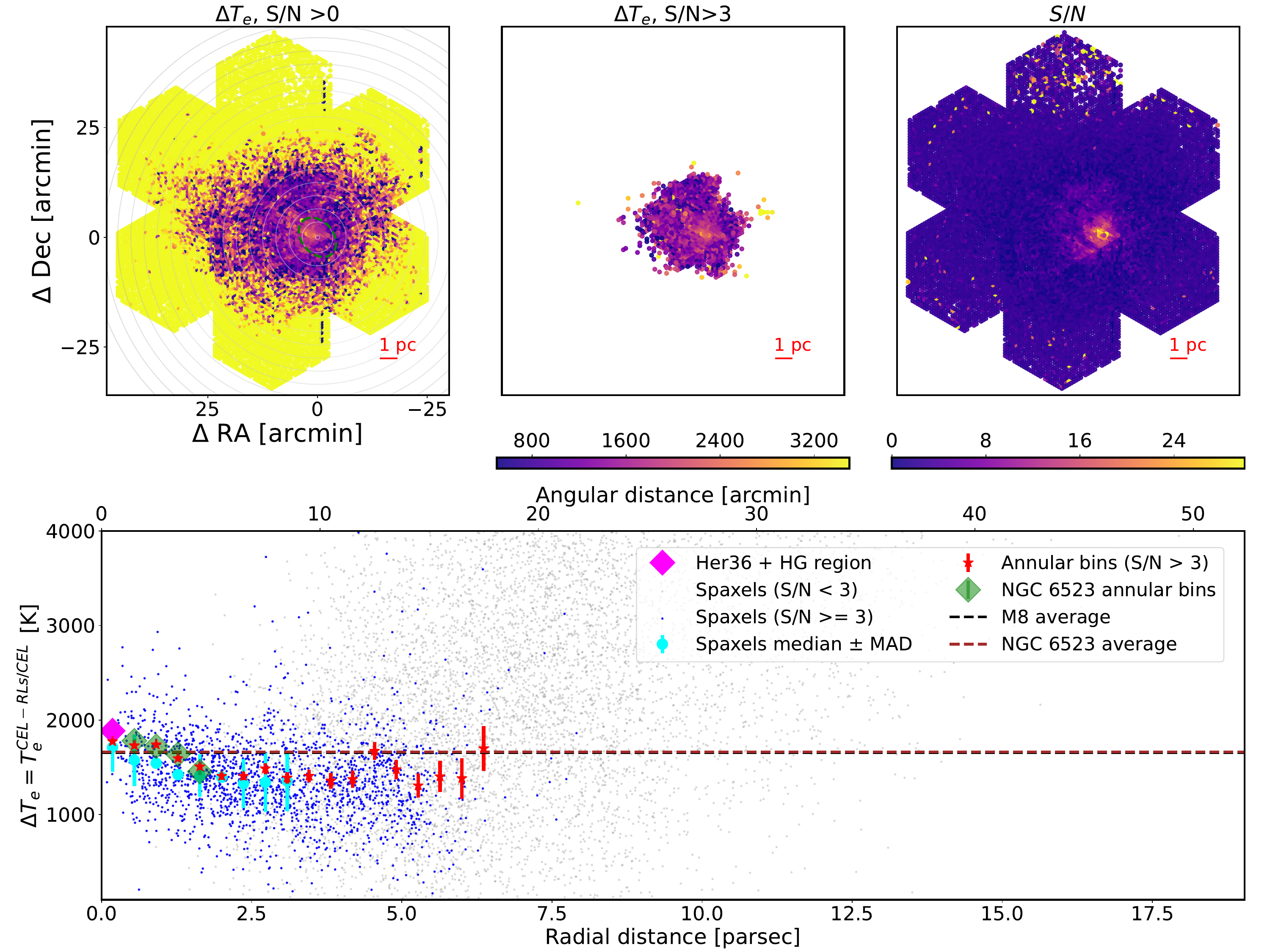}}
    \caption{Spatially resolved map and radial profile of $\Delta T_e = T_e^{CEL} - T_e^{RLs/CEL}$. Labels and $S/N$ cut on this plot are the same as in figure~\ref{fig:figure14} but for $\Delta T_e$.}
    \label{fig:figure15}
\end{figure*}

\subsubsection{Electron Temperatures}

Figures \ref{fig:figure12} and \ref{fig:figure13} present the spatial distributions and radial profiles of the CEL-based electron temperature for the high ($T_e^{CEL}(O^{2+})$) and low-ionization zones ($T_e^{CEL}(N^{+})$). The $T_e^{CEL}(O^{2+})$ map and its corresponding radial profile reveal an enhancement near the ionization center, with a peak temperature in the central bin that covers the HG region, followed by a decline to a minimum temperature of 7834$\pm$44 K at a radial distance of $\sim$2~pc. The temperature then slowly rises towards the outskirts of the nebula, reaching a constant value of $\simeq8000$~K and staying flat beyond a radius of 6~pc.

The  spatial distribution of low-ionization gas temperature ($T_e^{CEL}(N^{+})$) shows an increase in the bin that covers the HG region, dropping to values closer to $\simeq9000$~K within the central pc of the nebula. It is then followed by a plateau of constant temperature, which remains uniform up to 2~pc, and a slow decline towards the edge of the nebula ($\sim$17~pc), reaching values as low as $\simeq7000$~K in the outer regions. 

Figure~\ref{fig:figure14} presents the spatially resolved map and radial profile of RLs+CEL based hybrid $T_e^{RLs/CEL}(O^{2+})$. This quantity exhibits a smooth and almost constant profile, with only a small variation of $<200$K away from the center, all the way out to $\sim7$~pc, beyond which RLs are no longer detected at sufficient S/N. Note that the values of $T_e^{RLs/CEL}(O^{2+})$ are systematically lower than $T_e^{CEL}(O^{2+})$. These temperatures are $\sim1700$K lower than those measured from CELs alone, and the $T_e^{RLs/CEL}(O^{2+})$ profile lacks the central peak in temperature seen in the $T_e^{CEL}(O^{2+})$ profile.

As I(\oiii $\lambda$4363/$\lambda$5007) and I(\orl V1/$\lambda$5007) depend on $T_e$ in different ways (see Section~\ref{subsec:terl} for details), the observed discrepancy between $T_e$ values derived from CELs and RLs/CEL may be indicative of the presence of underlying thermal inhomogeneities. Figure~\ref{fig:figure15} presents the map and radial profile of this temperature discrepancy, defined as

\begin{equation}
\Delta T_e= T_e^{CEL}(O^{2+})-T_e^{RLs/CEL}(O^{2+})
\end{equation}

It is evident that this temperature discrepancy is higher in the high-ionization zones towards the center of NGC~6523. The temperature discrepancy peaks at $\sim$1900 K in the central bin, drops steadily across the inner 2~pc, reaching a value of $\sim$1400, and stays roughly flat all the way out to 7~pc.

\subsubsection{Ionic And Elemental O Abundances}
\label{subsubsection:results_abund_adf}

\begin{figure*}
    \centering
    \includegraphics[width=0.75\textwidth]{\detokenize{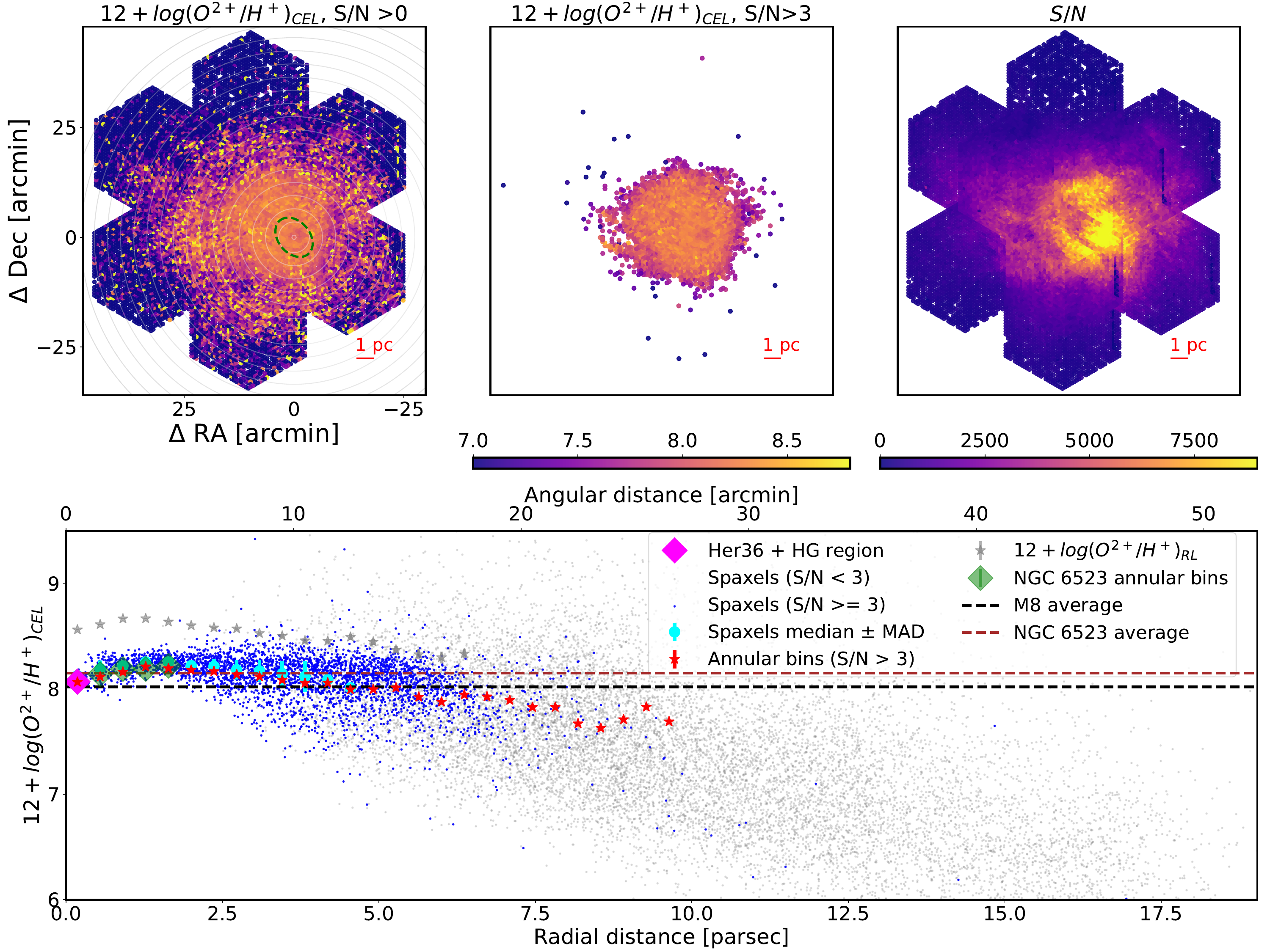}}
    \caption{Spatially resolved map and radial profile of CEL based $12+\log(O^{2+}/H^{+})$. Labels on the plot are the same as in figure~\ref{fig:figure12} but for $12+\log(O^{2+}/H^{+})$, with a $S/N> 3$ cut on \oiii$\lambda$4363 in this plot.}
    \label{fig:figure16}
\end{figure*}

\begin{figure*}
    \centering
    \includegraphics[width=0.75\textwidth]{\detokenize{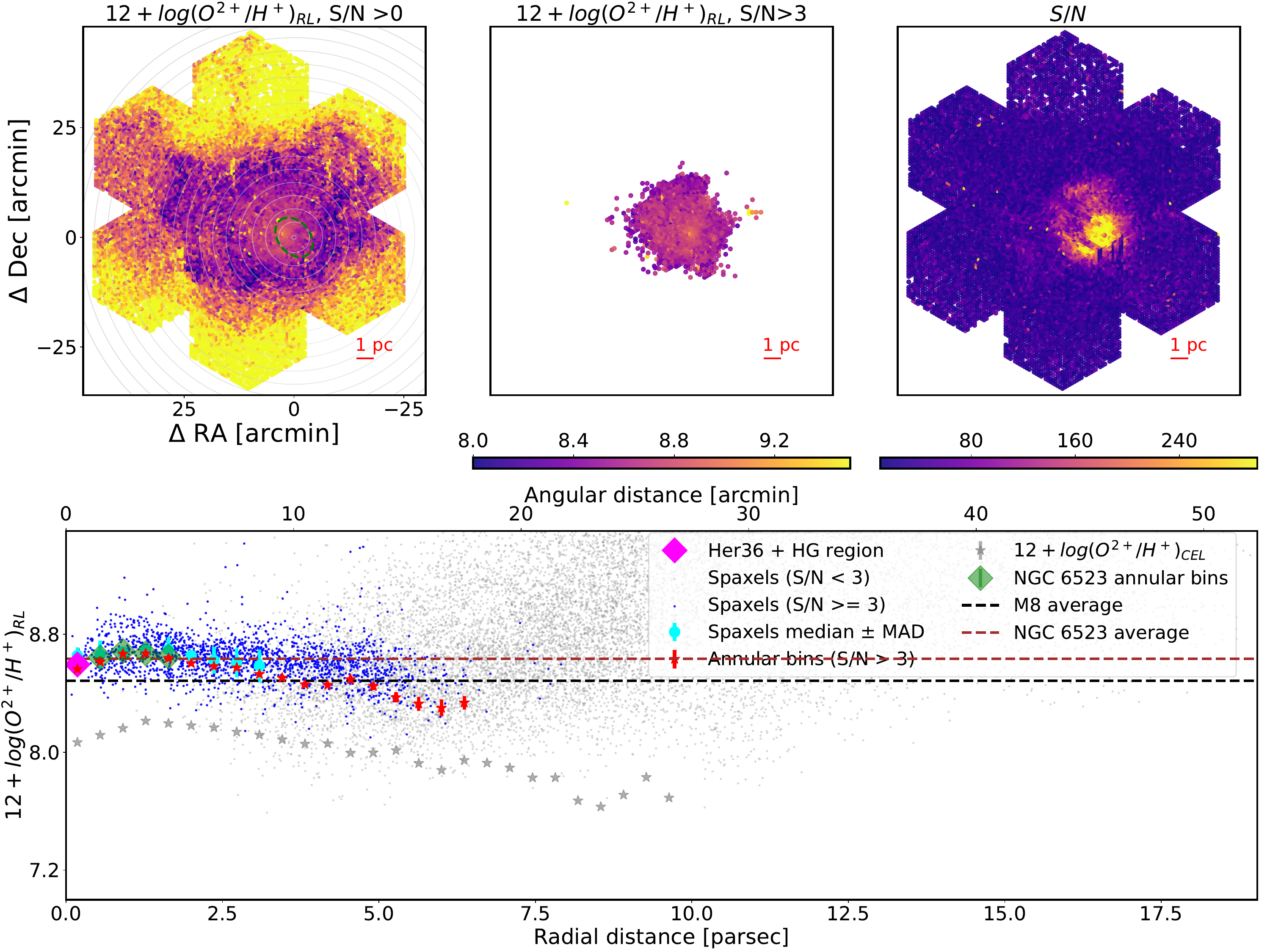} }
    \caption{Spatially resolved map and radial profile of RL-based $12+\log(O^{2+}/H^{+})$. Labels on the plot are the same as in Figure~\ref{fig:figure14} but for RL-based ionic abundance of $O^{2+}$, with a $S/N> 3$ cut on both \orl $\lambda\lambda$4649+4650 and \oiii $\lambda$4363 in this plot.}
    \label{fig:figure17}
\end{figure*}

\begin{figure*}
    \centering
    \includegraphics[width=0.75\textwidth]{\detokenize{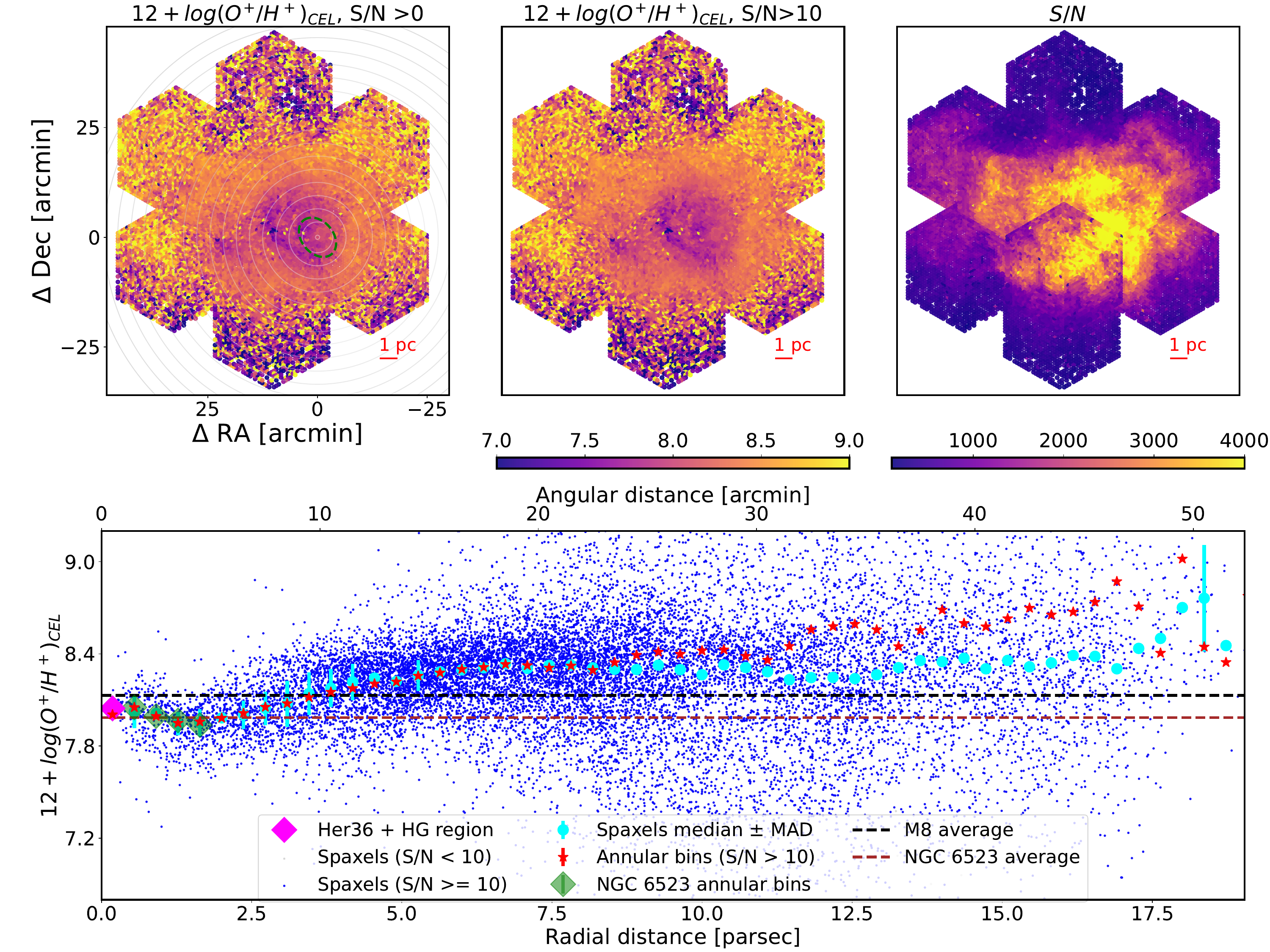}}
    \caption{Spatially resolved map and radial profile of CEL based $12+\log(O^{+}/H^{+})$. Labels on the plot are the same as in figure~\ref{fig:figure7} but for $12+\log(O^{+}/H^{+})$, with a $S/N> 10$ cut on \oii$\lambda\lambda$3726, 3729 in this plot.
    }
    \label{fig:figure18}
\end{figure*}

\begin{figure*}
    \centering
    \includegraphics[width=0.75\textwidth]{\detokenize{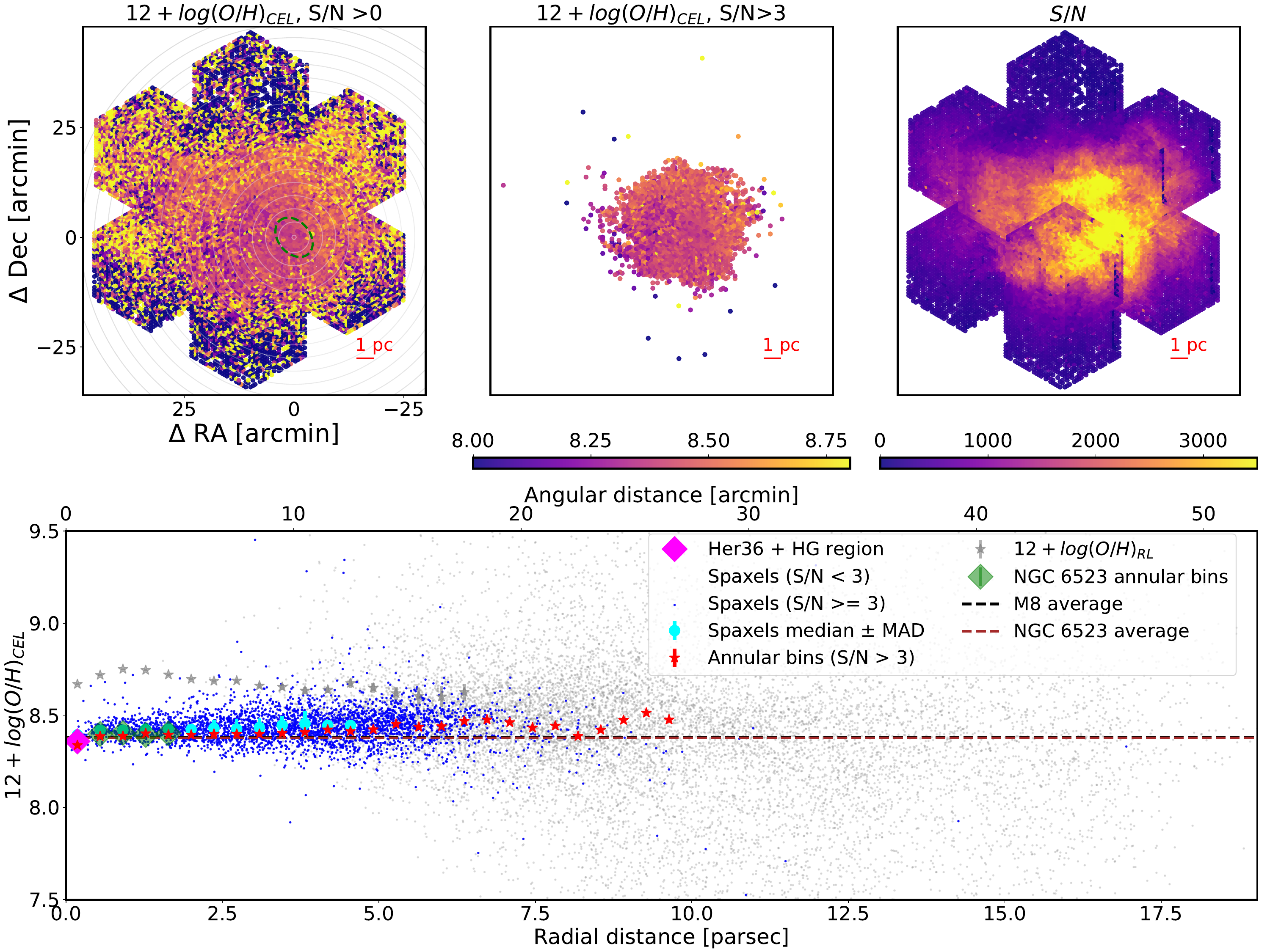}}
    \caption{Spatially resolved map and radial profile of CEL based $12+\log(O/H)$. Labels and $S/N$ cut are all the same as in Figure~\ref{fig:figure16}, but for the CEL-based elemental abundance of oxygen.}
    \label{fig:figure19}
\end{figure*}

\begin{figure*}
    \centering
    \includegraphics[width=0.75\textwidth]{\detokenize{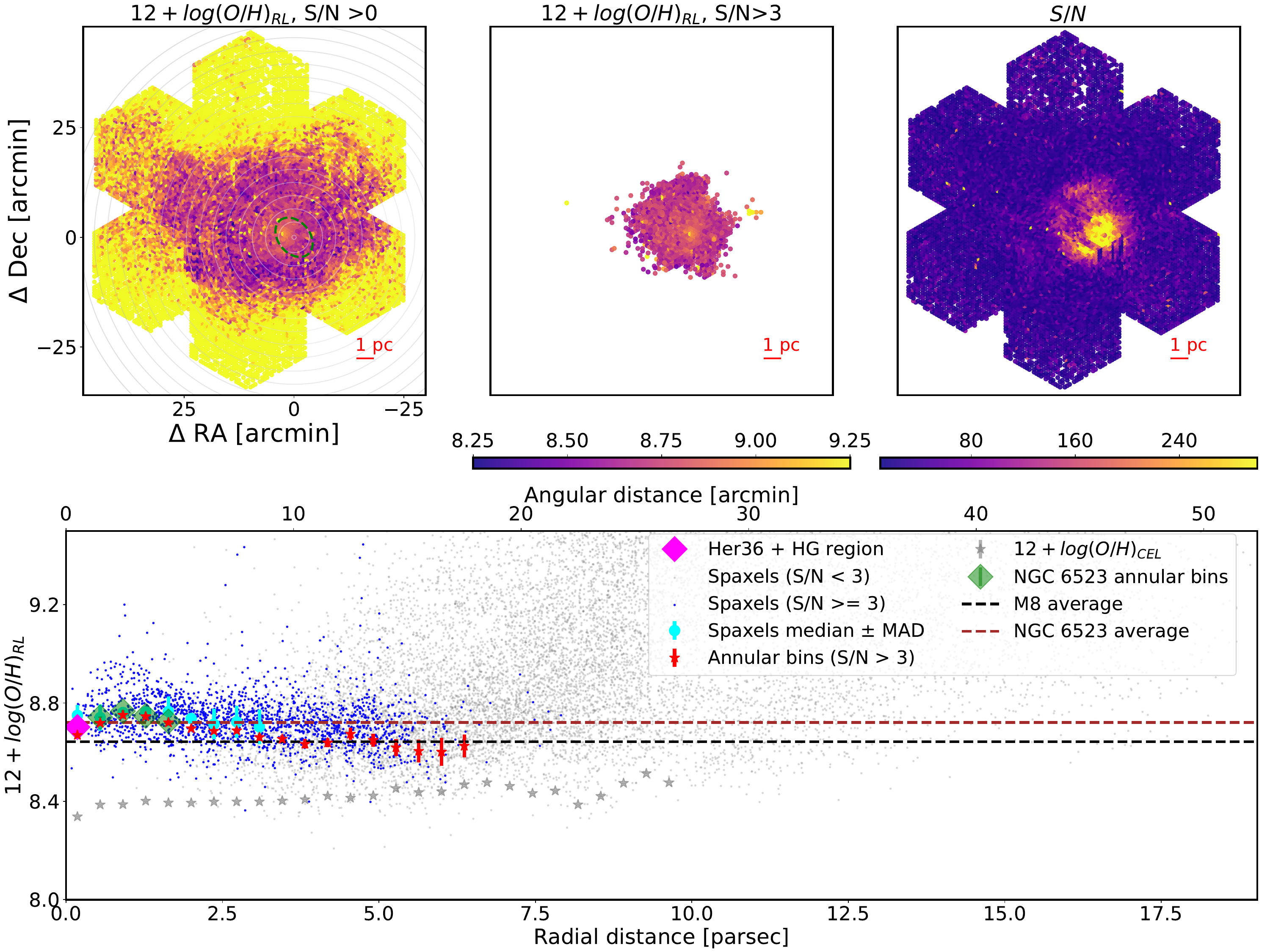}}
    \caption{Spatially resolved map and radial profile of hybrid oxygen elemental abundance $12+\log(O/H)$. The elemental abundance is calculated from RL-based $12+\log(O^{2+}/H^{+})$ and CEL based $12+\log(O^{+}/H^{+})$. Labels and $S/N$ cut are the same as in the previous Figure~\ref{fig:figure17}.}
    \label{fig:figure20}
\end{figure*}

\begin{figure*}[t]
    \centering
    \includegraphics[width=0.75\textwidth]{\detokenize{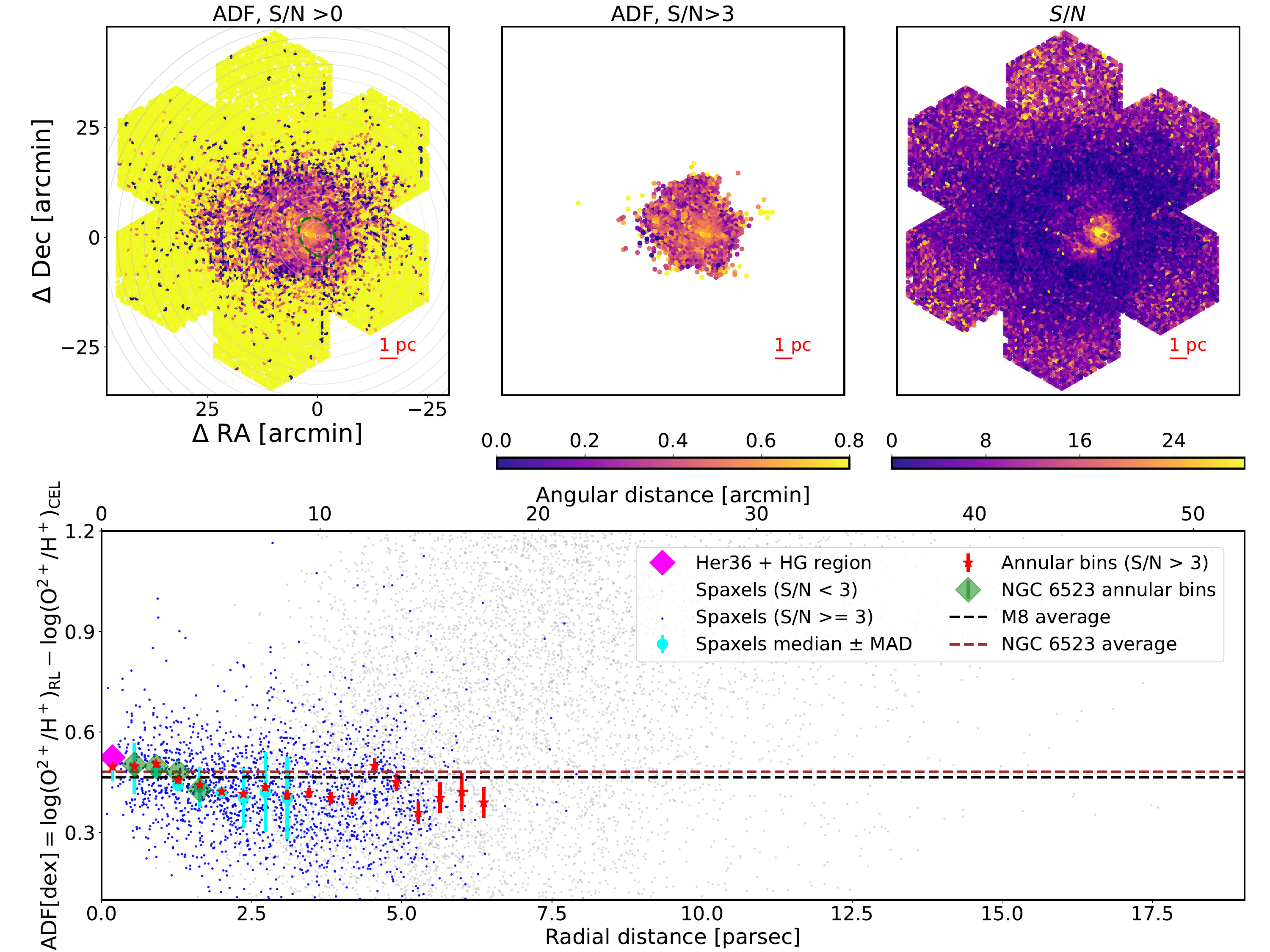}}
    \caption{Spatially resolved map and radial profile of the Abundance discrepancy factor (ADF). The color bar is in logarithmic scale, in units of dex. Labels on the plot are the same as in Figure~\ref{fig:figure7} but for ADF($O^{2+}$), with a $S/N> 3$ cut on both \orl $\lambda\lambda$4649+4650 and \oiii $\lambda$4363 in this plot.}
    \label{fig:figure21}
\end{figure*}

Figure~\ref{fig:figure16} presents the spatially resolved map and radial profile for the $O^{2+}$ ionic abundance derived from CELs. We observe that the value of $12+ \log(O^{2+}/H^{+})_{CEL}$ is lower in the ionization center (8.0699$\pm$0.0001), consistent with a possible overestimation of $T_e^{CEL}(O^{2+})$ in this region. The  CEL derived $O^{2+}$ abundance increases monotonically by $\sim0.2$ dex out to 2~pc, and then declines steadily towards the outer regions of M\,8. 

Figure~\ref{fig:figure17} shows the spatially resolved map and the radial profile of the RL-based $O^{2+}/ H^+$ abundance of M\,8. The RL-based ionic abundance of $O^{2+}$, which is systematically higher than its CEL-based counterpart, shows an enhancement of $\sim$0.1 dex at $\sim$1~pc, before declining in outer regions where the \orl $\lambda\lambda$4649+4651 line approaches a $S/N$ threshold of 3 around 7~pc. This enhancement may reflect more efficient dust destruction, releasing depleted oxygen into the gas phase. In both CEL- and RL-based measurements, $O^{2+}/H^+$ abundance decreases towards larger radii, where the oxygen budget becomes increasingly dominated by $O^{+}/H^+$. 

For the low ionization zone, Figure~\ref{fig:figure18} shows the spatially resolved map and the corresponding radial profile of the CEL-based $O^{+}$ abundance. The map reveals a pronounced depression of $O^{+}$ in the central region, where the ionization parameter is high and oxygen is predominantly in the $O^{2+}$ state; the $O^{+}$ distribution also displays significant substructures that closely trace the complex morphology of the ionization fronts and the transition layers between the ionized, atomic and molecular component of the nebula.

The radial trend reflects this ionization stratification. In agreement with the thermal structure traced by $T_e^{CEL}(N^{+})$, we find that $12 + \log(O^{+}/H^{+})$ exhibit a decrease in NGC~6523, where the gas is dominated by $O^{2+}$ out to $\sim$2~pc. Beyond this radius, the $O^{+}$ abundance increases and then remains approximately constant in the outer nebula. Due to its lower ionization potential, $O^{+}$ becomes the dominant ion in the outer regions of M\,8, allowing us to trace $O^{+}$ reliably across the entire FOV.

The elemental abundance measured by combining the CEL-based $O^{+}/H^+$ and $O^{2+}/H^+$ ionic abundances in Figure~\ref{fig:figure19}, is nearly uniform across the FOV, with a central decrease of $\sim 0.05$~dex in the central bin. In contrast, the hybrid (RLs+CEL) elemental abundance in Figure~\ref{fig:figure20}, shows a clear spatial structure: it increases from the center to $\sim$1~pc, declining by $\sim 0.1$~dex to the central value by $\sim$3~pc, and then remains approximately uniform. This behavior is driven by the RL-based $O^{2+}/H^+$ abundance, which reveals spatial variations that are absent in the CEL-based $O^{2+}/H^+$ abundance profile, resulting in the apparent uniformity of the latter. This may suggest that $T_e$ fluctuations suppress spatial structure in CEL-based abundances, while RL-based abundance, being largely insensitive to $T_e$ variations, preserves it. Consequently, RL-based abundances likely provide a more representative view of the true abundance structure of the nebula.

Studying the origin of the observed chemical abundance structure goes beyond the scope of this work. We highlight that the structure in the oxygen abundance of \hii regions can be expected from both injection and depletion of oxygen, associated with dust destruction and growth processes, respectively.

Both CEL and hybrid oxygen elemental abundances are broadly consistent with previous determinations for the brightest part of the Hourglass region. In particular, they agree with CEL-based abundance of 8.44$\pm$0.03 reported by \citet{rodriguez10}, and with the RL-based abundance of 8.74$\pm$0.06 and 8.75 reported by \citet{peimbert93} and \citet{esteban99}, respectively.\\

The systematic offset between the CEL- and RL-based $O^{2+}$ ionic abundances provides a direct measure of ADF. Figure~\ref{fig:figure21} presents the first spatially resolved map of ADF and its associated uncertainties in M\,8. Contrary to earlier findings of a uniform ADF within \hii regions \citep{rojas07}, our results show a distinct spatial gradient, with elevated values concentrated toward the ionizing source that decline outward. The radial profile of ADF in the annular bins drops from $\sim0.5$ to $\sim0.35$~dex, from the vicinity of the ionization center out to $\sim7$~pc, where the brightest pair of \orl V1 RLs are detected at $3\sigma$ level. In the NGC~6523 region, which is chiefly ionized by Her\,36, we observe a similar gradual decrease towards the outer radii. 

These values are significantly higher than previously reported for M\,8 and are indeed among the highest measured for Galactic \hii regions. However,they remain consistent with previous determinations focused on the central $\sim 10\arcsec-20\arcsec$ of the HG region \citep[see][]{esteban99, rojas06, rodriguez10} within the scatter observed in our spatially resolved data (ADF $\sim0.35-0.5$ dex). These literature measurements probe substantially smaller spatial scales than our adopted bin size of $1\arcmin$, the dispersion observed within our bins reflect underlying ADF substructure sampled at the spaxel scale (0.21 pc per spaxel), consistent with earlier studies. This radial trend indicates that the ADF is not uniform across the nebula, but instead varies with local physical conditions such as small-scale variations in temperature structure, particularly near the central ionizing source. It is also interesting to note that the radial variation of the ADF (Figure~\ref{fig:figure21}) closely mirrors that of $\Delta T_e$ (Figure~\ref{fig:figure15}). We discuss this in detail in Section~\ref{sec:discussion}.

\section{Discussion}
\label{sec:discussion}

Our spatially resolved analysis of the Lagoon nebula reveals key insights into the nature of the long-standing abundance discrepancy between RL- and CEL-based abundances. Despite decades of investigation and multiple proposed explanations (see Section~\ref{sec:intro}), the physical root cause of this discrepancy remains an active area of research. Among the leading interpretations, the presence of $T_e$ fluctuations within the ionized plasma is currently the most widely accepted explanation \citep[e.g.,][]{peimbert67, esteban99, rojas07, delgado08, skillman20, eduardo23Natur}, which favors the accuracy of RL-based abundances as the least biased diagnostic. Conversely, studies such as \citet{yuguang23} support CEL-based abundances as a less biased candidate, noting their consistency with far-infrared indicators, which are insensitive to $T_e$ variations. 

Spatially resolved studies, such as the one presented here, are essential to examine how the ADF varies across nebular environments and to assess potential correlations between the ADF and local physical conditions of the ionized gas, including $T_e$ and $n_e$. Such investigations provide a direct pathway to identifying the underlying physical drivers of the abundance discrepancy.

The correlation of ADF with these physical conditions has been exhaustively studied for PNe across their full extents. Several works have investigated spatially varying ADF in PNe, and interpreted their results within the framework of the two-phase or "bi-abundance" model \citep{torres-peimbert90, liu00, stasinska07, tsamis08, ali&dopita19, rojas22, llanos24}. This model suggests that \orl RLs and \oiii CELs originate, predominantly, from physically distinct gas phases: a warm, chemically well-mixed component responsible for most of CEL emission, and a colder, metal-rich component dominating the RL emission. Similar studies of Galactic \hii regions have been severely limited. Our work presents the first comprehensive spatially resolved study of an entire \hii region focused on this issue, reaching a spatial resolution of $\sim0.21$~pc/spaxel.


\subsection{Evidence against a two-phase ionized media in M\,8}

\label{subsec:two_phase_media}

Figures \ref{fig:figure9} and \ref{fig:figure10} present the first spatially resolved maps of the \oiii $\lambda$4363 auroral line and the composite map of the \orl V1 RLs. The individual maps of the extremely faint \orl V1 RLs, together with \oiii$\lambda$ 5007 nebular line map, are provided as a part of the online figure set in the journal. 

 The nebula is a complex of three \hii regions surrounded by atomic and molecular gas and dust. We detect \orl RLs in two of the three \hii regions with a $S/N>3:$ NGC~6523, within a radius of $5\arcmin$ (which contains the HG region in its center), and the NGC 6526 complex ($15\arcmin$ SE of the HG). The faintest of the three \hii regions, ionized by the HD 165052 binary, shows no detectable \orl RL and \oiii $\lambda$4363 emission at a $S/N>3$. We detect \oiii $\lambda$5007 in all three regions with a $S/N>3$. 

In contrast to observations in PNe studies, where RLs and CELs often originate in spatially segregated regions \citep[e.g.,][]{stasinska07, tsamis08, ali&dopita19, rojas22, llanos24}, our findings indicate that in M\,8, \orl RLs (Figure~\ref{fig:figure10}), \oiii $\lambda$4363 (Figure~\ref{fig:figure9}), and \oiii$\lambda$5007 CELs (in online figure set) intensity maps exhibit very similar spatial distributions and radial trends across the nebula. While \oiii$\lambda$ 5007 is detected at $S/N>3$ across nearly the entire FOV, detections of \oiii $\lambda$4363 and the \orl RLs are restricted to regions of high surface brightness. 

In a two-phase media scenario, we would expect RL emission to trace colder, and metal-rich inclusions, while CELs emission would be suppressed in such regions due to its strong temperature dependence. However, we find that both the CELs and the RL intensity maps exhibit practically identical spatial distributions and radial trends. 

Consistent with this, the hybrid $T_e^{RLs/CEL}$ derived directly from the ratio of \orl RLs to \oiii$\lambda$5007 CEL is remarkably uniform across the field (see Figure~\ref{fig:figure14}), despite the presence of a clear structure in the CEL-based $T_e(O^{2+})$ map. This behavior suggests that both the CEL and the RL emission predominantly arise from the same emitting regions, while the spatial structure seen  in the CEL-based temperatures could be reflecting biases in the diagnostic, induced by local temperature variations, rather than the presence of distinct gas components with different temperatures.

\begin{figure}
    \centering
    \includegraphics[width=1.0\columnwidth]{\detokenize{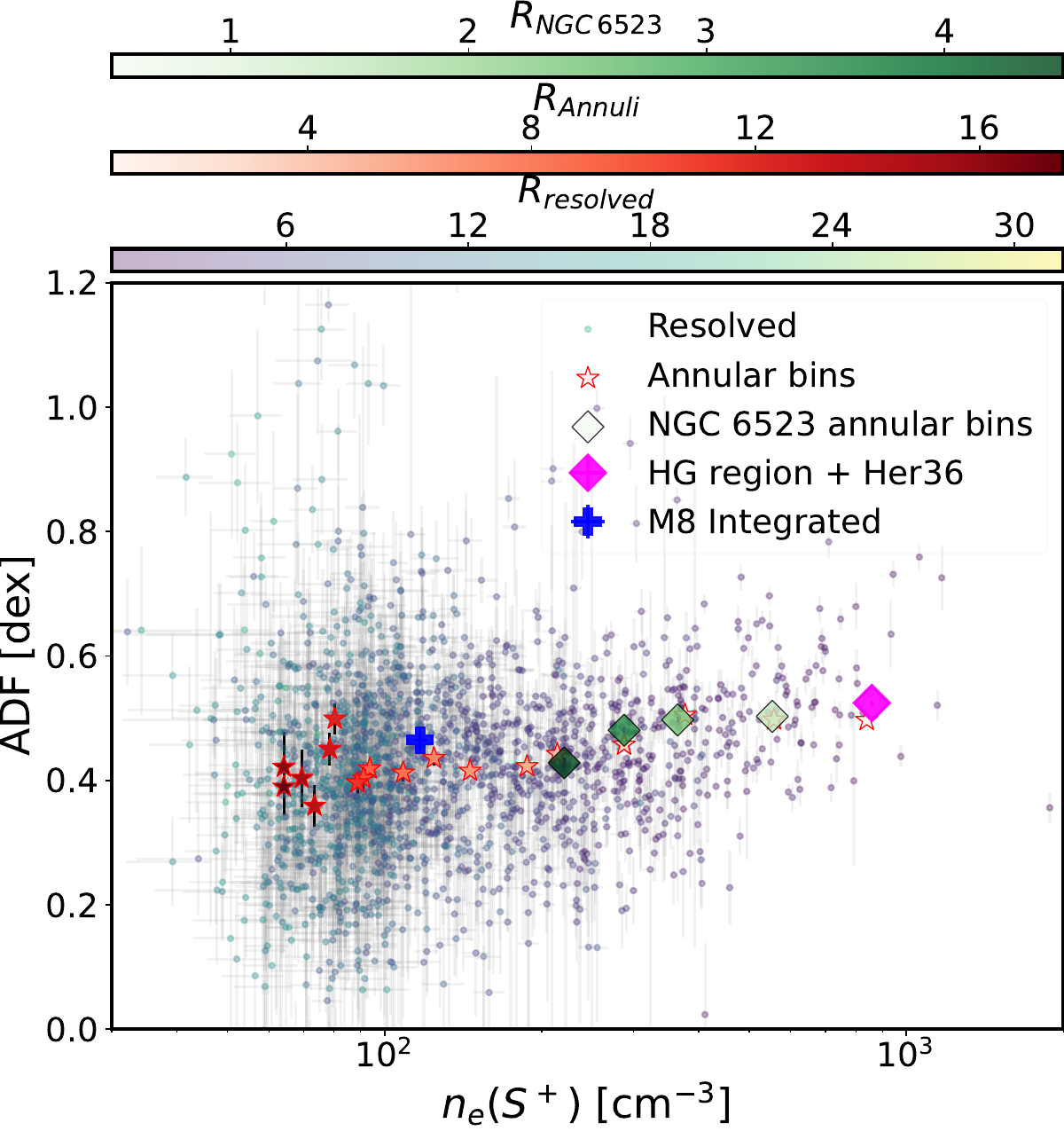}}
    \caption{ADF vs $n_e(S^+)$. The plot is color coded with radial distance, R, in arcmins, the three separate color bars are for the spatially resolved (viridis); the annular binned measurements (Reds, star markers), and the NGC\,6523 bins (Greens, rhombus markers), respectively. The pink '+' denotes the average measurement and the magenta rhombus highlights the central bin which contains the HG region + Her\,36. Only spaxels with a $S/N > 3$ on \orl$\lambda\lambda$ 4649+4651 are included in the plot.}
    \label{fig:figure22}
\end{figure}

In Figure~\ref{fig:figure22}, we present the ADF as a function of $n_e$ for spatially resolved, binned and integrated spectra in M\,8. We find evidence for a weak positive correlation between the ADF and $n_e$. At spaxel level, we detect a weak but statistically significant positive trend (Kendall rank correlation coefficient $\tau = 0.11$, $p = 3 \times 10^{-13}$). This correlation becomes stronger for annular-binned measurements ($\tau = 0.50$ $p = 0.002$). For NGC 6523, using five radial bins, we obtain $\tau = 1$ with $p = 0.016$, although the small sample size limits the statistical power of this measurement. The $p$-value represents the probability that such a correlation could arise by chance.

Density, temperature, and ionization structure vary systematically with radius in M\,8, making it difficult to interpret this trend as a direct causal relation. The $n_e(S^{+})$ diagnostic is sensitive only up to $n_e \sim 10^4$ cm$^{-3}$, and therefore does not allow us to directly probe the presence of very high-density clumps. 

The observed trend is consistent with the fact that lines-of-sight that intersect the central higher density regions of the nebula, better sample the large-scale temperature gradient of the region. While lines-of-sight on the periphery, where the $n_e$ is lower, sample more thermally uniform gas. This implies larger temperature fluctuations for the former, which can induce a larger ADF.

At the spatial resolution of the LVM data ($\sim0.21$~pc/spaxel), the spatial behavior of RLs, CELs, and nebular diagnostics does not favor a two-phase, bi-abundance origin of the observed ADF. Nevertheless, we cannot rule out the presence of very small clumpy substructures below the LVM resolution. Observations at  higher spatial resolution will be required to further investigate this scenario more stringently.


\subsection{Evidence for $T_e$ fluctuations in M\,8}
\label{subsec:te_fluctuations}

The hypothesis of the presence of temperature fluctuations in the ionized plasma was first formalized by \citet{peimbert67}. The mathematical formalism uses a Taylor expansion approximation scheme to introduce a temperature fluctuation parameter, $t^2$, to quantify variations around a mean temperature $T_0$. Temperature fluctuations introduce a bias in the CEL-based $T_e$ estimates, which typically leads to an overestimation of $T_e^{CEL}$, and an underestimation of the oxygen ionic abundances.

On the other hand, the hybrid $T_e^{RLs/CEL}$ diagnostic is significantly less sensitive to the presence of temperature fluctuations than $T_e^{CEL}$. Therefore, we expect the difference between these two temperatures, $\Delta T_e$ (see Figure \ref{fig:figure15}), to increase in the presence of larger temperature fluctuations within the gas. 

We note that the $\Delta T_e$ radial profile decreases when moving away from the main ionizing source (Her\,36), and flattens beyond $\sim2$~pc radius, closely mirroring the spatial variations seen in the ADF radial profile. In particular, we observe an increase in the ADF at smaller projected distances from Her\,36, in the same regions where we observe an enhancement in $\Delta T_e$.

This seems consistent with the classical $t^2$ paradigm proposed by \citet{peimbert67}, as well as with more recent results for other Galactic \hii regions \citep[e.g.,][]{eduardo23Natur}, suggesting a link between the ADF and the presence of temperature fluctuations, most likely associated with the effects of stellar feedback from the ionizing sources, reinforcing the interpretation that temperature fluctuations could be the primary driver of the ADF in this nebula. We will conduct a detailed quantitative analysis of the effects of temperature fluctuations in above temperature and abundance diagnostics, as a part of a follow-up study, which will be presented in a separate publication.


Taken together, our results suggest that temperature fluctuations, rather than a two-phase ionized media, may be the principal cause of the observed ADF in  M\,8. By combining spatially resolved RL and CEL diagnostics, we are, for the first time, being able to study these correlations in \hii regions. This work highlights the importance of wide-field, high-resolution spectroscopic observations for addressing fundamental questions in nebular physics and ISM chemical abundances. 

\section{Conclusions}
\label{sec:conclusion}

We present the first ultra-wide-field, high-resolution IFS study of the Lagoon nebula (M\,8, see Panel b of Figure~\ref{fig:figure1}). M\,8 is observed as part of the SDSS-V LVM project, covering the full optical spectral range of $3600-9800$~\AA, with a spectral resolution of $\approx 4000$ at \ha, and a spatial resolution of $0.21$~pc/ spaxel. The entire FOV is covered by a total of 10 LVM-I tiles, and a combined exposure time of $\sim 6.25$ hours (see Table \ref{tab:table2} for details). 

To our knowledge, this is the first IFU dataset to fully cover M\,8, with such high spatial resolution and depth. The LVM deep IFU spectroscopy enables spatially resolved measurements and mapping of HI RLs and strong nebular CELs at extremely high $S/N$, as well as the detection and mapping of faint auroral CELs, and extremely faint \orl V1 RLs (up to 8000 times fainter than \ha, see Figure~\ref{fig:figure10}) at $S/N>3$. An atlas of emission line maps for all these transitions across M\,8 is presented in the online figure set. 

We use the data to investigate spatial variations in the physical and chemical conditions across M\,8, with the primary goal of studying the potential causes behind the so called "abundance discrepancy problem", which manifests itself as a logarithmic difference between the abundances measured from metal RLs and CELs of heavy ions.

We measure emission line intensities by fitting Gaussian profiles, and apply dust extinction corrections, to then derive key physical conditions such as the electron density and temperature, and the ionic and elemental oxygen abundances using different diagnostics. From the analysis of the observed emission line maps and the derived physical and chemical parameters, we conclude the following:

\begin{enumerate}

    \item  M\,8 presents a spatially varying reddening law, as parameterized by the $R_V$ value, which peaks at $R_V\simeq6$ around the center, and declines towards the outer regions reaching values close to the standard $R_V=3.1$ for diffuse sight-lines in the MW, at the edge of the nebula. These variations in the reddening law are most likely associated with changes in the physical properties and the size distribution of dust grains, which can be caused by the destruction of dust grains due to stellar feedback effects from the main ionizing sources in the nebula (most prominently Her\,36).

    \item The electron density, as measured from the \sii doublet ($n_e(S^{+})$), changes by more than an order of magnitude across the nebula. The radial density profile exhibits a central peak near Her\,36 with values as high as $\sim850$cm$^{-3}$, dropping to values closer to $\sim30$cm$^{-3}$ in the outskirts of the nebula (see Figure.~\ref{fig:figure11}). On top of this gradient, filamentary and bubble-like electron density structures are revealed in the spatially resolved map.

    

    \item The direct method electron temperature of the high ionization gas, derived from CELs of $O^{2+}$, exhibits spatial structure in the central regions of the nebula, with an increase of several hundred K towards the main ionization source (Her\,36). This spatial structure is absent in the electron temperature map derived from a hybrid $O^{2+}$ RLs/CEL diagnostic, which is less sensitive to the presence of thermal fluctuations. Furthermore, the direct method electron temperatures are overestimated with respect to the RLs/CEL electron temperatures by $\sim1500-2000$~K. The origin of these discrepancies is likely to be associated with biases in the CEL based direct method, caused by the presence of temperature fluctuations along the line of sight.



    \item Having measured the $O^{2+}$ abundance with both the CELs based direct method, and the RLs method, we construct the first spatially resolved ADF map of an \hii region (see Figure~\ref{fig:figure21}). We observe a radial variation in the ADF across the central regions of the nebula, with higher values towards the the main ionizing source (Her\,36). This is consistent with previous works which have linked the presence of strong thermal fluctuations in the gas, and the strong biases they induce in direct method abundances, with the effects of stellar feedback in the vicinity of massive stars, where stellar winds and shocks can impact the thermal balance of the gas \citep[see][]{eduardo23Natur}. 

    \item At the spatial resolution of the LVM data, the co-spatial distributions and radial profiles of both $O^{2+}$ RLs and CELs intensities, together with the uniformity of RLs/CEL based $T_e$, and the weak observed correlation between the ADF and the electron density, do not favor the presence of a two-phase, bi-abundance ionized medium, as the origin of the observed ADF (see Section~\ref{subsec:two_phase_media}). On the other hand, the fact that the difference between the CEL based and the RL/CEL based electron temperatures ($\Delta T_e$) is expected to be a proxy for the presence of thermal fluctuations, and that the observed ADF variations largely mirror the variations seen in $\Delta T_e$, support the idea that inhomogeneities in the thermal structure of the ionized gas are the most probable cause of the observed abundance discrepancies in M\,8 (see Section~\ref{subsec:te_fluctuations}). 
        
    \item The average ADF value across M8 is $\sim0.47$, implying that the direct method abundances for doubly ionized oxygen suffer from systematic errors at the level of a factor of $\sim3$ in this region. This has important implications in what regards to our ability to constrain the chemical abundance of metals in galaxies, using methods that are either based on, or calibrated against the direct method, and which do not take into account the presence of thermal fluctuations in the gas.


\end{enumerate}

The unprecedented combination of field-of-view, spatial resolution, and depth of the LVM data presented in this study, underscore the power of the SDSS-V LVM survey to resolve fine-scale variations in the physical conditions of Galactic \hii regions. Our results reveal the complexity of the internal nebular structure of these objects, and emphasize the importance of spatially resolved observations for the accurate characterization and interpretation of the physical conditions and chemical abundance patterns, and discrepancies in \hii regions. Expanding this analysis to a larger number of \hii regions within the LVM survey footprint, will enable a statistical characterization of the ADF across a range of environments and evolutionary states, providing new constraints on the physical mechanisms behind the CEL vs RL abundance mismatch. These results will be critical for improving the accuracy of chemical abundance determinations in unresolved extragalactic \hii regions across all redshifts, where such diagnostics are routinely used to trace galactic chemical evolution.

\section*{Acknowledgments}

We are grateful for financial support from the STScI Director’s Research Fund. A.S. acknowledges the support from the ANID Basal projects FB21250718 and FB210003. G.A.B. acknowledges the support from the ANID Basal project FB210003. C.M. acknowledges the support of grant UNAM/DGAPA/PAPIIT IG101224. J.E. M-D., C. M., R de J. Z. thank the support by SECIHTI CBF-2025-I-2048 project ``Resolviendo la Física Interna de las Galaxias: De las Escalas Locales a la Estructura Global con el SDSS-V Local Volume Mapper'' (PI: M\'endez Delgado). KK, EE, NS gratefully acknowledge funding from the Deutsche Forschungsgemeinschaft (DFG, German Research Foundation) in the form of an Emmy Noether Research Group (grant number KR4598/2-1, PI Kreckel) and the European Research Council’s starting grant ERC StG-101077573 (“ISM-METALS"). OE acknowledges funding from the Deutsche Forschungsgemeinschaft (DFG, German Research Foundation) -- project-ID 541068876. D.M. acknowledges support by ANID Fondecyt Regular grant No. 1220724, and by the BASAL Center for Astrophysics and Associated Technologies (CATA) through ANID grants ACE210002 and FB210003. J.G.F-T gratefully acknowledges the grants support provided by ANID Fondecyt Postdoc No. 3230001 (Sponsoring researcher), the Joint Committee ESO-Government of Chile under the agreement 2023 ORP 062/2023, and the support of the Doctoral Program in Artificial Intelligence, DISC-UCN.

Funding for the Sloan Digital Sky Survey V has been provided by the Alfred P. Sloan Foundation, the Heising-Simons Foundation, the National Science Foundation, and the Participating Institutions. SDSS acknowledges support and resources from the Center for High-Performance Computing at the University of Utah. SDSS telescopes are located at Apache Point Observatory, funded by the Astrophysical Research Consortium and operated by New Mexico State University, and at Las Campanas Observatory, operated by the Carnegie Institution for Science. The SDSS web site is \url{www.sdss.org}.

SDSS is managed by the Astrophysical Research Consortium for the Participating Institutions of the SDSS Collaboration, including the Carnegie Institution for Science, Chilean National Time Allocation Committee (CNTAC) ratified researchers, Caltech, the Gotham Participation Group, Harvard University, Heidelberg University, The Flatiron Institute, The Johns Hopkins University, L'Ecole polytechnique f\'{e}d\'{e}rale de Lausanne (EPFL), Leibniz-Institut f\"{u}r Astrophysik Potsdam (AIP), Max-Planck-Institut f\"{u}r Astronomie (MPIA Heidelberg), Max-Planck-Institut f\"{u}r Extraterrestrische Physik (MPE), Nanjing University, National Astronomical Observatories of China (NAOC), New Mexico State University, The Ohio State University, Pennsylvania State University, Smithsonian Astrophysical Observatory, Space Telescope Science Institute (STScI), the Stellar Astrophysics Participation Group, Universidad Nacional Aut\'{o}noma de M\'{e}xico, University of Arizona, University of Colorado Boulder, University of Illinois at Urbana-Champaign, University of Toronto, University of Utah, University of Virginia, Yale University, and Yunnan University.

\bibliography{references}

\begin{thebibliography}{}
\expandafter\ifx\csname natexlab\endcsname\relax\def\natexlab#1{#1}\fi
\providecommand{\url}[1]{\href{#1}{#1}}
\providecommand{\dodoi}[1]{doi:~\href{http://doi.org/#1}{\nolinkurl{#1}}}
\providecommand{\doeprint}[1]{\href{http://ascl.net/#1}{\nolinkurl{http://ascl.net/#1}}}
\providecommand{\doarXiv}[1]{\href{https://arxiv.org/abs/#1}{\nolinkurl{https://arxiv.org/abs/#1}}}

\bibitem[{{Aggarwal} \& {Keenan}(1999)}]{aggarwal99}
{Aggarwal}, K.~M., \& {Keenan}, F.~P. 1999, \apjs, 123, 311, \dodoi{10.1086/313232}

\bibitem[{{Ali} \& {Dopita}(2019)}]{ali&dopita19}
{Ali}, A., \& {Dopita}, M.~A. 2019, \mnras, 484, 3251, \dodoi{10.1093/mnras/stz201}

\bibitem[{{Andrade} {et~al.}(2025){Andrade}, {Saviane}, {Monaco}, \& {Gullieuszik}}]{andrade25}
{Andrade}, A., {Saviane}, I., {Monaco}, L., \& {Gullieuszik}, M. 2025, \aap, 699, A281, \dodoi{10.1051/0004-6361/202553678}

\bibitem[{{Arias} {et~al.}(2002){Arias}, {Morrell}, {Barb{\'a}}, {Bosch}, {Grosso}, \& {Corcoran}}]{arias02}
{Arias}, J.~I., {Morrell}, N.~I., {Barb{\'a}}, R.~H., {et~al.} 2002, \mnras, 333, 202, \dodoi{10.1046/j.1365-8711.2002.05404.x}

\bibitem[{{Arias} {et~al.}(2010){Arias}, {Barb{\'a}}, {Gamen}, {Morrell}, {Ma{\'\i}z Apell{\'a}niz}, {Alfaro}, {Sota}, {Walborn}, \& {Moni Bidin}}]{arias10}
{Arias}, J.~I., {Barb{\'a}}, R.~H., {Gamen}, R.~C., {et~al.} 2010, \apjl, 710, L30, \dodoi{10.1088/2041-8205/710/1/L30}

\bibitem[{{Bacon} {et~al.}(2010){Bacon}, {Accardo}, {Adjali}, {Anwand}, {Bauer}, {Biswas}, {Blaizot}, {Boudon}, {Brau-Nogue}, {Brinchmann}, {Caillier}, {Capoani}, {Carollo}, {Contini}, {Couderc}, {Daguis{\'e}}, {Deiries}, {Delabre}, {Dreizler}, {Dubois}, {Dupieux}, {Dupuy}, {Emsellem}, {Fechner}, {Fleischmann}, {Fran{\c{c}}ois}, {Gallou}, {Gharsa}, {Glindemann}, {Gojak}, {Guiderdoni}, {Hansali}, {Hahn}, {Jarno}, {Kelz}, {Koehler}, {Kosmalski}, {Laurent}, {Le Floch}, {Lilly}, {Lizon}, {Loupias}, {Manescau}, {Monstein}, {Nicklas}, {Olaya}, {Pares}, {Pasquini}, {P{\'e}contal-Rousset}, {Pell{\'o}}, {Petit}, {Popow}, {Reiss}, {Remillieux}, {Renault}, {Roth}, {Rupprecht}, {Serre}, {Schaye}, {Soucail}, {Steinmetz}, {Streicher}, {Stuik}, {Valentin}, {Vernet}, {Weilbacher}, {Wisotzki}, \& {Yerle}}]{bacon10}
{Bacon}, R., {Accardo}, M., {Adjali}, L., {et~al.} 2010, in Society of Photo-Optical Instrumentation Engineers (SPIE) Conference Series, Vol. 7735, Ground-based and Airborne Instrumentation for Astronomy III, ed. I.~S. {McLean}, S.~K. {Ramsay}, \& H.~{Takami}, 773508, \dodoi{10.1117/12.856027}

\bibitem[{{Berg} {et~al.}(2020){Berg}, {Pogge}, {Skillman}, {Croxall}, {Moustakas}, {Rogers}, \& {Sun}}]{berg20}
{Berg}, D.~A., {Pogge}, R.~W., {Skillman}, E.~D., {et~al.} 2020, \apj, 893, 96, \dodoi{10.3847/1538-4357/ab7eab}

\bibitem[{{Berg} {et~al.}(2015){Berg}, {Skillman}, {Croxall}, {Pogge}, {Moustakas}, \& {Johnson-Groh}}]{berg15}
{Berg}, D.~A., {Skillman}, E.~D., {Croxall}, K.~V., {et~al.} 2015, \apj, 806, 16, \dodoi{10.1088/0004-637X/806/1/16}

\bibitem[{Bergerud {et~al.}(2019)Bergerud, Spangler, \& Beauchamp}]{bergerud19}
Bergerud, B.~M., Spangler, S.~R., \& Beauchamp, K.~M. 2019, Monthly Notices of the Royal Astronomical Society, 492, 1142–1153, \dodoi{10.1093/mnras/stz3515}

\bibitem[{{Blanc}(in prep)}]{blanc25}
{Blanc}, G. in prep, {LVM Technical Overview}

\bibitem[{{Blanc} {et~al.}(2015){Blanc}, {Kewley}, {Vogt}, \& {Dopita}}]{blanc15}
{Blanc}, G.~A., {Kewley}, L., {Vogt}, F. P.~A., \& {Dopita}, M.~A. 2015, \apj, 798, 99, \dodoi{10.1088/0004-637X/798/2/99}

\bibitem[{{Cardelli} {et~al.}(1989){Cardelli}, {Clayton}, \& {Mathis}}]{cardelli89}
{Cardelli}, J.~A., {Clayton}, G.~C., \& {Mathis}, J.~S. 1989, \apj, 345, 245, \dodoi{10.1086/167900}

\bibitem[{{Castellanos} {et~al.}(2002){Castellanos}, {D{\'\i}az}, \& {Tenorio-Tagle}}]{castellanos02}
{Castellanos}, M., {D{\'\i}az}, {\'A}.~I., \& {Tenorio-Tagle}, G. 2002, \apjl, 565, L79, \dodoi{10.1086/339367}

\bibitem[{{Chen} {et~al.}(2007){Chen}, {de Grijs}, \& {Zhao}}]{chen07}
{Chen}, L., {de Grijs}, R., \& {Zhao}, J.~L. 2007, \aj, 134, 1368, \dodoi{10.1086/521022}

\bibitem[{{Chen} {et~al.}(2023){Chen}, {Jones}, {Sanders}, {Fadda}, {Sutter}, {Minchin}, {Huntzinger}, {Senchyna}, {Stark}, {Spilker}, {Weiner}, \& {Roberts-Borsani}}]{yuguang23}
{Chen}, Y., {Jones}, T., {Sanders}, R., {et~al.} 2023, Nature Astronomy, 7, 771, \dodoi{10.1038/s41550-023-01953-7}

\bibitem[{{Corradi} {et~al.}(2015){Corradi}, {Garc{\'\i}a-Rojas}, {Jones}, \& {Rodr{\'\i}guez-Gil}}]{corradi15}
{Corradi}, R. L.~M., {Garc{\'\i}a-Rojas}, J., {Jones}, D., \& {Rodr{\'\i}guez-Gil}, P. 2015, \apj, 803, 99, \dodoi{10.1088/0004-637X/803/2/99}

\bibitem[{{Curti} {et~al.}(2017){Curti}, {Cresci}, {Mannucci}, {Marconi}, {Maiolino}, \& {Esposito}}]{curti17}
{Curti}, M., {Cresci}, G., {Mannucci}, F., {et~al.} 2017, \mnras, 465, 1384, \dodoi{10.1093/mnras/stw2766}

\bibitem[{{Damiani, F.} {et~al.}(2017){Damiani, F.}, {Bonito, R.}, {Prisinzano, L.}, {Zwitter, T.}, {Bayo, A.}, {Kalari, V.}, {Jiménez-Esteban, F. M.}, {Costado, M. T.}, {Jofré, P.}, {Randich, S.}, {Flaccomio, E.}, {Lanzafame, A. C.}, {Lardo, C.}, {Morbidelli, L.}, \& {Zaggia, S.}}]{damiani17}
{Damiani, F.}, {Bonito, R.}, {Prisinzano, L.}, {et~al.} 2017, A\&A, 604, A135, \dodoi{10.1051/0004-6361/201730986}

\bibitem[{{Dom{\'\i}nguez-Guzm{\'a}n} {et~al.}(2022){Dom{\'\i}nguez-Guzm{\'a}n}, {Rodr{\'\i}guez}, {Garc{\'\i}a-Rojas}, {Esteban}, \& {Toribio San Cipriano}}]{guzman22}
{Dom{\'\i}nguez-Guzm{\'a}n}, G., {Rodr{\'\i}guez}, M., {Garc{\'\i}a-Rojas}, J., {Esteban}, C., \& {Toribio San Cipriano}, L. 2022, \mnras, 517, 4497, \dodoi{10.1093/mnras/stac2974}

\bibitem[{{Draine}(2003)}]{draine03}
{Draine}, B.~T. 2003, \araa, 41, 241, \dodoi{10.1146/annurev.astro.41.011802.094840}

\bibitem[{{Draine}(2011)}]{draine11}
---. 2011, {Physics of the Interstellar and Intergalactic Medium}

\bibitem[{{Drory} {et~al.}(2024){Drory}, {Blanc}, {Kreckel}, {Sanchez}, {Mejia-Narvaez}, {Johnston}, {Jones}, {Pellegrini}, {Konidaris}, {Herbst}, {Sanchez-Gallego}, {Kollmeier}, {de Almeida}, {Barrera-Ballesteros}, {Bizyaev}, {Brownstein}, {Saguer}, {Cherinka}, {Cioni}, {Congiu}, {Cosens}, {Dias}, {Donor}, {Egorov}, {Egorova}, {Froning}, {Garcia}, {Glover}, {Greve}, {Haeberle}, {Hoy}, {Ibarra}, {Li}, {Klessen}, {Krishnarao}, {Kumari}, {Long}, {Mendez-Delgado}, {Popa}, {Ramirez}, {Rix}, {Mata Sanchez}, {Sankrit}, {Sattler}, {Sayres}, {Singh}, {Stringfellow}, {Wachter}, {Watkins}, {Wong}, \& {Wofford}}]{drory24}
{Drory}, N., {Blanc}, G.~A., {Kreckel}, K., {et~al.} 2024, arXiv e-prints, arXiv:2405.01637, \dodoi{10.48550/arXiv.2405.01637}

\bibitem[{{Dyson}(1968)}]{dyson68}
{Dyson}, J.~E. 1968, \apss, 1, 388, \dodoi{10.1007/BF00656009}

\bibitem[{{Esteban} {et~al.}(2020){Esteban}, {Bresolin}, {Garc{\'\i}a-Rojas}, \& {Toribio San Cipriano}}]{esteban20}
{Esteban}, C., {Bresolin}, F., {Garc{\'\i}a-Rojas}, J., \& {Toribio San Cipriano}, L. 2020, \mnras, 491, 2137, \dodoi{10.1093/mnras/stz3134}

\bibitem[{{Esteban} {et~al.}(2009){Esteban}, {Bresolin}, {Peimbert}, {Garc{\'\i}a-Rojas}, {Peimbert}, \& {Mesa-Delgado}}]{esteban09}
{Esteban}, C., {Bresolin}, F., {Peimbert}, M., {et~al.} 2009, \apj, 700, 654, \dodoi{10.1088/0004-637X/700/1/654}

\bibitem[{{Esteban} {et~al.}(2004){Esteban}, {Garc{\'\i}a L{\'o}pez}, {Herrero}, \& {S{\'a}nchez}}]{esteban04}
{Esteban}, C., {Garc{\'\i}a L{\'o}pez}, R., {Herrero}, A., \& {S{\'a}nchez}, F., eds. 2004, {Cosmochemistry. The melting pot of the elements}

\bibitem[{{Esteban} {et~al.}(1995){Esteban}, {Peimbert}, {Torres-Peimbert}, \& {Escalante}}]{esteban95}
{Esteban}, C., {Peimbert}, M., {Torres-Peimbert}, S., \& {Escalante}, V. 1995, in Revista Mexicana de Astronomia y Astrofisica Conference Series, Vol.~3, Revista Mexicana de Astronomia y Astrofisica Conference Series, ed. M.~{Pena} \& S.~{Kurtz}, 241

\bibitem[{{Esteban} {et~al.}(1999){Esteban}, {Peimbert}, {Torres-Peimbert}, {Garc{\'\i}a-Rojas}, \& {Rodr{\'\i}guez}}]{esteban99}
{Esteban}, C., {Peimbert}, M., {Torres-Peimbert}, S., {Garc{\'\i}a-Rojas}, J., \& {Rodr{\'\i}guez}, M. 1999, \apjs, 120, 113, \dodoi{10.1086/313172}

\bibitem[{{Fang} {et~al.}(2011){Fang}, {Storey}, \& {Liu}}]{fang11}
{Fang}, X., {Storey}, P.~J., \& {Liu}, X.~W. 2011, \aap, 530, A18, \dodoi{10.1051/0004-6361/201116511}

\bibitem[{{Ferland} {et~al.}(2017){Ferland}, {Chatzikos}, {Guzm{\'a}n}, {Lykins}, {van Hoof}, {Williams}, {Abel}, {Badnell}, {Keenan}, {Porter}, \& {Stancil}}]{ferland17}
{Ferland}, G.~J., {Chatzikos}, M., {Guzm{\'a}n}, F., {et~al.} 2017, \rmxaa, 53, 385, \dodoi{10.48550/arXiv.1705.10877}

\bibitem[{{Fitzpatrick}(1999)}]{fitzpatrick99}
{Fitzpatrick}, E.~L. 1999, \pasp, 111, 63, \dodoi{10.1086/316293}

\bibitem[{{Froese Fischer} \& {Tachiev}(2004)}]{fischer04}
{Froese Fischer}, C., \& {Tachiev}, G. 2004, Atomic Data and Nuclear Data Tables, 87, 1, \dodoi{10.1016/j.adt.2004.02.001}

\bibitem[{{Garc{\'\i}a-Rojas} {et~al.}(2005){Garc{\'\i}a-Rojas}, {Esteban}, {Peimbert}, {Peimbert}, {Rodr{\'\i}guez}, \& {Ruiz}}]{rojas05}
{Garc{\'\i}a-Rojas}, J., {Esteban}, C., {Peimbert}, A., {et~al.} 2005, \mnras, 362, 301, \dodoi{10.1111/j.1365-2966.2005.09302.x}

\bibitem[{{Garc{\'\i}a-Rojas} {et~al.}(2007){Garc{\'\i}a-Rojas}, {Esteban}, {Peimbert}, {Rodr{\'\i}guez}, {Peimbert}, \& {Ruiz}}]{rojas07}
---. 2007, \rmxaa, 43, 3, \dodoi{10.48550/arXiv.astro-ph/0610065}

\bibitem[{{Garc{\'\i}a-Rojas} {et~al.}(2006){Garc{\'\i}a-Rojas}, {Esteban}, {Peimbert}, {Costado}, {Rodr{\'\i}guez}, {Peimbert}, \& {Ruiz}}]{rojas06}
{Garc{\'\i}a-Rojas}, J., {Esteban}, C., {Peimbert}, M., {et~al.} 2006, \mnras, 368, 253, \dodoi{10.1111/j.1365-2966.2006.10105.x}

\bibitem[{{Garc{\'\i}a-Rojas} {et~al.}(2022){Garc{\'\i}a-Rojas}, {Morisset}, {Jones}, {Wesson}, {Boffin}, {Monteiro}, {Corradi}, \& {Rodr{\'\i}guez-Gil}}]{rojas22}
{Garc{\'\i}a-Rojas}, J., {Morisset}, C., {Jones}, D., {et~al.} 2022, \mnras, 510, 5444, \dodoi{10.1093/mnras/stab3523}

\bibitem[{{Garnett}(1992)}]{garnett92}
{Garnett}, D.~R. 1992, \aj, 103, 1330, \dodoi{10.1086/116146}

\bibitem[{{G{\'o}mez-Llanos} {et~al.}(2024{\natexlab{a}}){G{\'o}mez-Llanos}, {Garc{\'\i}a-Rojas}, {Morisset}, {Monteiro}, {Jones}, {Wesson}, {Boffin}, \& {Corradi}}]{gomez+llanos24}
{G{\'o}mez-Llanos}, V., {Garc{\'\i}a-Rojas}, J., {Morisset}, C., {et~al.} 2024{\natexlab{a}}, \aap, 689, A228, \dodoi{10.1051/0004-6361/202450822}

\bibitem[{{G{\'o}mez-Llanos} {et~al.}(2024{\natexlab{b}}){G{\'o}mez-Llanos}, {Garc{\'\i}a-Rojas}, {Morisset}, {Monteiro}, {Jones}, {Wesson}, {Boffin}, \& {Corradi}}]{llanos24}
---. 2024{\natexlab{b}}, \aap, 689, A228, \dodoi{10.1051/0004-6361/202450822}

\bibitem[{{Griffith} {et~al.}(2019){Griffith}, {Martini}, \& {Conroy}}]{griffith19}
{Griffith}, E., {Martini}, P., \& {Conroy}, C. 2019, \mnras, 484, 562, \dodoi{10.1093/mnras/sty3405}

\bibitem[{{Groves} {et~al.}(2012){Groves}, {Brinchmann}, \& {Walcher}}]{groves12}
{Groves}, B., {Brinchmann}, J., \& {Walcher}, C.~J. 2012, \mnras, 419, 1402, \dodoi{10.1111/j.1365-2966.2011.19796.x}

\bibitem[{{Hecht} {et~al.}(1982){Hecht}, {Helfer}, {Wolf}, {Donn}, \& {Pipher}}]{hecht82}
{Hecht}, J., {Helfer}, H.~L., {Wolf}, J., {Donn}, B., \& {Pipher}, J.~L. 1982, \apjl, 263, L39, \dodoi{10.1086/183919}

\bibitem[{{Herbst} {et~al.}(2024){Herbst}, {Bizenberger}, {Blanc}, {Briegel}, {Drory}, {Froning}, {Gaessler}, {H{\"a}berle}, {Konidaris}, {Kuhlberg}, {Lanz}, {Mathar}, {Mohr}, {Ramirez}, {Ritz}, {Rohloff}, {S{\'a}nchez-Gallego}, {Wachter}, {Ahn}, {Besser}, {Case}, {Feger}, {Hebert}, {Kollmeier}, {Pak}, {Rix}, {Robertson}, {St{\k{e}}pie{\'n}}, \& {Zhelem}}]{herbst24}
{Herbst}, T.~M., {Bizenberger}, P., {Blanc}, G.~A., {et~al.} 2024, \aj, 168, 267, \dodoi{10.3847/1538-3881/ad7948}

\bibitem[{{Hester} {et~al.}(1996){Hester}, {Scowen}, {Sankrit}, {Lauer}, {Ajhar}, {Baum}, {Code}, {Currie}, {Danielson}, {Ewald}, {Faber}, {Grillmair}, {Groth}, {Holtzman}, {Hunter}, {Kristian}, {Light}, {Lynds}, {Monet}, {O'Neil}, {Shaya}, {Seidelmann}, \& {Westphal}}]{hester96}
{Hester}, J.~J., {Scowen}, P.~A., {Sankrit}, R., {et~al.} 1996, \aj, 111, 2349, \dodoi{10.1086/117968}

\bibitem[{{Hilder} {et~al.}(2025){Hilder}, {Casey}, {Dalcanton}, {Kreckel}, {Stutz}, {Singh}, {Blanc}, {S{\'a}nchez}, {M{\'e}ndez-Delgado}, {Saydjari}, {Vargas-Herrera}, {Drory}, {Bizyaev}, {Fern{\'a}ndez-Trincado}, {Rom{\'a}n-Z{\'u}{\~n}iga}, {Kollmeier}, \& {Johnston}}]{hilder25}
{Hilder}, T., {Casey}, A.~R., {Dalcanton}, J.~J., {et~al.} 2025, arXiv e-prints, arXiv:2510.07395, \dodoi{10.48550/arXiv.2510.07395}

\bibitem[{{Jin} {et~al.}(2022){Jin}, {Kewley}, \& {Sutherland}}]{Jin22}
{Jin}, Y., {Kewley}, L.~J., \& {Sutherland}, R.~S. 2022, \apjl, 934, L8, \dodoi{10.3847/2041-8213/ac80f3}

\bibitem[{{Kahle} {et~al.}(2024){Kahle}, {Wyrowski}, {K{\"o}nig}, {Christensen}, {Tiwari}, \& {Menten}}]{kahle24}
{Kahle}, K.~A., {Wyrowski}, F., {K{\"o}nig}, C., {et~al.} 2024, \aap, 687, A162, \dodoi{10.1051/0004-6361/202349009}

\bibitem[{{Kehrig} {et~al.}(2012){Kehrig}, {Monreal-Ibero}, {Papaderos}, {V{\'\i}lchez}, {Gomes}, {Masegosa}, {S{\'a}nchez}, {Lehnert}, {Cid Fernandes}, {Bland-Hawthorn}, {Bomans}, {Marquez}, {Mast}, {Aguerri}, {L{\'o}pez-S{\'a}nchez}, {Marino}, {Pasquali}, {Perez}, {Roth}, {S{\'a}nchez-Bl{\'a}zquez}, \& {Ziegler}}]{kehrig12}
{Kehrig}, C., {Monreal-Ibero}, A., {Papaderos}, P., {et~al.} 2012, \aap, 540, A11, \dodoi{10.1051/0004-6361/201118357}

\bibitem[{{Kewley} {et~al.}(2019){Kewley}, {Nicholls}, \& {Sutherland}}]{kewley19}
{Kewley}, L.~J., {Nicholls}, D.~C., \& {Sutherland}, R.~S. 2019, \araa, 57, 511, \dodoi{10.1146/annurev-astro-081817-051832}

\bibitem[{{Kisielius} {et~al.}(2009){Kisielius}, {Storey}, {Ferland}, \& {Keenan}}]{kisielius09}
{Kisielius}, R., {Storey}, P.~J., {Ferland}, G.~J., \& {Keenan}, F.~P. 2009, \mnras, 397, 903, \dodoi{10.1111/j.1365-2966.2009.14989.x}

\bibitem[{{Kobayashi} {et~al.}(2020){Kobayashi}, {Karakas}, \& {Lugaro}}]{kobayashi20}
{Kobayashi}, C., {Karakas}, A.~I., \& {Lugaro}, M. 2020, \apj, 900, 179, \dodoi{10.3847/1538-4357/abae65}

\bibitem[{{Kollmeier} {et~al.}(2019){Kollmeier}, {Anderson}, {Blanc}, {Blanton}, {Covey}, {Crane}, {Drory}, {Frinchaboy}, {Froning}, {Johnson}, {Kneib}, {Kreckel}, {Merloni}, {Pellegrini}, {Pogge}, {Ramirez}, {Rix}, {Sayres}, {S{\'a}nchez-Gallego}, {Shen}, {Tkachenko}, {Trump}, {Tuttle}, {Weijmans}, {Zasowski}, {Barbuy}, {Beaton}, {Bergemann}, {Bochanski}, {Brandt}, {Casey}, {Cherinka}, {Eracleous}, {Fan}, {Garc{\'\i}a}, {Green}, {Hekker}, {Lane}, {Longa-Pe{\~n}a}, {Mathur}, {Meza}, {Minchev}, {Myers}, {Nidever}, {Nitschelm}, {O'Connell}, {Price-Whelan}, {Raddick}, {Rossi}, {Sankrit}, {Simon}, {Stutz}, {Ting}, {Trakhtenbrot}, {Weaver}, {Willmer}, \& {Weinberg}}]{kollmeier19}
{Kollmeier}, J., {Anderson}, S.~F., {Blanc}, G.~A., {et~al.} 2019, in Bulletin of the American Astronomical Society, Vol.~51, 274

\bibitem[{{Kollmeier} {et~al.}(2026){Kollmeier}, {Rix}, {Aerts}, {Aird}, {Alfaro}, {Almeida}, {Anderson}, {Arseneau}, {Assef}, {Aviram}, {Aydar}, {Badenes}, {Bandyopadhyay}, {Barger}, {Barkhouser}, {Bauer}, {Behmard}, {Bender}, {Besser}, {Bhattarai}, {Bilgi}, {Bird}, {Bizyaev}, {Blanc}, {Blanton}, {Bochanski}, {Bovy}, {Brandon}, {Brandt}, {Brownstein}, {Buchner}, {Burchett}, {Carlberg}, {Casey}, {Castaneda-Carlos}, {Chakraborty}, {Chanam{\'e}}, {Chandra}, {Cherinka}, {Chilingarian}, {Comparat}, {Cosens}, {Covey}, {Crane}, {Crumpler}, {Cruz-Gonzalez}, {Cunha}, {Cunningham}, {Dai}, {Darling}, {Davidson}, {Davis}, {De Lee}, {Deacon}, {M{\'e}ndez Delgado}, {Demasi}, {Demianenko}, {Derwent}, {D'Onghia}, {Di Mille}, {Dias}, {Donor}, {Dow}, {Drory}, {Dwelly}, {Egorov}, {Egorova}, {El-Badry}, {Engelman}, {Eracleous}, {Fan}, {Farr}, {Fries}, {Frinchaboy}, {Froning}, {G{\"a}nsicke}, {Garc{\'\i}a}, {Gelfand}, {Gentile Fusillo}, {Glover}, {Grabowski}, {Grebel}, {Green}, {Grier}, {Gupta}, {Gray}, {H{\"a}berle}, {Hall},
  {Hammond}, {Hawkins}, {Harding}, {Heged{\H{u}}s}, {Herbst}, {Hermes}, {Rodr{\'\i}guez Hidalgo}, {Hilder}, {Hogg}, {Holtzman}, {Horta}, {Huang}, {Hwang}, {Ibarra-Medel}, {Imig}, {Inight}, {Jana}, {Ji}, {Jim{\'e}nez-Arranz}, {Jofre}, {Johns}, {Johnson}, {Johnson}, {Johnston}, {Jones}, {Katkov}, {Knapp}, {Koekemoer}, {Kounkel}, {Kreckel}, {Krishnarao}, {Krumpe}, {Kumari}, {Kupfer}, {Lacerna}, {Laporte}, {Lepine}, {Li}, {Liu}, {Loebman}, {Long}, {Roman-Lopes}, {Lu}, {Majewski}, {Maoz}, {McKinnon}, {Medan}, {Merloni}, {Minniti}, {Morrison}, {Myers}, {M{\'e}sz{\'a}ros}, {Nandra}, {Nayak}, {Ness}, {Nidever}, {O'Brien}, {Oeur}, {Oravetz}, {Oravetz}, {Otto}, {Pallathadka}, {Palunas}, {Pan}, {Pappalardo}, {Pandey}, {Pe{\~n}aloza}, {Pinsonneault}, {Pogge}, {Taghizadeh Popp}, {Price-Whelan}, {Pulatova}, {Qiu}, {Ramirez}, {Rankine}, {Ricci}, {Runnoe}, {Sanchez}, {Salvato}, {Sarbadhicary}, {Sattler}, {Saydjari}, {Sayres}, {Schinnerer}, {Schlaufman}, {Schneider}, {Schreiber}, {Schwope}, {Serna}, {Shen}, {Sif{\'o}n},
  {Singh}, {Sinha}, {Smee}, {Song}, {Souto}, {Stassun}, {Steinmetz}, {Stone-Martinez}, {Stringfellow}, {Stutz}, {S{\'a}nchez-Gallego}, {Tan}, {Tayar}, {Thai}, {Thakar}, {Ting}, {Tkachenko}, {Tovmassian}, {Trakhtenbrot}, {Fern{\'a}ndez-Trincado}, \& {Troup}}]{kollmeier26}
{Kollmeier}, J.~A., {Rix}, H.-W., {Aerts}, C., {et~al.} 2026, \aj, 171, 52, \dodoi{10.3847/1538-3881/ae0576}

\bibitem[{{Konidaris} {et~al.}(2024){Konidaris}, {Herbst}, {Froning}, {Wachter}, {Ram{\'\i}rez}, {Kollmeier}, {Rix}, {Drory}, {Blanc}, {S{\'a}nchez-Gallego}, {Godoy Alfaro}, {Ahn}, {Aslan}, {Besser}, {Bilgi}, {Bizenberger}, {Bizyaev}, {Brady}, {Brown}, {Briegel}, {Flores Cabrales}, {Case}, {Donor}, {Feger}, {Gaessler}, {H{\"a}berle}, {Houston}, {Hull}, {Jones}, {Kim}, {Kowal}, {Kripak}, {Kreckel}, {Kuhlberg}, {Lanz}, {Lesser}, {Mathar}, {Mohr}, {Morales}, {Pak}, {Palunas}, {Pogge}, {Ritz}, {Robertson}, {Rodriguez}, {Rohloff}, {Uomoto}, {Yeong}, {Zapata}, \& {Zhelem}}]{konidaris24}
{Konidaris}, N.~P., {Herbst}, T., {Froning}, C., {et~al.} 2024, in Society of Photo-Optical Instrumentation Engineers (SPIE) Conference Series, Vol. 13096, Ground-based and Airborne Instrumentation for Astronomy X, ed. J.~J. {Bryant}, K.~{Motohara}, \& J.~R.~D. {Vernet}, 130961Z, \dodoi{10.1117/12.3019892}

\bibitem[{{Kreckel} {et~al.}(2016){Kreckel}, {Blanc}, {Schinnerer}, {Groves}, {Adamo}, {Hughes}, \& {Meidt}}]{kreckel16}
{Kreckel}, K., {Blanc}, G.~A., {Schinnerer}, E., {et~al.} 2016, \apj, 827, 103, \dodoi{10.3847/0004-637X/827/2/103}

\bibitem[{{Kreckel} {et~al.}(2024){Kreckel}, {Egorov}, {Egorova}, {Blanc}, {Drory}, {Kounkel}, {M{\'e}ndez-Delgado}, {Rom{\'a}n-Z{\'u}{\~n}iga}, {S{\'a}nchez}, {Stringfellow}, {Stutz}, {Zari}, {Barrera-Ballesteros}, {Bizyaev}, {Brownstein}, {Congiu}, {Fern{\'a}ndez-Trincado}, {Garc{\'\i}a}, {Hillenbrand}, {Ibarra-Medel}, {Jin}, {Johnston}, {Jones}, {Kim}, {Kollmeier}, {Kong}, {Krishnarao}, {Kumari}, {Li}, {Long}, {Mata-S{\'a}nchez}, {Mej{\'\i}a-Narv{\'a}ez}, {Popa}, {Rix}, {Sattler}, {Serna}, {Singh}, {S{\'a}nchez-Gallego}, {Wofford}, \& {Wong}}]{kreckel24}
{Kreckel}, K., {Egorov}, O.~V., {Egorova}, E., {et~al.} 2024, \aap, 689, A352, \dodoi{10.1051/0004-6361/202449943}

\bibitem[{{Kumar} \& {Anandarao}(2010)}]{kumar10}
{Kumar}, D.~L., \& {Anandarao}, B.~G. 2010, \mnras, 407, 1170, \dodoi{10.1111/j.1365-2966.2010.16964.x}

\bibitem[{{Levin}(2018)}]{nist}
{Levin}, I. 2018, {NIST Inorganic Crystal Structure Database (ICSD) 1.13.1}, \dodoi{https://doi.org/10.18434/M32147}

\bibitem[{{Liu} {et~al.}(2006){Liu}, {Barlow}, {Zhang}, {Bastin}, \& {Storey}}]{liu06}
{Liu}, X.~W., {Barlow}, M.~J., {Zhang}, Y., {Bastin}, R.~J., \& {Storey}, P.~J. 2006, \mnras, 368, 1959, \dodoi{10.1111/j.1365-2966.2006.10283.x}

\bibitem[{{Liu} {et~al.}(2000){Liu}, {Storey}, {Barlow}, {Danziger}, {Cohen}, \& {Bryce}}]{liu00}
{Liu}, X.~W., {Storey}, P.~J., {Barlow}, M.~J., {et~al.} 2000, \mnras, 312, 585, \dodoi{10.1046/j.1365-8711.2000.03167.x}

\bibitem[{{Luridiana} {et~al.}(2015){Luridiana}, {Morisset}, \& {Shaw}}]{luridiana15}
{Luridiana}, V., {Morisset}, C., \& {Shaw}, R.~A. 2015, \aap, 573, A42, \dodoi{10.1051/0004-6361/201323152}

\bibitem[{{Luridiana} {et~al.}(2009){Luridiana}, {Sim{\'o}n-D{\'\i}az}, {Cervi{\~n}o}, {Gonz{\'a}lez Delgado}, {Porter}, \& {Ferland}}]{luridiana09}
{Luridiana}, V., {Sim{\'o}n-D{\'\i}az}, S., {Cervi{\~n}o}, M., {et~al.} 2009, \apj, 691, 1712, \dodoi{10.1088/0004-637X/691/2/1712}

\bibitem[{{Maheson} {et~al.}(2024){Maheson}, {Maiolino}, {Curti}, {Sanders}, {Tacchella}, \& {Sandles}}]{maheson24}
{Maheson}, G., {Maiolino}, R., {Curti}, M., {et~al.} 2024, \mnras, 527, 8213, \dodoi{10.1093/mnras/stad3685}

\bibitem[{{Maiolino} \& {Mannucci}(2019)}]{maiolino19}
{Maiolino}, R., \& {Mannucci}, F. 2019, \aapr, 27, 3, \dodoi{10.1007/s00159-018-0112-2}

\bibitem[{{Mathis}(1990)}]{mathis90}
{Mathis}, J.~S. 1990, \araa, 28, 37, \dodoi{10.1146/annurev.aa.28.090190.000345}

\bibitem[{{Mejia}(in prep)}]{mejia25}
{Mejia}, A. in prep, {LVM Data Reduction pipeline}

\bibitem[{{M{\'e}ndez-Delgado} {et~al.}(2022){M{\'e}ndez-Delgado}, {Amayo}, {Arellano-C{\'o}rdova}, {Esteban}, {Garc{\'\i}a-Rojas}, {Carigi}, \& {Delgado-Inglada}}]{eduardo22}
{M{\'e}ndez-Delgado}, J.~E., {Amayo}, A., {Arellano-C{\'o}rdova}, K.~Z., {et~al.} 2022, \mnras, 510, 4436, \dodoi{10.1093/mnras/stab3782}

\bibitem[{{M{\'e}ndez-Delgado} {et~al.}(2023{\natexlab{a}}){M{\'e}ndez-Delgado}, {Esteban}, {Garc{\'\i}a-Rojas}, {Kreckel}, \& {Peimbert}}]{eduardo23Natur}
{M{\'e}ndez-Delgado}, J.~E., {Esteban}, C., {Garc{\'\i}a-Rojas}, J., {Kreckel}, K., \& {Peimbert}, M. 2023{\natexlab{a}}, \nat, 618, 249, \dodoi{10.1038/s41586-023-05956-2}

\bibitem[{{M{\'e}ndez-Delgado} {et~al.}(2023{\natexlab{b}}){M{\'e}ndez-Delgado}, {Esteban}, {Garc{\'\i}a-Rojas}, {Arellano-C{\'o}rdova}, {Kreckel}, {G{\'o}mez-Llanos}, {Egorov}, {Peimbert}, \& {Orte-Garc{\'\i}a}}]{eduardo23}
{M{\'e}ndez-Delgado}, J.~E., {Esteban}, C., {Garc{\'\i}a-Rojas}, J., {et~al.} 2023{\natexlab{b}}, \mnras, 523, 2952, \dodoi{10.1093/mnras/stad1569}

\bibitem[{{M{\'e}ndez-Delgado} {et~al.}(2024){M{\'e}ndez-Delgado}, {Kreckel}, {Esteban}, {Garc{\'\i}a-Rojas}, {Carigi}, {Sander}, {Palla}, {Chru{\'s}li{\'n}ska}, {De Looze}, {Rela{\~n}o}, {van der Giessen}, {Reyes-Rodr{\'\i}guez}, \& {S{\'a}nchez}}]{mendez-delgado24}
{M{\'e}ndez-Delgado}, J.~E., {Kreckel}, K., {Esteban}, C., {et~al.} 2024, \aap, 690, A248, \dodoi{10.1051/0004-6361/202450928}

\bibitem[{{Mesa-Delgado} \& {Esteban}(2010)}]{delgado10}
{Mesa-Delgado}, A., \& {Esteban}, C. 2010, \mnras, 405, 2651, \dodoi{10.1111/j.1365-2966.2010.16664.x}

\bibitem[{{Mesa-Delgado} {et~al.}(2009){Mesa-Delgado}, {Esteban}, {Garc{\'\i}a-Rojas}, {Luridiana}, {Bautista}, {Rodr{\'\i}guez}, {L{\'o}pez-Mart{\'\i}n}, \& {Peimbert}}]{delgado09}
{Mesa-Delgado}, A., {Esteban}, C., {Garc{\'\i}a-Rojas}, J., {et~al.} 2009, \mnras, 395, 855, \dodoi{10.1111/j.1365-2966.2009.14554.x}

\bibitem[{Mesa‐Delgado {et~al.}(2008)Mesa‐Delgado, Esteban, \& García‐Rojas}]{delgado08}
Mesa‐Delgado, A., Esteban, C., \& García‐Rojas, J. 2008, The Astrophysical Journal, 675, 389–404, \dodoi{10.1086/524296}

\bibitem[{{Morisset} {et~al.}(2020){Morisset}, {Luridiana}, {Garc{\'\i}a-Rojas}, {G{\'o}mez-Llanos}, {Bautista}, \& {Mendoza}}]{morisset20}
{Morisset}, C., {Luridiana}, V., {Garc{\'\i}a-Rojas}, J., {et~al.} 2020, Atoms, 8, 66, \dodoi{10.3390/atoms8040066}

\bibitem[{{Nicholls} {et~al.}(2012){Nicholls}, {Dopita}, \& {Sutherland}}]{nicholls12}
{Nicholls}, D.~C., {Dopita}, M.~A., \& {Sutherland}, R.~S. 2012, \apj, 752, 148, \dodoi{10.1088/0004-637X/752/2/148}

\bibitem[{{Oey} \& {Kennicutt}(1997)}]{oey97}
{Oey}, M.~S., \& {Kennicutt}, Jr., R.~C. 1997, \mnras, 291, 827, \dodoi{10.1093/mnras/291.4.827}

\bibitem[{{Osterbrock} \& {Ferland}(2006)}]{osterbrock06}
{Osterbrock}, D.~E., \& {Ferland}, G.~J. 2006, {Astrophysics of gaseous nebulae and active galactic nuclei}

\bibitem[{{Pathak} {et~al.}(2025){Pathak}, {Leroy}, {Thompson}, {Lopez}, {Barnes}, {Dale}, {Blackstone}, {Glover}, {Menon}, {Sutter}, {Williams}, {Baron}, {Belfiore}, {Bigiel}, {Bolatto}, {Boquien}, {Chandar}, {Chevance}, {Chown}, {Grasha}, {Groves}, {Klessen}, {Kreckel}, {Li}, {M{\'e}ndez-Delgado}, {Rosolowsky}, {Sandstrom}, {Sarbadhicary}, {Sun}, \& {{\'U}beda}}]{pathak25}
{Pathak}, D., {Leroy}, A.~K., {Thompson}, T.~A., {et~al.} 2025, \apj, 982, 140, \dodoi{10.3847/1538-4357/adb484}

\bibitem[{{Peimbert} \& {Peimbert}(2013)}]{peimbert13}
{Peimbert}, A., \& {Peimbert}, M. 2013, \apj, 778, 89, \dodoi{10.1088/0004-637X/778/2/89}

\bibitem[{{Peimbert}(1967)}]{peimbert67}
{Peimbert}, M. 1967, \apj, 150, 825, \dodoi{10.1086/149385}

\bibitem[{{Peimbert} {et~al.}(2017){Peimbert}, {Peimbert}, \& {Delgado-Inglada}}]{peimbert17}
{Peimbert}, M., {Peimbert}, A., \& {Delgado-Inglada}, G. 2017, \pasp, 129, 082001, \dodoi{10.1088/1538-3873/aa72c3}

\bibitem[{{Peimbert} {et~al.}(1991){Peimbert}, {Sarmiento}, \& {Fierro}}]{peimbert91}
{Peimbert}, M., {Sarmiento}, A., \& {Fierro}, J. 1991, \pasp, 103, 815, \dodoi{10.1086/132886}

\bibitem[{{Peimbert} {et~al.}(1993{\natexlab{a}}){Peimbert}, {Storey}, \& {Torres-Peimbert}}]{peimbert_storey93}
{Peimbert}, M., {Storey}, P.~J., \& {Torres-Peimbert}, S. 1993{\natexlab{a}}, \apj, 414, 626, \dodoi{10.1086/173108}

\bibitem[{{Peimbert} {et~al.}(1993{\natexlab{b}}){Peimbert}, {Torres-Peimbert}, \& {Dufour}}]{peimbert93}
{Peimbert}, M., {Torres-Peimbert}, S., \& {Dufour}, R.~J. 1993{\natexlab{b}}, \apj, 418, 760, \dodoi{10.1086/173433}

\bibitem[{{P{\'e}rez-Montero}(2017)}]{montero17}
{P{\'e}rez-Montero}, E. 2017, \pasp, 129, 043001, \dodoi{10.1088/1538-3873/aa5abb}

\bibitem[{{Podobedova} {et~al.}(2009){Podobedova}, {Kelleher}, \& {Wiese}}]{podobedova09}
{Podobedova}, L.~I., {Kelleher}, D.~E., \& {Wiese}, W.~L. 2009, Journal of Physical and Chemical Reference Data, 38, 171, \dodoi{10.1063/1.3032939}

\bibitem[{{Pottasch} \& {Bernard-Salas}(2013)}]{pottasch13}
{Pottasch}, S.~R., \& {Bernard-Salas}, J. 2013, \aap, 550, A35, \dodoi{10.1051/0004-6361/201219647}

\bibitem[{{Prisinzano} {et~al.}(2005){Prisinzano}, {Damiani}, {Micela}, \& {Sciortino}}]{prisinzano05}
{Prisinzano}, L., {Damiani}, F., {Micela}, G., \& {Sciortino}, S. 2005, \aap, 430, 941, \dodoi{10.1051/0004-6361:20040432}

\bibitem[{{Ramambason} {et~al.}(2022){Ramambason}, {Lebouteiller}, {Bik}, {Richardson}, {Galliano}, {Schaerer}, {Morisset}, {Polles}, {Madden}, {Chevance}, \& {De Looze}}]{ramambason22}
{Ramambason}, L., {Lebouteiller}, V., {Bik}, A., {et~al.} 2022, \aap, 667, A35, \dodoi{10.1051/0004-6361/202243866}

\bibitem[{{Rodr{\'\i}guez} \& {Garc{\'\i}a-Rojas}(2010)}]{rodriguez10}
{Rodr{\'\i}guez}, M., \& {Garc{\'\i}a-Rojas}, J. 2010, \apj, 708, 1551, \dodoi{10.1088/0004-637X/708/2/1551}

\bibitem[{{Ruiz} {et~al.}(2003){Ruiz}, {Peimbert}, {Peimbert}, \& {Esteban}}]{ruiz03}
{Ruiz}, M.~T., {Peimbert}, A., {Peimbert}, M., \& {Esteban}, C. 2003, \apj, 595, 247, \dodoi{10.1086/377255}

\bibitem[{{Sanchez} \& {Peimbert}(1991)}]{sanchez91}
{Sanchez}, L.~J., \& {Peimbert}, M. 1991, \rmxaa, 22, 285

\bibitem[{{S{\'a}nchez} {et~al.}(2025){S{\'a}nchez}, {Mej{\'\i}a-Narv{\'a}ez}, {Egorov}, {Kreckel}, {Drory}, {Blanc}, {M{\'e}ndez-Delgado}, {Barrera-Ballesteros}, {Ibarra-Medel}, {Bizyaev}, {Garc{\'\i}a}, {Wofford}, \& {Lugo-Aranda}}]{sanchez25}
{S{\'a}nchez}, S.~F., {Mej{\'\i}a-Narv{\'a}ez}, A., {Egorov}, O.~V., {et~al.} 2025, \aj, 169, 52, \dodoi{10.3847/1538-3881/ad93bb}

\bibitem[{{Sarbadhicary} {et~al.}(2025){Sarbadhicary}, {Long}, {Raymond}, {Sankrit}, {Egorov}, {Roman-Lopes}, {Blanc}, {Gelfand}, {Badenes}, {Drory}, {Fern{\'a}ndez-Trincado}, {Garc{\'\i}a}, {Johnston}, {Jones}, {Katkov}, {Kreckel}, {Li}, {Mej{\'\i}a-Narv{\'a}ez}, {M{\'e}ndez-Delgado}, {Orozco-Duarte}, {Sanchez}, \& {Wong}}]{sumit25}
{Sarbadhicary}, S.~K., {Long}, K.~S., {Raymond}, J.~C., {et~al.} 2025, arXiv e-prints, arXiv:2507.08257, \dodoi{10.48550/arXiv.2507.08257}

\bibitem[{{Sattler} {et~al.}(2025){Sattler}, {M{\'e}ndez-Delgado}, {Kreckel}, {Morisset}, {Egorov}, {Egorova}, {Nemer}, {Liang}, {Sander}, {Roman-Lopes}, {Rom{\'a}n-Z{\'u}{\~n}iga}, {Johnston}, {S{\'a}nchez}, {Fern{\'a}ndez-Trincado}, {Drory}, {Singh}, {Bizyaev}, {Sarbadhicary}, {Garc{\'\i}a}, {Mej{\'\i}a-Narv{\'a}ez}, \& {Blanc}}]{sattler25}
{Sattler}, N., {M{\'e}ndez-Delgado}, J.~E., {Kreckel}, K., {et~al.} 2025, arXiv e-prints, arXiv:2512.02802, \dodoi{10.48550/arXiv.2512.02802}

\bibitem[{{Schlafly} \& {Finkbeiner}(2011)}]{schlafly11}
{Schlafly}, E.~F., \& {Finkbeiner}, D.~P. 2011, \apj, 737, 103, \dodoi{10.1088/0004-637X/737/2/103}

\bibitem[{{Siebenmorgen} {et~al.}(2023){Siebenmorgen}, {Smoker}, {Kre{\l}owski}, {Gordon}, \& {Chini}}]{siebenmorgen23}
{Siebenmorgen}, R., {Smoker}, J., {Kre{\l}owski}, J., {Gordon}, K., \& {Chini}, R. 2023, \aap, 676, A132, \dodoi{10.1051/0004-6361/202244594}

\bibitem[{{Skillman} {et~al.}(2020){Skillman}, {Berg}, {Pogge}, {Moustakas}, {Rogers}, \& {Croxall}}]{skillman20}
{Skillman}, E.~D., {Berg}, D.~A., {Pogge}, R.~W., {et~al.} 2020, \apj, 894, 138, \dodoi{10.3847/1538-4357/ab86ae}

\bibitem[{{Stasinska}(2023)}]{stasinska23}
{Stasinska}, G. 2023, arXiv e-prints, arXiv:2312.01873, \dodoi{10.48550/arXiv.2312.01873}

\bibitem[{{Stasi{\'n}ska} {et~al.}(2007){Stasi{\'n}ska}, {Tenorio-Tagle}, {Rodr{\'\i}guez}, \& {Henney}}]{stasinska07}
{Stasi{\'n}ska}, G., {Tenorio-Tagle}, G., {Rodr{\'\i}guez}, M., \& {Henney}, W.~J. 2007, \aap, 471, 193, \dodoi{10.1051/0004-6361:20065675}

\bibitem[{{Storey} \& {Hummer}(1995)}]{storey95}
{Storey}, P.~J., \& {Hummer}, D.~G. 1995, \mnras, 272, 41, \dodoi{10.1093/mnras/272.1.41}

\bibitem[{{Storey} {et~al.}(2017){Storey}, {Sochi}, \& {Bastin}}]{storey17}
{Storey}, P.~J., {Sochi}, T., \& {Bastin}, R. 2017, \mnras, 470, 379, \dodoi{10.1093/mnras/stx1189}

\bibitem[{{Storey} \& {Zeippen}(2000)}]{storey00}
{Storey}, P.~J., \& {Zeippen}, C.~J. 2000, \mnras, 312, 813, \dodoi{10.1046/j.1365-8711.2000.03184.x}

\bibitem[{{Sutherland} {et~al.}(2013){Sutherland}, {Dopita}, {Binette}, \& {Groves}}]{sutherland13}
{Sutherland}, R., {Dopita}, M., {Binette}, L., \& {Groves}, B. 2013, {MAPPINGS III: Modelling And Prediction in PhotoIonized Nebulae and Gasdynamical Shocks}, Astrophysics Source Code Library, record ascl:1306.008

\bibitem[{{Tayal}(2011)}]{tayal11}
{Tayal}, S.~S. 2011, \apjs, 195, 12, \dodoi{10.1088/0067-0049/195/2/12}

\bibitem[{{Tayal} \& {Zatsarinny}(2010)}]{tayal10}
{Tayal}, S.~S., \& {Zatsarinny}, O. 2010, \apjs, 188, 32, \dodoi{10.1088/0067-0049/188/1/32}

\bibitem[{{Tiwari} {et~al.}(2020){Tiwari}, {Menten}, {Wyrowski}, {Giannetti}, {Lee}, {Kim}, \& {P{\'e}rez-Beaupuits}}]{tiwari20}
{Tiwari}, M., {Menten}, K.~M., {Wyrowski}, F., {et~al.} 2020, \aap, 644, A25, \dodoi{10.1051/0004-6361/202038886}

\bibitem[{{Toribio San Cipriano} {et~al.}(2016){Toribio San Cipriano}, {Garc{\'\i}a-Rojas}, {Esteban}, {Bresolin}, \& {Peimbert}}]{toribio16}
{Toribio San Cipriano}, L., {Garc{\'\i}a-Rojas}, J., {Esteban}, C., {Bresolin}, F., \& {Peimbert}, M. 2016, \mnras, 458, 1866, \dodoi{10.1093/mnras/stw397}

\bibitem[{{Torres-Peimbert} {et~al.}(1980){Torres-Peimbert}, {Peimbert}, \& {Daltabuit}}]{torres-peimbert80}
{Torres-Peimbert}, S., {Peimbert}, M., \& {Daltabuit}, E. 1980, \apj, 238, 133, \dodoi{10.1086/157966}

\bibitem[{{Torres-Peimbert} {et~al.}(1990){Torres-Peimbert}, {Peimbert}, \& {Pena}}]{torres-peimbert90}
{Torres-Peimbert}, S., {Peimbert}, M., \& {Pena}, M. 1990, \aap, 233, 540

\bibitem[{Tothill {et~al.}(2008)Tothill, Gagné, Stecklum, \& Kenworthy}]{tothill08}
Tothill, N. F.~H., Gagné, M., Stecklum, B., \& Kenworthy, M.~A. 2008, The Lagoon Nebula and its Vicinity.
\newblock \doarXiv{0809.3380}

\bibitem[{{Tsamis} {et~al.}(2003){Tsamis}, {Barlow}, {Liu}, {Danziger}, \& {Storey}}]{tsamis03}
{Tsamis}, Y.~G., {Barlow}, M.~J., {Liu}, X.~W., {Danziger}, I.~J., \& {Storey}, P.~J. 2003, \mnras, 338, 687, \dodoi{10.1046/j.1365-8711.2003.06081.x}

\bibitem[{{Tsamis} {et~al.}(2004){Tsamis}, {Barlow}, {Liu}, {Storey}, \& {Danziger}}]{tsamis04}
{Tsamis}, Y.~G., {Barlow}, M.~J., {Liu}, X.-W., {Storey}, P.~J., \& {Danziger}, I.~J. 2004, \mnras, 353, 953, \dodoi{10.1111/j.1365-2966.2004.08140.x}

\bibitem[{{Tsamis} {et~al.}(2008){Tsamis}, {Walsh}, {P{\'e}quignot}, {Barlow}, {Danziger}, \& {Liu}}]{tsamis08}
{Tsamis}, Y.~G., {Walsh}, J.~R., {P{\'e}quignot}, D., {et~al.} 2008, \mnras, 386, 22, \dodoi{10.1111/j.1365-2966.2008.13051.x}

\bibitem[{{Viegas} \& {Clegg}(1994)}]{viegas94}
{Viegas}, S.~M., \& {Clegg}, R.~E.~S. 1994, \mnras, 271, 993, \dodoi{10.1093/mnras/271.4.993}

\bibitem[{{Villa-Durango} {et~al.}(2025){Villa-Durango}, {Barrera-Ballesteros}, {Rom{\'a}n-Z{\'u}{\~n}iga}, {Moran}, {Ybarra}, {M{\'e}ndez-Delgado}, {Drory}, {Kreckel}, {Ibarra-Medel}, {S{\'a}nchez}, {Johnston}, {Roman-Lopes}, {Hernandez}, {Fern{\'a}ndez-Trincado}, {Stutz}, {Henney}, {Ghosh}, {Sarbadhicary}, {Lugo-Aranda}, {Bizyaev}, {Jones}, \& {Blanc}}]{villa25}
{Villa-Durango}, M.~A., {Barrera-Ballesteros}, J., {Rom{\'a}n-Z{\'u}{\~n}iga}, C.~G., {et~al.} 2025, \mnras, 543, 1196, \dodoi{10.1093/mnras/staf1530}

\bibitem[{{Wesson} {et~al.}(2003){Wesson}, {Liu}, \& {Barlow}}]{wesson03}
{Wesson}, R., {Liu}, X.-W., \& {Barlow}, M.~J. 2003, \mnras, 340, 253, \dodoi{10.1046/j.1365-8711.2003.06289.x}

\bibitem[{{Wright} {et~al.}(2019){Wright}, {Jeffries}, {Jackson}, {Bayo}, {Bonito}, {Damiani}, {Kalari}, {Lanzafame}, {Pancino}, {Parker}, {Prisinzano}, {Randich}, {Vink}, {Alfaro}, {Bergemann}, {Franciosini}, {Gilmore}, {Gonneau}, {Hourihane}, {Jofr{\'e}}, {Koposov}, {Lewis}, {Magrini}, {Micela}, {Morbidelli}, {Sacco}, {Worley}, \& {Zaggia}}]{wright19}
{Wright}, N.~J., {Jeffries}, R.~D., {Jackson}, R.~J., {et~al.} 2019, \mnras, 486, 2477, \dodoi{10.1093/mnras/stz870}

\bibitem[{{Wyse}(1942)}]{wyse42}
{Wyse}, A.~B. 1942, \apj, 95, 356, \dodoi{10.1086/144409}

\bibitem[{{Xiao} {et~al.}(2012){Xiao}, {Wang}, {Wang}, {Zhou}, {Lu}, \& {Dong}}]{xiao12}
{Xiao}, T., {Wang}, T., {Wang}, H., {et~al.} 2012, \mnras, 421, 486, \dodoi{10.1111/j.1365-2966.2011.20327.x}

\bibitem[{{Xing} {et~al.}(2026){Xing}, {Choustikov}, {Katz}, \& {Cameron}}]{Xing26}
{Xing}, L., {Choustikov}, N., {Katz}, H., \& {Cameron}, A.~J. 2026, arXiv e-prints, arXiv:2602.02949, \dodoi{10.48550/arXiv.2602.02949}

\bibitem[{{Zeippen}(1982)}]{zeippen82}
{Zeippen}, C.~J. 1982, \mnras, 198, 111, \dodoi{10.1093/mnras/198.1.111}

\end{thebibliography}

\clearpage
\appendix

\section{Emission line intensity maps and radial profiles}
\label{sec:line_maps}
In this section, we present the spatially resolved maps and radial profile of emission line intensities (in units of erg s$^{-1}$ cm$^{-2}$ arcsec$^{-2}$). In all plots, the upper panel shows: (a) the line intensity map overlaid with annular bins (each $3\arcmin$ wide for visualization; original bin width is $1\arcmin$), (b) the line intensity map with a detection threshold of  $3\sigma$ (spaxels below this threshold are masked, in white), and (c) the corresponding S/N map. The lower panel displays the radial intensity profile: blue points mark $3\sigma$ detection of the measured quantity(same as b), spaxels below this threshold (if any) are shown in grey, cyan points represent the median $\pm$ median absolute deviation (MAD) in radial bins of $1\arcmin$ in which at least $50\%$ spaxels have $S/N$ higher than the threshold, red stars correspond to annular binned measurements in the entire FOV, and the green rhombuses show annular binned measurements in NGC~6523, the elliptical region. The black dashed line shows the average measurement across the FOV, and the brown dashed line indicates the average measurement in the elliptical region. The central bin that contains the HG region and the Her\,36 star is marked in magenta.

\begin{figure*}[!ht]
\centering
\includegraphics[width=0.78\textwidth]{\detokenize{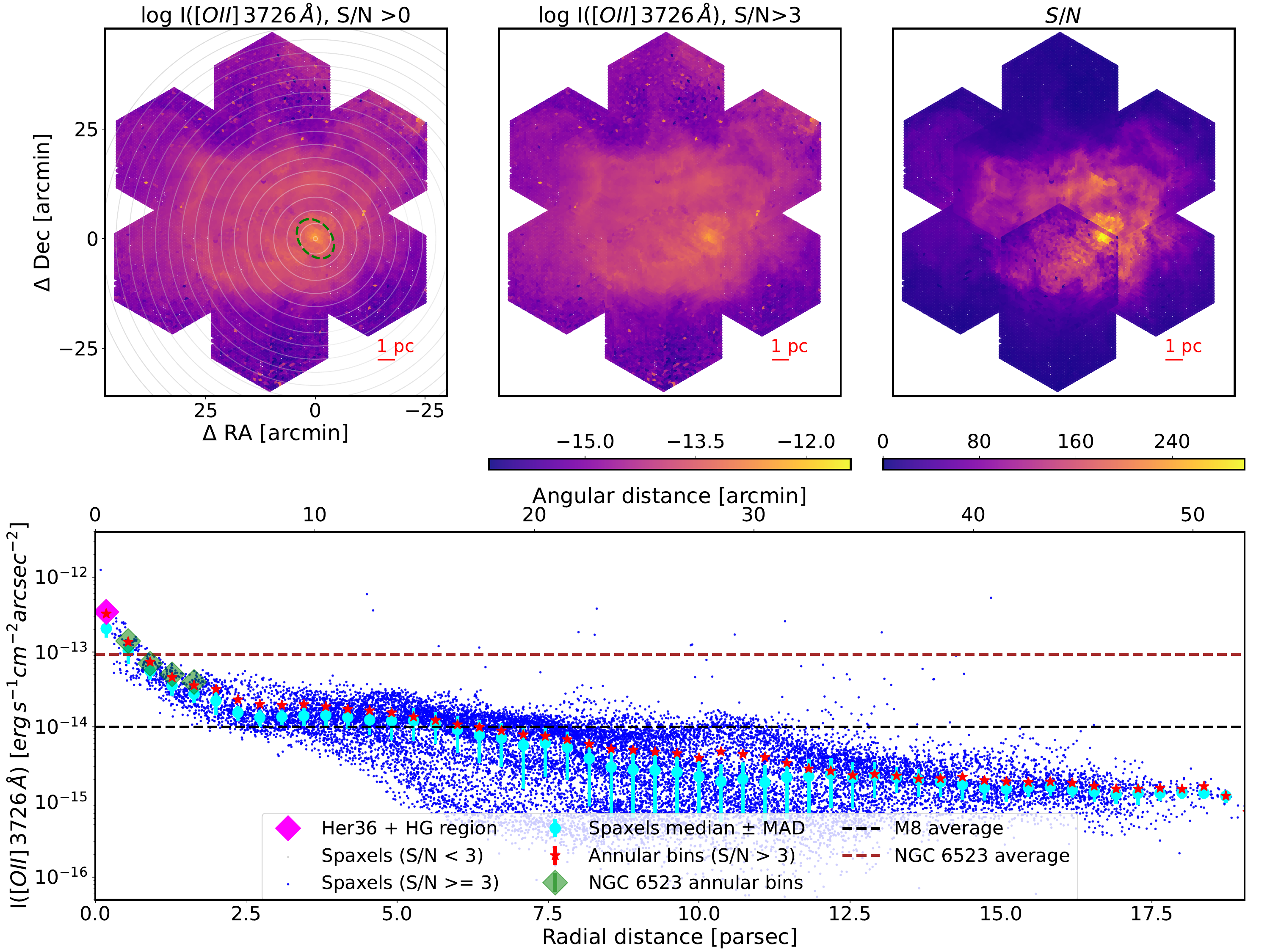}}

\caption{This figure shows \oii\wave3726 map and radial profiles as an example. 
The complete atlas of emission-line maps (27 images) is available as a figure set 
in the online journal (see \href{https://doi.org/10.5281/zenodo.19165622}{Zenodo, doi:10.5281/zenodo.19165622}).}
\label{fig:figure23}
\end{figure*}

\figsetstart

\figsetgrpstart
\figsetgrpnum{24.1}
\figsetgrptitle{\oii $\lambda3726$}
\figsetplot{OII3726radial_profile_snr_3_mean_radbin_for_paper_mar12.pdf}
\figsetgrpnote{Spatially resolved intensity map and radial profile of  \oii\wave3726.}
\figsetgrpend

\figsetgrpstart
\figsetgrpnum{24.2}
\figsetgrptitle{\oii $\lambda3729$}
\figsetplot{OII3729_radial_profile_snr_3_mean_radbin_for_paper_mar12.pdf}
\figsetgrpend

\figsetgrpstart
\figsetgrpnum{24.3}
\figsetgrptitle{\hk}
\figsetplot{HI3750radial_profile_snr_3_mean_radbin_for_paper_mar12.pdf}
\figsetgrpend

\figsetgrpstart
\figsetgrpnum{24.4}
\figsetgrptitle{\hi}
\figsetplot{HI3771radial_profile_snr_3_mean_radbin_for_paper_mar12.pdf}
\figsetgrpend

\figsetgrpstart
\figsetgrpnum{24.5}
\figsetgrptitle{\he}
\figsetplot{HI3835radial_profile_snr_3_mean_radbin_for_paper_mar12.pdf}
\figsetgrpend

\figsetgrpstart
\figsetgrpnum{24.6}
\figsetgrptitle{\hd}
\figsetplot{HI4102radial_profile_snr_3_mean_radbin_for_paper_mar12.pdf}
\figsetgrpend

\figsetgrpstart
\figsetgrpnum{24.7}
\figsetgrptitle{\hg}
\figsetplot{Hgm4340radial_profile_snr_3_mean_radbin_for_paper_mar12.pdf}
\figsetgrpend

\figsetgrpstart
\figsetgrpnum{24.8}
\figsetgrptitle{\oiii\wave4363}
\figsetplot{OIII4363_radial_profile_snr_3_mean_radbin_for_paper_mar12.pdf}
\figsetgrpend

\figsetgrpstart
\figsetgrpnum{24.9}
\figsetgrptitle{\orl\wave4638}
\figsetplot{OII4638radial_profile_snr_3_mean_radbin_for_paper_mar12.pdf}
\figsetgrpend

\figsetgrpstart
\figsetgrpnum{24.10}
\figsetgrptitle{\orl\wave4641}
\figsetplot{OII4641radial_profile_snr_3_mean_radbin_for_paper_mar12.pdf}
\figsetgrpend

\figsetgrpstart
\figsetgrpnum{24.11}
\figsetgrptitle{\orl\wave4649}
\figsetplot{OII4649radial_profile_snr_3_mean_radbin_for_paper_mar12.pdf}
\figsetgrpend

\figsetgrpstart
\figsetgrpnum{24.12}
\figsetgrptitle{\orl\wave4651}
\figsetplot{OII4651radial_profile_snr_3_mean_radbin_for_paper_mar12.pdf}
\figsetgrpend

\figsetgrpstart
\figsetgrpnum{24.13}
\figsetgrptitle{\orl\wave4662}
\figsetplot{OII4662radial_profile_snr_3_mean_radbin_for_paper_mar12.pdf}
\figsetgrpend

\figsetgrpstart
\figsetgrpnum{24.14}
\figsetgrptitle{\orl\wave4676}
\figsetplot{OII4676radial_profile_snr_3_mean_radbin_for_paper_mar12.pdf}
\figsetgrpend

\figsetgrpstart
\figsetgrpnum{24.15}
\figsetgrptitle{\orl V1 sum}
\figsetplot{oii_sumradial_profile_snr_3_mean_radbin_for_paper_mar12.pdf}
\figsetgrpend

\figsetgrpstart
\figsetgrpnum{24.16}
\figsetgrptitle{\hb}
\figsetplot{Hb4861radial_profile_snr_3_mean_radbin_for_paper_mar12.pdf}
\figsetgrpend

\figsetend

\end{document}